\documentclass[10pt]{article}
\usepackage[utf8]{inputenc}
\usepackage[a4paper, margin=0.8in]{geometry}
\usepackage[style=nature,doi=false,isbn=false,eprint=false,url=false,maxbibnames=99, ]{biblatex}
\DeclareFieldFormat{sentencecase}{\MakeSentenceCase{#1}}
\AtEveryCitekey{\clearfield{month}}
\AtEveryBibitem{\clearlist{language}}
\AtEveryBibitem{\clearlist{urldate}}

\DeclareUnicodeCharacter{2212}{-}
\DeclareUnicodeCharacter{0327}{}
\DeclareUnicodeCharacter{03B1}{$\alpha$}
\DeclareUnicodeCharacter{03B3}{$\gamma$}
\DeclareUnicodeCharacter{03BA}{$\kappa$}
\DeclareUnicodeCharacter{03B2}{$\beta$}

\DeclareUnicodeCharacter{2061}{}

\usepackage{tabularx}
\usepackage{array}
\usepackage{amsmath,amssymb}
\usepackage{booktabs}
\usepackage{makecell}
\usepackage{multirow}
\usepackage{tabulary}
\usepackage{mhchem}
\usepackage{tabularray}
\UseTblrLibrary{booktabs}
\usepackage{xcolor}
\usepackage[normalem]{ulem}
\usepackage{ulem}
\usepackage{siunitx}
\usepackage{diagbox}
\DeclareSIUnit\Molar{\textsc{m}}

\usepackage{graphicx}
\usepackage{caption}
\usepackage{subcaption}
\usepackage{float}
\usepackage{hyperref}
\usepackage[protrusion=true]{microtype}
\usepackage[capitalise,noabbrev]{cleveref}
\usepackage{fancyhdr}
\usepackage{authblk}
\usepackage{setspace}

\providecommand{\keywords}[1]
{
  \small	
  \textbf{\textit{Keywords---}} #1
}

\title{Mechanistic Modeling of Lipid Nanoparticle Formation for the Delivery of Nucleic Acid Therapeutics}

\author[1]{Pavan K. Inguva}
\author[1]{Saikat Mukherjee}
\author[2]{Pierre J. Walker}
\author[1]{Vico Tenberg}
\author[1]{Cedric Devos}
\author[1]{Sunkyu Shin}
\author[1]{Yanchen~Wu}
\author[1]{Srimanta Santra}
\author[1]{Jie Wang}
\author[1]{Shalini Singh}
\author[1]{Mona A. Kanso}
\author[3]{Shin Hyuk Kim}
\author[1]{Bernhardt~L.~Trout}
\author[1]{Martin~Z.~Bazant}
\author[1]{Allan S. Myerson}
\author[1]{Richard~D.~Braatz\thanks{Corresponding author. Email: braatz@mit.edu}}
\affil[1]{Department of Chemical Engineering, Massachusetts Institute of Technology, 77~Massachusetts Avenue, Cambridge, MA 02139, United States}
\affil[2]{Division of Chemistry and Chemical Engineering, California Institute of Technology, Pasedena, CA 91125, United States}
\affil[3]{Department of Biological and Chemical Engineering, Hanbat National University, Daejeon, Republic of Korea}

\date{}
\begin{document}

\maketitle

\begin{abstract}
Nucleic acids such as mRNA have emerged as a promising therapeutic modality with the capability of addressing a wide range of diseases. Lipid nanoparticles (LNPs) as a delivery platform for nucleic acids were used in the COVID-19 vaccines and have received much attention. While modern manufacturing processes which involve rapidly mixing an organic stream containing the lipids with an aqueous stream containing the nucleic acids are conceptually straightforward, detailed understanding of LNP formation and structure is still limited and scale-up can be challenging. Mathematical and computational methods are a promising avenue for deepening scientific understanding of the LNP formation process and facilitating improved process development and control. This article describes strategies for the mechanistic modeling of LNP formation, starting with strategies to estimate and predict important physicochemical properties of the various species such as diffusivities and solubilities. Subsequently, a framework is outlined for constructing mechanistic models of reactor- and particle-scale processes. Insights gained from the various models are mapped back to product quality attributes and process insights. Lastly, the use of the models to guide development of advanced process control and optimization strategies is discussed.
\end{abstract}


\section{Introduction}

Nucleic acid-based therapeutics (NATs) have emerged as an exciting modality with the capability of addressing a wide variety of indications such as genetic and oncological conditions, and for use in vaccines \parencite{damase_limitless_2021,wang_rna_2020,buck_lipid-based_2019}. {A diverse range of nucleic acid constructs with varying lengths and molecular structures have been explored such as plasmid DNA (pDNA), single- and double-stranded DNA (ss/dsDNA), anti-sense oligonucleotides (ASO), small interfering RNA (siRNA), and messenger RNA (mRNA) \parencite{damase_limitless_2021,ibraheem_gene_2014}}, with each construct type providing therapeutic effect in a specific way e.g., RNA interference or expression of an encoded protein. In many cases, the nucleic acid construct may also include various modifications either native (e.g., 5' capping and 3' polyadenylation for mRNA constructs) or non-native (e.g., chemical modifications of the sugar / nucleobase) at the level of individual nucleotides to the whole construct \parencite{kim_modifications_2022,mckenzie_recent_2021}. It is helpful to have an appreciation of the structure and chemistry of the construct, including modifications thereof, for modeling and analysis. 

Direct delivery of naked nucleic acid constructs remains challenging for multiple reasons: They tend to be unstable and prone to enzymatic degradation, the physicochemical properties of constructs (i.e., their large size and negative charge) impede cellular uptake, and exogenous nucleic acids can provoke an undesirable immunogenic effect \parencite{hamilton_biotechnology_2023,kulkarni_current_2021,damase_limitless_2021}. In combination with some of the construct-level modifications previously mentioned, several platform technologies for nucleic acid delivery such as \textit{N}-acetylgalactosamine (GalNAc)-RNA conjugation, viral vectors, lipid-based nanoparticles, and polymeric nanoparticles have been explored to improve efficacy and mitigate undesirable effects \parencite{byun_advances_2022,bulcha_viral_2021,kulkarni_current_2021,gupta_nucleic_2021}. Of the various platforms, non-viral vectors and in particular, lipid nanoparticles (LNPs) have received significant attention due to several advantages: comparative maturity of technology\parencite{buck_lipid-based_2019}, feasibility of rational design and modification of excipients and adjuvants to achieve desirable properties \parencite{damase_limitless_2021,de_jesus_solid_2015}, lower immunogenicty compared to viral vectors \parencite{buck_lipid-based_2019,yin_non-viral_2014}, and scalable and cost-efficient manufacturing \parencite{buck_lipid-based_2019,kulkarni_lipid_2018}. LNPs, while a mature technology, have undergone substantial development in the last decades. The interested reader is referred to  \parencite{xu_lipid_2022,buck_lipid-based_2019,cullis_lipid_2017} for an overview of the development of LNPs and the current state of the art. Modern LNP formulations for NATs are multi-component and typically consist of four types of lipids: an ionizable lipid, a PEGylated lipid, a helper lipid, and cholesterol. The design of the various lipids, in particular the ionizable lipid, is an active area of research as the physicochemical properties of the various can have significant impact on the safety and efficacy of the final product \parencite{hald_albertsen_role_2022,kon_principles_2022}. 

A range of manufacturing processes have been explored for the production of LNPs. Conventional techniques such as high pressure homogenization or the thin-film hydration method have been found to be unsatisfactory for NAT-LNPs for various reasons such as poor encapsulation efficiency, scalability, and potential damage to the sensitive nucleic acid construct \parencite{xu_lipid_2022,evers_stateart_2018}. The current state-of-the-art set-ups employs rapid mixing of an ethanolic stream containing the lipids with an aqueous stream containing the nucleic acids in microfluidic devices at lab scales and T-junctions / confined-impinging jets \parencite{evers_stateart_2018,erfle_stabilized_2019} at larger scales. These rapid mixing methods have been able to generate suitably high-quality LNPs with high encapsulation efficiencies \parencite{evers_stateart_2018}, finding use in the production of the LNPs for approved products e.g., the Pfizer--BioNTech Covid-19 vaccine \parencite{thorn_journey_2022}. However, rapid mixing processes are strongly impacted by the fluid dynamics in the mixer (which is a function of the geometry, formulation, and operating conditions) and need to be carefully engineered during process development \parencite{evers_stateart_2018,thorn_journey_2022,devos_impinging_2025}.

Considering the current and growing importance of LNP-based NATs \parencite{verma_landscape_2023}, understanding the LNP formation and manufacturing process is of immense importance to facilitate the production of high-quality drug products in a cost-efficient and scalable manner. To that end, computational modeling and simulations, in particular first-principles and mechanistic modeling, can play a significant role in advancing biomanufacturing. Three key benefits of incorporating modeling and simulations into the process development workflow are (1) improve fundamental scientific understanding of the process, (2) augment process development through guiding scale-up, process transfer, and optimization, and (3) improve quality control and process operation \parencite{destro_advanced_2024,narayanan_bioprocessing_2020,hong_challenges_2018,rantanen_future_2015,rogers_challenges_2015}. To our knowledge, there are limited studies in the literature applying computational methods to the manufacturing of LNP-based NATs with most relevant works applying molecular dynamics in the context of product design and formulation rather than to manufacturing \parencite{cardenas_review_2023,de_jesus_solid_2015}. 

This article aims to consolidate various computational modeling techniques that can be used to analyze LNP formation and manufacturing in the context of RNA therapeutics. { The methods and approaches discussed in this article can be extended to other nucleic acid constructs and similar non-viral delivery platforms e.g., polymeric nanoparticles. The specific nuances of the product/process should be considered when carrying out these extensions.} Many of the techniques draw upon expertise established in adjacent fields such as polymer precipitation and crystallization.  We show how different modeling approaches, ranging from comparatively simple methods at the length- and time-scales of the mixer, to more complex meso- and molecular-scale methods can provide valuable insights, and in some cases, even be used to predict important product and process characteristics. The use of these modeling approaches to inform the development of advanced process monitoring and control strategies is also discussed.  

\section{Product and Process Description}
\subsection{Product and Process Overview}

Modern NAT-LNP formulations employ multiple different lipids, each with specific physicochemical properties, that contribute to the efficacy of the LNPs. A summary of the key components present in the LNPs and their function(s) can be found in Table~\ref{tbl:species}. The structure of LNPs is complex and highly dependent on many factors such as the formulation (i.e., types of lipids and proportions) \parencite{mendonca_design_2023,eygeris_deconvoluting_2020}, the size of the nucleic acid fragment and its loading in the LNP \parencite{leung_microfluidic_2015}, and the manufacturing process and operating conditions \parencite{cheng_induction_2023,daniel_quality_2022,hassett_impact_2021,hu_kinetic_2019}. While the structure of the NAT-LNP and its formation is not fully understood, the current consensus of NAT-LNP structures is that there is an electron-dense lipid core in the LNP where the nucleic acid, majority of the ionizable lipid, and some water are present while the surface is rich in PEGylated and helper lipids \parencite{schoenmaker_mrna-lipid_2021}. Some NAT-LNP formulations, in particular older generation cationic/neutral liposomes, may incorporate fewer types of lipids in the final product. LNPs for other drug classes such as small molecules also typically  consist of fewer lipid components \parencite{ickenstein_lipid-based_2019}. These simpler systems can be used as a foundation for model development and validation as they are comparatively more established in the literature and are easier to model as there are fewer components with complex physicochemical properties and may have simpler structures.

\begin{table}[ht]
\centering
\caption{Summary of species present in LNPs. The interested reader is referred to \parencite{hald_albertsen_role_2022,cheng_role_2016} for excellent reviews on the function of the different lipid components in the LNP.}
\vspace{-0.2cm}
\scriptsize
    \begin{tabular}{cccc}
    \toprule
     Species & Function(s) & Examples \\
     \hline 
     Ionizable lipid & \makecell[c]{Enables efficient nucleic acid encapsulation, \\ Facilitates cellular uptake of cargo, \\ Adjuvant} & \makecell[c]{DLin-MC3-DMA, SM-102, \\ ALC-0315} \\
     \hline
     PEGylated lipid & \makecell[c]{Controls LNP particle size, \\Enhances product stability, \\ Enhance in vivo circulation time}  & \makecell[c]{PEG-DMG, PEG-DSPE,\\ PEG-DSG, PEG-DMPE} \\
     \hline
     Helper lipid & \makecell[c]{Improves LNP structural integrity, \\ Improves encapsulation efficiency} & DSPC, DOPE \\
     \hline
     Cholesterol &  \makecell[c]{Improves LNP stability, \\ Promotes membrane fusion} & \makecell[c]{Cholesterol, $\beta$-Sitosterol, \\ Stigmastanol, Vitamin D3}\\
     \hline
     Nucleic acid cargo &  Active pharmaceutical ingredient  & \makecell[c]{BNT-162b2 (4284 base pairs)}  \\
     \hline
     Water & Residual & -- \\
     \bottomrule
    \end{tabular}
    \label{tbl:species}
\end{table}

\begin{figure}[htb]
    \centering
    \begin{subfigure}{0.3\textwidth}
        \centering
        \includegraphics[width=\textwidth,trim = 0cm 1cm 0cm 0cm,clip]{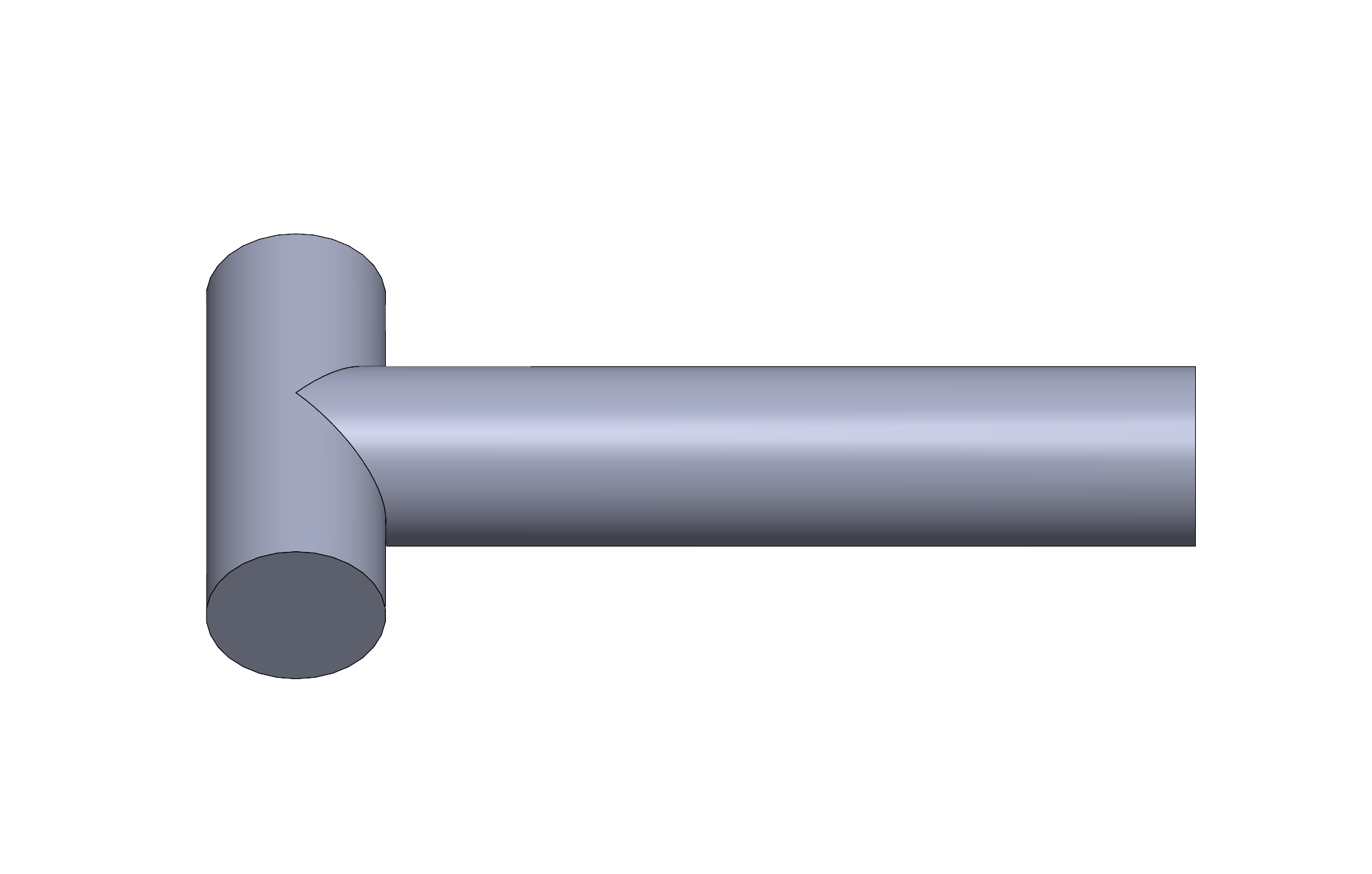} 
        \vspace{-0.2cm}
        \caption{T-Junction}
    \end{subfigure}
    \hfill
    \begin{subfigure}{0.3\textwidth}
        \centering
        \includegraphics[width=\textwidth,trim = 0cm 1cm 0cm 0cm,clip]{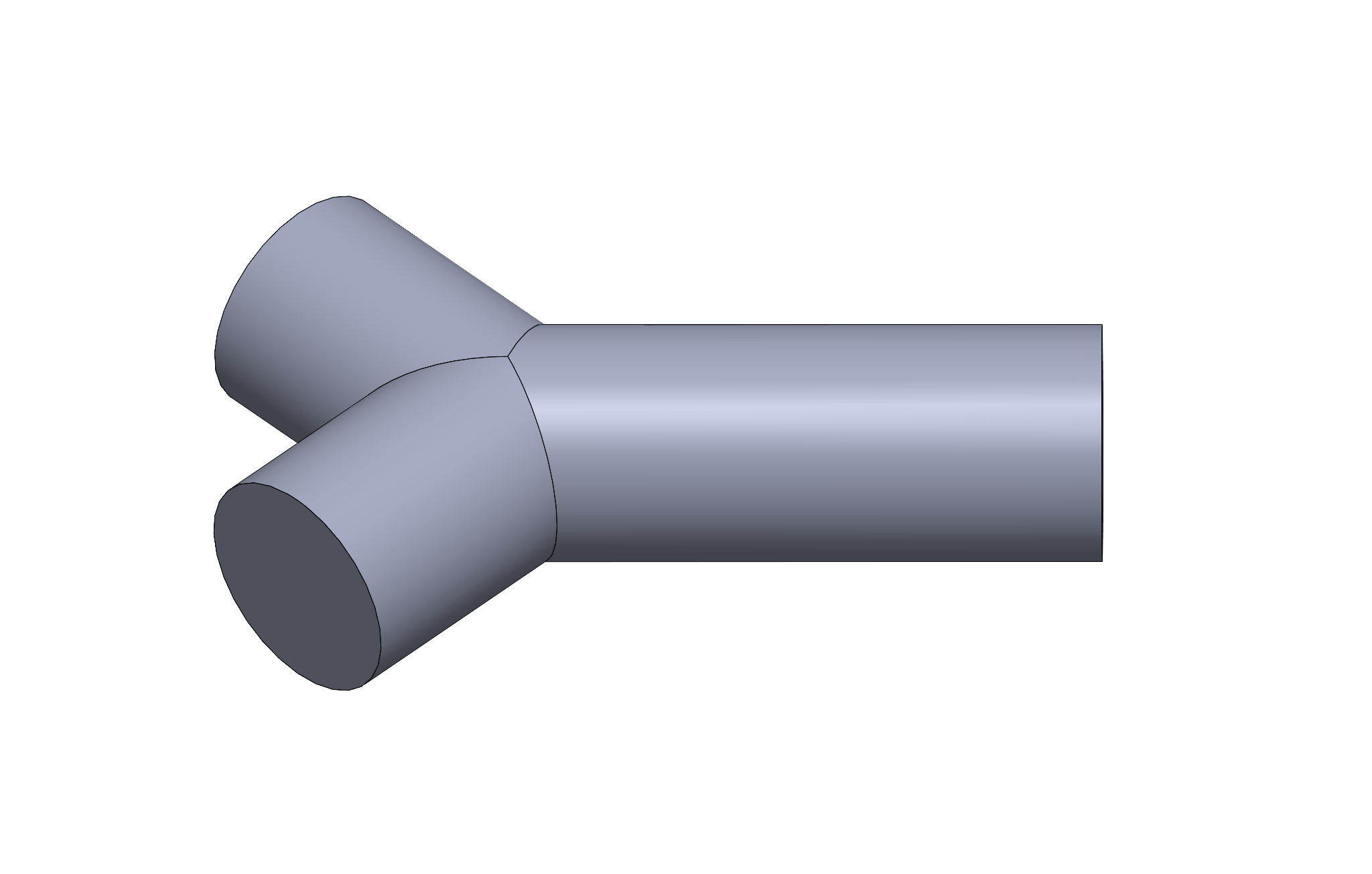}
        \vspace{-0.2cm}
        \caption{Y-Junction}
    \end{subfigure}
    \hfill
    \begin{subfigure}{0.3\textwidth}
        \centering
        \includegraphics[width=\textwidth,trim = 0cm 1cm 0cm 0cm,clip]{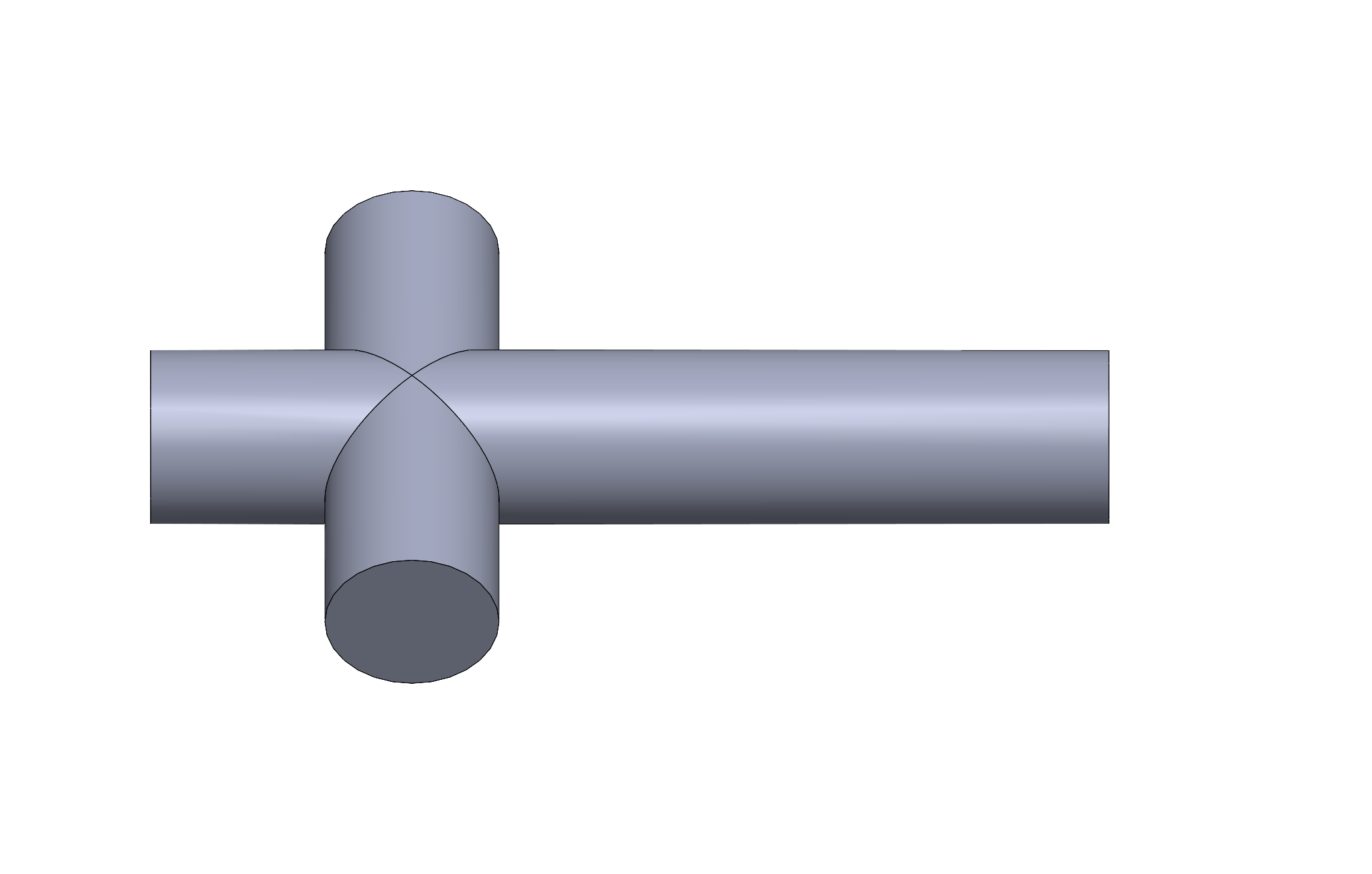}
        \vspace{-0.2cm}
        \caption{Cross-Junction}
    \end{subfigure}
    \hfill
    \begin{subfigure}{0.3\textwidth}
        \centering
        \includegraphics[width=\textwidth,trim = 0cm 1cm 0cm 0cm,clip]{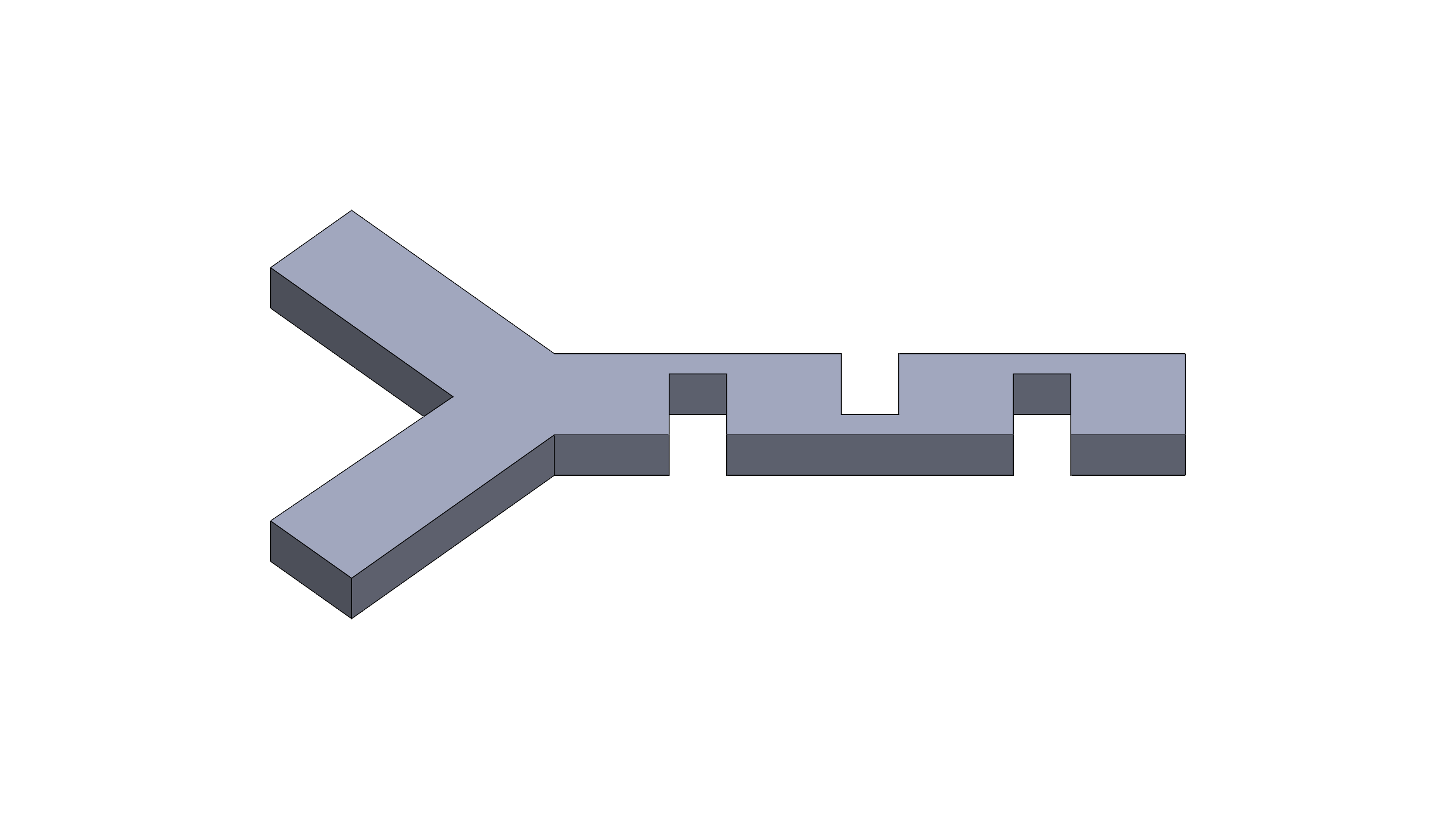}
        \vspace{-0.2cm}
        \caption{Baffled Mixer}
    \end{subfigure}
    \begin{subfigure}{0.3\textwidth}
        \centering
        \includegraphics[width=\textwidth,trim = 0cm 1cm 0cm 0cm,clip]{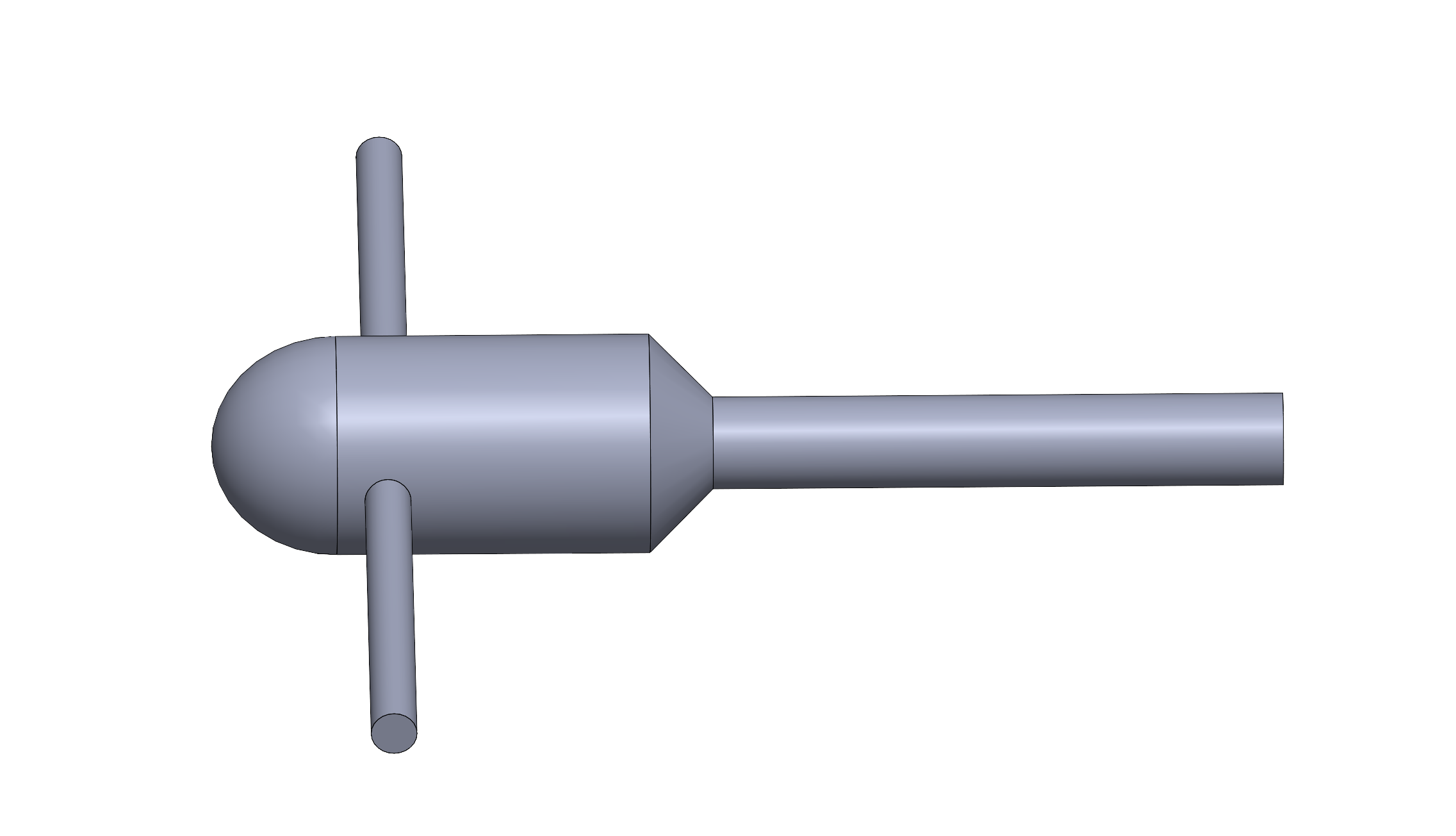}
        \vspace{-0.2cm}
        \caption{Confined Impinging Jet Mixer}
    \end{subfigure}
    \begin{subfigure}{0.3\textwidth}
        \centering
        \includegraphics[width=\textwidth]{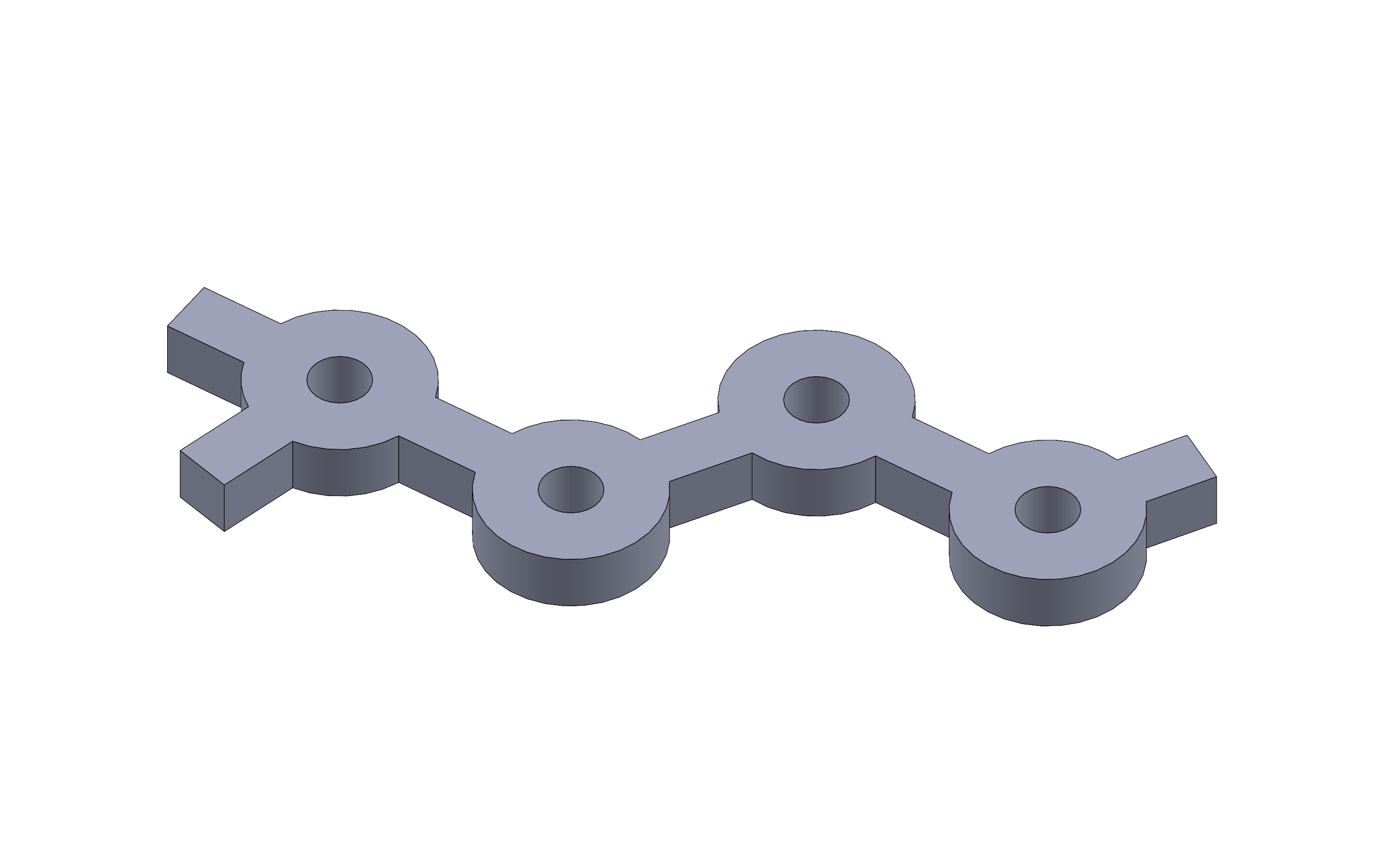}
        \vspace{-0.2cm}
        \caption{Ring Mixer}
    \end{subfigure}
    \caption{{Exemplar mixer geometries for rapid mixing of organic and aqueous streams for LNP manufacturing. Various modifications, such as flow constrictions and baffles, can be made upstream and/or downstream of the mixing point. For many microfluidic systems, rectangular geometries are typically employed due to the mixer manufacturing process.}}
    \label{fig:mixer_geom}
\end{figure}

Current approaches to designing mixing processes for LNP production aim for rapid mixing which helps to achieve a smaller and more uniform LNP size distribution with less aggregation \parencite{zhigaltsev_bottom-up_2012}. The typical rapid mixing process involves combining two streams: (1) an aqueous acidic buffer stream containing the nucleic acids and (2) an organic stream (typically ethanol) containing the lipids. { Upon mixing, not only does the solvent polarity change which decreases the solubility of the various species, the various species associate, thus causing the formation of the LNPs which are precipitated out of solution. Precipitation in this context refers to the formation of the LNPs, which constitute a new phase in the system.} A range of mixer configurations and geometries have been explored in the literature, e.g., a T-junction, Y-junction, and cross junction, at various scales ranging from pipe fittings ($\mathcal{O}$(\SI{1}{\centi\metre})) for large-scale production, to microfluidic devices ($\mathcal{O}$(\SI{100}{\micro\metre})) for lab-scale production \parencite{evers_stateart_2018,maeki_advances_2018}. Exemplar mixer geometries can be found in Figure~\ref{fig:mixer_geom}. In many cases, geometrical elements such as baffles, bends, and internal structures (e.g., staggered herringbone) can be incorporated upstream and/or downstream of the mixing point. These elements enhance mixing by introducing flow phenomena such as flow turning, flow splitting, and vortex generation, which result in chaotic/turbulent flows even at low Reynolds numbers \parencite{inguva_computer_2018,evers_stateart_2018}. 

{Immediately after the rapid mixing step, a buffer exchange is carried out to achieve four purposes: 1) remove the ethanol which can destabilize the LNPs \parencite{hardianto_effect_2023,kimura_development_2020}, 2) increase the pH of the system to physiological pH (typically around 7.4) which neutralizes the ionizable lipids, 3) concentrate the bulk product \parencite{wu_process_2025}, and 4) remove excess lipids / unencapsulated cargo. Physically, three key variables change during this process, namely the ethanol concentration decreases, the pH increases, and the ionic strength increases \parencite{cheng_induction_2023}. The buffer exchange step can be performed by various types of methods such as dialysis and tangential flow filtration, with process scale influencing the choice of method. The buffer exchange step can influence various aspects of LNPs -- including their size distribution, surface charge, and internal structure -- via a fusion process whereby smaller LNPs fuse together to form larger particles. This fusion process is moderated by the PEGylated lipids occurring during the buffer exchange \parencite{kamanzi_quantitative_2024,gilbert_evolution_2024,vargas_dialysis_2023,kulkarni_fusion-dependent_2019}. Although this perspective focuses on the rapid mixing step and discusses the formulation of mechanistic models in that context, it is likely that many of the same methods are applicable to understanding the buffer exchange process, with suitable adaptations.}

\begin{table}[ht]
\scriptsize
\centering

\caption{Exemplar NAT-LNP formulations and process operation conditions. The following abbreviations are used: FRR (Flowrate ratio of aqueous to organic streams), IL (Ionizable lipid), PL (PEGylated lipid), Ch (Cholesterol), HL (Helper lipid).}

\vspace{-0.2cm}

    \begin{tabular}{cccc}
    \toprule
     \makecell[c]{Process and\\product description} & \makecell[c]{Mixer\\(length scale)} & \makecell[c]{Stream compositions\\and flowrates} & Other comments \\
     \hline  
     \makecell[c]{Production of siRNA\\LNPs for luciferase\\expression  \\ \parencite{chen_rapid_2012}.} &  \makecell[c]{Rectangular\\staggered herringbone\\(\SI{70}{\micro \metre} $\times$ \SI{200}{\micro \metre})} & \makecell[c]{Organic (\SI{}{\milli\gram \per \milli \litre}):\\ IL(2), HL(0.28), Ch(0.52), PL(0.13)  \\ Aqueous: \SI{0.4}{\milli\gram \per \milli \litre} siRNA  \\ FRR: 1 \\ Total flowrate: \SIrange{0.1}{1}{\milli \litre \per \minute}} & \makecell[c]{An additional PBS buffer\\stream was fed further\\downstream in the mixer}   \\
     \hline
     \makecell[c]{Production of siRNA LNPs\\for FVII suppression\\ \parencite{kimura_development_2018}.} &  \makecell[c]{Rectangular\\baffled mixer\\(\SI{100}{\micro \metre} $\times$ \SI{200}{\micro \metre})} & \makecell[c]{Organic (\SI{}{\milli \Molar}):\\ IL(3.96), Ch(3.96), PL(0.08)  \\ Aqueous: \SI{0.071}{\milli\gram \per \milli \litre} siRNA  \\ FRR: 3--9 \\ Total flowrate: \SIrange{0.05}{0.5}{\milli \litre \per \minute}} & \makecell[c]{--}  \\
     \hline
     \makecell[c]{Production of LNPs\\with mLuc mRNA\\ \parencite{strelkova_petersen_mixing_2023}.} & \makecell[c]{NanoAssemblr\textsuperscript{\textregistered}\\Benchtop}  & \makecell[c]{Organic (Molar ratio):\\ IL(35), Ch(46.5), HL(16), PL(2.5)  \\ Aqeuous: \SI{0.05}{\milli\gram \per \milli \litre} mRNA  \\ FRR: 1--3 \\ Total flowrate: \SIrange{4}{14}{\milli \litre \per \minute}} & \makecell[c]{IL--mRNA ratio\\is 10:1(w/w)}  \\
     \bottomrule
    \end{tabular}
    \renewcommand{\arraystretch}{1}
    \label{tbl:stream_composition}
\end{table}

\subsection{Analytical Techniques and Quality Attributes}

A necessary component of formulating models for any manufacturing process is an understanding of the analytical tools available for characterizing the product/process and the features of the information these tools are able to provide. Not only are these analytical tools valuable for generating experimental data for model parameter estimation and validation, they also provide the ability to study the process/product to gain deeper physical insights which guide model development. In addition, process analytical technologies (PAT) is essential for developing monitoring and control strategies to ensure product quality (see \textcite{narayanan_bioprocessing_2020} and Section~\ref{sec:control}). 

\begin{figure}[htbp]
    \centering
    \includegraphics[width= 0.8\linewidth]{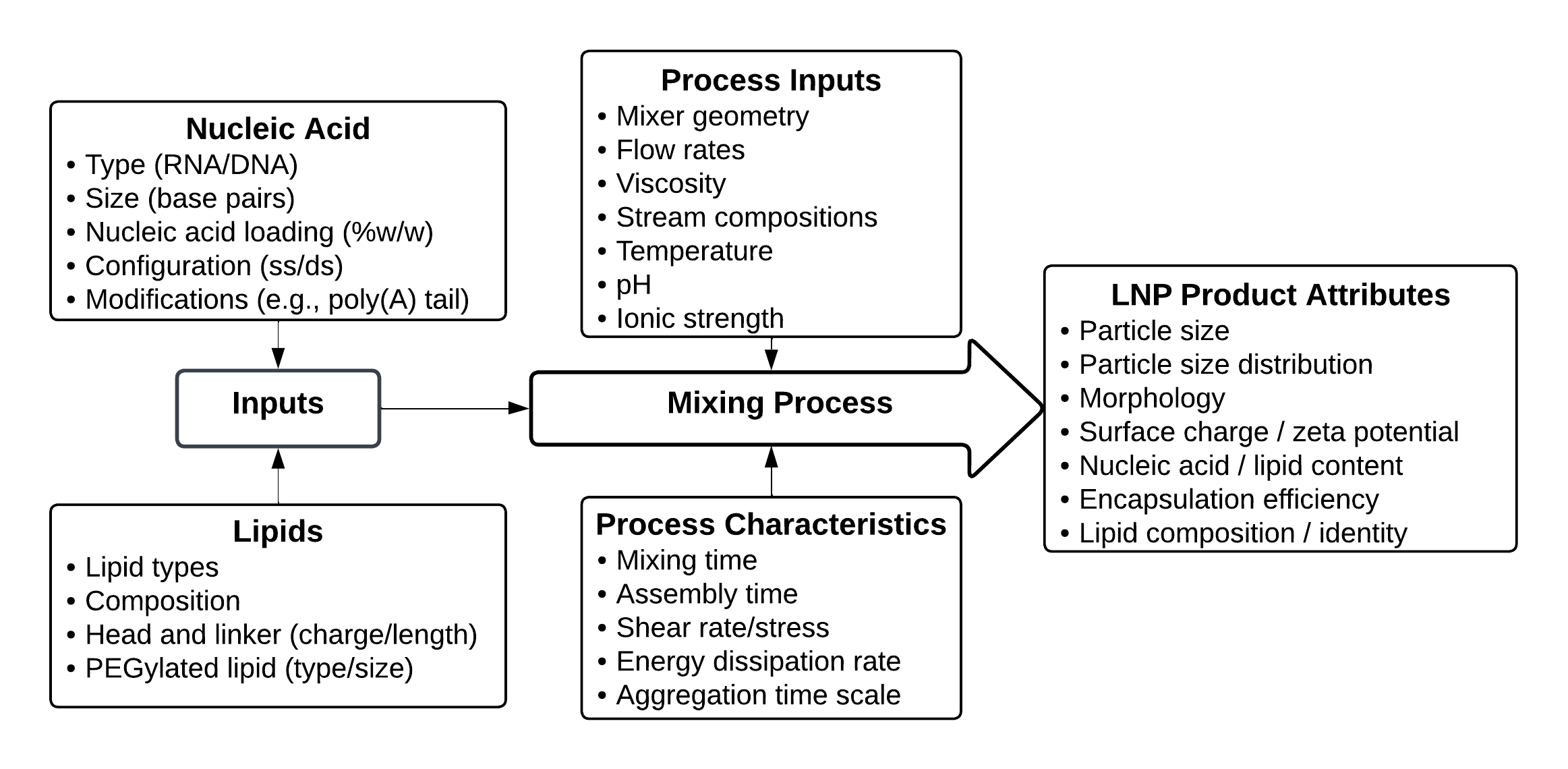}

    \vspace{-0.5cm}
    
    \caption{Summary of important material, process, and product variables and attributes affecting LNP manufacturing and product quality.}
    \label{fig:cqa_schematic}
\end{figure}

A summary of material and process inputs/variables, and product quality attributes that are regarded as significant in the production of LNPs is outlined in Figure~\ref{fig:cqa_schematic}. The mapping of available analytical techniques to characterize LNP product quality attributes is presented in Table~\ref{tbl:cqa}. Currently, with the exception of particle size measurements, almost all analytical techniques used for characterizing LNPs and their formation are offline (i.e., a sample is taken for analysis, often performed manually and away from the production line). The cost and/or complexity of many of the measurement techniques currently employed result in limited data availability (due to limited/infrequent sampling) and large uncertainties in measured data. Advances in both sensor technology and application are going to be vital in facilitating progress for both developing a better physical understanding of LNP formation and for model formulation. One particular technology that we believe has potential for rapid mixing systems is hyperspectral imaging which can be used to provide inline spatially resolved chemical spectral data, potentially enabling a non-contact approach to probing physicochemical changes in the mixer in real time as LNPs are being formed.

\begin{table}[ht]
\centering
\caption{{ Manufacturing-related product and process attributes of LNPs and corresponding analytical techniques available to measure relevant properties. Where possible, references are related to NAT-LNPs specifically. This table is not intended to be exhaustive and there are rapid advances currently happening in the analytical methods to characterize LNPs. The interested reader is referred to \textcite{nogueira_analytical_2024,daniel_quality_2022,malburet_size_2022,fan_analytical_2021} for additional information.}}
\scriptsize
\vspace{-0.2cm}

    \begin{tabular}{ccc}
    \toprule
     \makecell[c]{Quality attribute / \\ Process variable} & Analytical techniques \\
     \hline 
     \makecell[c]{Particle size / \\ Size distribution}   & \makecell[c]{Electron microscopy (EM) imaging (offline) \\ Dynamic Light Scattering (DLS) (offline) \\ Spatially resolved DLS (inline) \parencite{besseling_new_2019} \\ Taylor Dispersion Analysis (TDA) (offline) \parencite{malburet_taylor_2023} \\
     Nanoparticle Tracking Analysis (NTA) (offline) \parencite{dragovic_sizing_2011} \\ Convex Lens-induced Confinement microscopy (CLiC) microscopy (offline) \parencite{kamanzi_simultaneous_2021} } \\
     \hline
     \makecell[c]{Particle morphology \\ and structure}  & \makecell[c]{Cryo TEM (offline) \\ NMR spectroscopy (offline) \\ Differential scanning calorimetry (offline) \parencite{eygeris_deconvoluting_2020} \\ Small-Angle X-ray Scattering (SAXS) (offline) \parencite{uebbing_investigation_2020} \\ Small-Angle Neutron Scattering (offline) \parencite{gilbert_evolution_2024} }  \\
     \hline
     \makecell[c]{Nucleic acid content \ \\ encapsulation efficiency} & \makecell[c]{Ribogreen assay (offline) \\ Size-exclusion chromatography (offline) \\ Cylindrical Illumination Confocal Spectroscopy (CICS) (offline) \parencite{li_payload_2022} \\ CLiC (offline) \parencite{kamanzi_simultaneous_2021}} \\
     \hline
     Surface charge &  \makecell[c]{Electrophoretic light scattering (offline) \\ Capillary electrophoresis (offline) \parencite{franzen_physicochemical_2011}}  \\
      \hline
     Chemical composition &  \makecell[c]{Liquid Chromatography -- Mass Spectroscopy (offline) \parencite{parot_quality_2024} \\ Liquid Chromatography -- Corona Charged Aerosol Detection (offline) \parencite{kinsey_determination_2022} \\ Single Particle Automated Raman Trapping Analysis (SPARTA) (offline) \parencite{barriga_coupling_2022} }  \\
     \hline
     \makecell[c]{Physicochemical changes \\ in the mixer} & \makecell[c]{Spectroscopy (e.g., Raman) (inline/online) \\ Hyperspectral imaging (online) \parencite{kise_submillisecond_2014}}  \\ 
     
     \bottomrule
    \end{tabular}
    \label{tbl:cqa}
\end{table}

\clearpage
\section{Physical Properties and Thermodynamics}
\subsection{Transport Properties}

The physical properties (e.g., viscosity and diffusivity) of the various LNP constituents and of the whole particle itself are important for characterizing the behaviour of the different species and are important model parameters / inputs to the models described subsequently. This section reviews how transport properties, can be obtained/estimated for both the molecular LNP constituents and the LNP particle itself.

\subsubsection{Molecular Properties} 

The best way to obtain numerical values for various transport properties such as the viscosity, diffusivity, and radius of gyration would be through experimental measurements. While some of these values have been determined experimentally (e.g., see \parencite{vargas_mechanism_2005,tadakuma_imaging_2006,lifland_dynamics_2011,cui_impact_2014,gallud_time_2021,soong_lateral_2005,aliakbarinodehi_interaction_2022,ermilova_ionizable_2023,jeon_anomalous_2012,baumler_diffusion_2017,brake_formation_2005,pilz_studies_1972,perelman_spontaneous_2023,fee_prediction_2004,chen_formation_2019,sugiura_formation_2001, armstrong_nanosecond_2014,baker_dynamics_2014,kessel_interactions_2001,goncalves_pvt_2010,swindells_absolute_1952}), the context in which the values were obtained are not identical, making it difficult to predict the `true' value within the full LNP system, particularly when considering the composition and temperature-dependence \parencite{schoenmaker_mrna-lipid_2021} of these properties. Further, these values were measured in bulk systems and the transport properties within the LNP are expected to be quite different, due to the higher concentration of lipids \parencite{evans_surface_1980, hayashi_phase_1975}. For modeling the self-assembly of LNPs, bulk transport properties would be a good start. 

In the case where experimental values are lacking and the objective is to extrapolate to the operating conditions, we need an alternative, predictive approach to estimate the transport properties. It is here where theoretical approaches are invaluable in estimating values and providing some understanding. In conventional liquids, the diffusion coefficient ($D_{i}$) of isolated spherical particles governed by their Brownian motion is well-described by the Stokes−-Einstein relation\parencite{andreoli_membrane_1980},
\begin{equation}
\label{eq:SE}
D_{i} = \frac{kT}{6\pi r_i \eta_0},
\end{equation}
where $r_i$ is the particle hydrodynamic radius (which vastly exceeds that of solvent molecules), $k$ is the Maxwell--Boltzmann constant, $T$ is temperature, and $\eta_0$ is the solvent viscosity. While this equation can work well for spherical, neutral species, for many of the species involved in LNPs, it is unlikely that the Stokes--Einstein equation can be used reliably without empirically modifying the hydrodynamic radius. Numerous theoretical \parencite{onsagerZurTheorieElectrolyte1926,onsagerZurTheorieElectrolyte1927,bernardElectrophoreticMobilityPolyelectrolyte1991,muthukumarDynamicsPolyelectrolyteSolutions1997,liuEffectElectrostaticInteractions1998,skibinskaEffectElectrostaticInteractions1999} and computational \parencite{fongOnsagerTransportCoefficients2020,fongIonCorrelationsTheir2021} studies over the past few decades have been carried out to better understand the transport properties of such species. In particular, \textcite{muthukumarDynamicsPolyelectrolyteSolutions1997} provides an excellent summary of the expected scaling laws for various transport properties (diffusion coefficients, viscosity, radius of gyration, etc.) in different regimes (dilute, semi-dilute, and concentrated). Some of the relationships that could be applied to the LNP system are summarized in Table~\ref{tbl:scale}. These scaling laws provide an intuition for how the transport properties might be expected to scale as a function of the species' properties and during the LNP self-assembly process. Unfortunately, the scaling relationships are only applicable within the systems and regimes that they have been derived for \parencite{karatrantos_polymer_2017,kalathi_nanoparticle_2014,rudyak_molecular_2011,maldonado-camargo_scale-dependent_2017,yuan_effect_2022,cicuta_diffusion_2007,vaz_translational_1985}. As discussed earlier, while the self-assembly process will primarily occur in the bulk where the concentrations are dilute, once the LNP has formed, transport within the particle will be very different. Further, obtaining quantitatively accurate estimates of transport properties using these approaches is both computationally costly and, due to approximations made in these studies, unlikely.

Aside from performing fully atomistic molecular dynamics simulations, which carries its own challenges (which are discussed in Sec.~\ref{sec:md_sims}), another way to obtain an accurate estimate of transport properties would be to use empirical correlations. In the case of the LNP components, very few correlations exist \parencite{lopezDiffusionViscosityUnentangled2021}, particularly for the larger species (mRNA and some of the lipids). As such, more generalized correlation methods may need to be considered. In the case of transport properties, one approach would be to use entropy-scaling methods \parencite{rosenfeldRelationTransportCoefficients1977,bellModifiedEntropyScaling2019} which assume that any transport property can be related to the residual entropy of the system. Historically, these methods have been very effective for large alkanes \parencite{jagerResidualEntropyScaling2023}, some mixtures \parencite{lotgering-linPureSubstanceMixture2018}, and some charged systems \parencite{melfiViscosityImidazoliumIonic2024}. While extrapolating to the LNP system might be challenging, the true limitation to using such a method is a lack of experimental data. It may be possible to fit these correlations of structurally similar systems (such as dilute polyelectrolytes), from which extrapolation could be more reliable. Such an exploration is a deserving topic for future study.

\begin{table}[ht]
\scriptsize
\centering
\caption{ Scaling relationships of the radius of gyration ($R_g$), diffusion coefficient ($D$), and viscosity ($\eta$) of a polyelectrolyte within the dilute regime \parencite{muthukumarDynamicsPolyelectrolyteSolutions1997}: $l$ is the Kuhn length, $w$ is the excluded volume parameter, $l_B$ is the Bjerrum length of the solvent, $\kappa$ is the inverse Debye screening length, $N$ is the chain length, and $c$ is the polyelectrolyte concentration. Such scaling relationships can be applied to species within an LNP system.}
\label{tbl:scale}
\begin{tabular}{cc}
\toprule
Property & Scaling\\
\hline
\multirow{2}{*}{Radius of Gyration ($6R_g^2/L$)} & high salt: $\left(\frac{4}{3l^2}\sqrt{\frac{3}{2\pi}}\left(w+\frac{4\pi l_{\!B}}{\kappa^2}\right)\right)^{\!2/5}N^{1/5}l\nonumber$ \\       & low salt: $\left(\frac{4\pi l_{\!B}}{2\sqrt{6}\pi^{5/2}l}\right)^{\!2/3}Nl  $ \\
& \\
Diffusion coefficient ($D$) & $\frac{8}{3\sqrt{\pi}}\frac{kT}{6\pi\eta_0R_g}\nonumber$ \\
& \\
\multirow{2}{*}{Viscosity ($\frac{\eta-\eta_0}{\eta_0}$)} & high salt: $c\!\left(w+\frac{4\pi l_{\!B}}{\kappa^2}\right)^{\!3/5}l^{6/5}N^{4/5}\nonumber$ \\
           & low salt: $cl_{\!B}l^2N^2$ \\
\bottomrule
\end{tabular}
\end{table}

\subsubsection{Particle Properties} 

Particle-scale transport properties of an LNP particle can influence its transport and stability, which can then impact subsequent manufacturing steps, as well as its \textit{in vivo} behaviour. Two key particle-scale properties are particle diffusivity and surface charge. 

Particle diffusivity can be observed as both translational and rotational diffusivity. Translational diffusivity ($D_t$) is related to the translational motion of particles, while rotational diffusivity ($D_r$) is associated with the rotational motion of the particle, measuring how quickly the particle can rotate or reorient itself within a fluid \parencite{kittel_introduction_2005}. As LNPs are not motile, both translational and rotational diffusion are caused by Brownian motion in the fluid. The translational diffusivity of the LNP particle is important for reactor-scale transport simulations (i.e., computational fluid dynamics and population balance models) which require the diffusivity of all species being tracked, including the LNPs (e.g., see Sec.~\ref{sec:spec_tra}). For non-spherical/anisotropic LNPs, which have been experimentally observed in some LNP formulations \parencite{brader_encapsulation_2021,kloczewiak_biopharmaceutical_2022}, the orientation and orientability of the particle, which is characterized by the rotational diffusivity, can be significant. For example, the cellular attachment and uptake of anisotropic nanoparticles can be affected by the particle shape and orientation \parencite{lovegrove_flow_2023}.  

Limited experimental knowledge exists regarding LNP diffusivity, particularly rotational diffusivity. Techniques employed for similar systems, such as viral particles, could also be used for LNPs.{A range of experimental techniques have been able to measure translational diffusivity such as photon-correlation spectroscopy \parencite{oliver_diffusion_1976}, Taylor dispersion analysis \parencite{malburet_taylor_2023}, nanoparticle tracking analysis \parencite{dragovic_sizing_2011}, and CLiC microscopy \parencite{kamanzi_simultaneous_2021}.} Similarly, rotational diffusivity can be measured using techniques such as light scattering \parencite{king_translational_1973,lehner_determination_2000,wada_rotational_1971}, transient electric birefringence \parencite{okonski_characterization_1956}, and flow birefringence \parencite{boedtker_preparation_1958}. These experimental measurements can be challenging and costly, motivating theoretical approaches for property estimation. For spherical particles, the translational and rotational diffusivities can be calculated using the Stokes--Einstein equation \eqref{eq:SE} and the Stokes--Einstein--Debye equation ($D_r=\frac{kT}{8\pi\eta_0 r^3}$) respectively \parencite{unni_fast_2021}. In the case of non-spherical/anisotropic particles, these theoretical relationships are not directly applicable and more-complicated methods are necessary (e.g., see \parencite{de_kee_general_2022,kanso_macromolecular_2019,kanso_coronavirus_2020}). 

{LNP surface charge is a critical physicochemical property \parencite{us_food_and_drug_administration_liposome_2018} as it can impact the ability of the LNP to deliver its cargo, as well as influence the toxicity and immunogenicity of the final product, with high surface charges being undesirable \parencite{sharma_immunostimulatory_2024,gueguen_evaluating_2024}.} Most pertinently, the particle charge can also influence its transport properties (e.g., particle diffusivity and aggregation characteristics) which need to be accounted for during manufacturing, storage, and transportation \parencite{du_effect_2010}. The surface charge is a function of several factors, such as LNP composition, particle size, and buffer composition \parencite{cardellini_thermal_2016}. Direct measurement of the surface charge is typically not possible, and instead, the zeta ($\zeta$) potential, which represents the potential difference between the dispersion medium and the stationary layer of fluid attached to the particle \parencite{barba_polymeric_2019}, is commonly reported. Experimental techniques that can be used to evaluate the $\zeta$ potential / surface charge are outlined in Table~\ref{tbl:cqa}. Experimentally reported $\zeta$ potential values for mRNA-loaded LNPs range between $-2.5$ and $-20$ mV for pH values between 7 and 8, and between $+8$ and $+27$ mV for pH values between 4 to 6 \parencite{larson_ph-dependent_2022,malburet_size_2022}. Higher absolute zeta potential values, indicative of increased electrostatic repulsion, enhance stability by preventing particle aggregation \parencite{retamal_marin_zeta_2017}. This electrostatic force also plays a role in influencing particle size distribution, settling behavior, and agglomeration tendencies.

Theoretical models can be used to estimate the $\zeta$ potential of a particle. Electrokinetic models such as Smoluchowski's, Hückel's, and Henry's delve into how the electrophoretic mobility can be related to the $\zeta$ potential \parencite{retamal_marin_zeta_2017}. Unfortunately, theoretical models often rely on simplifications and may be more suitable for simple colloidal systems or conditions.

\subsection{Thermodynamic Properties}
\label{sec:thermo}

\begin{figure}
    \centering
    \includegraphics[width=\textwidth]{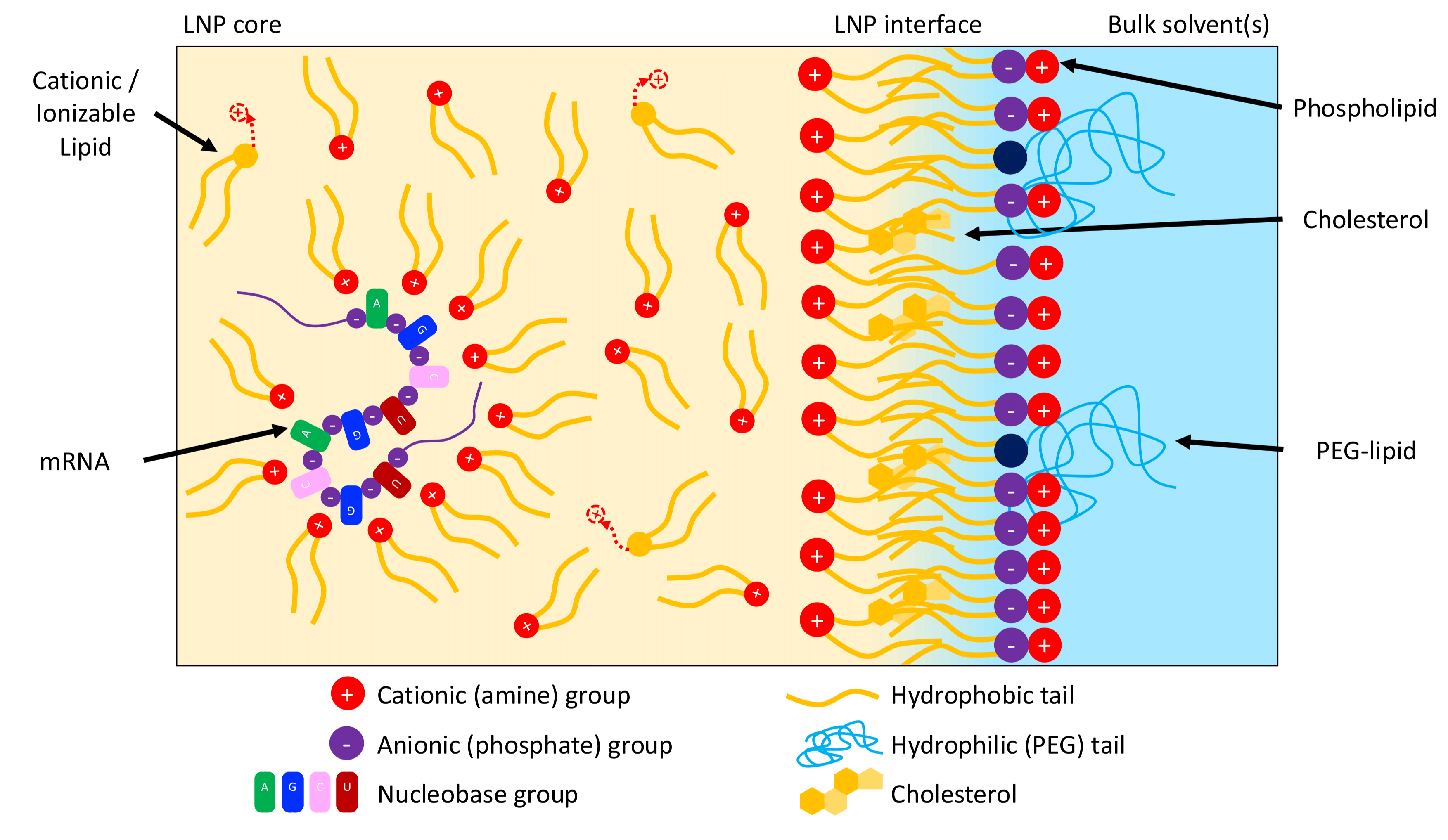} 
    \caption{{Simplified representation of LNP interface. The distribution of components along the interface is not representative of what might occur in a real system as it will depend strongly on the formulation.}}
    \label{fig:lnp_simple}
\end{figure}

Figure~\ref{fig:lnp_simple} is a simplified schematic of the structure of the LNP interface. { The ionizable lipid (IL) plays two key roles in the formation of LNPs: mRNA complexation and formation of the LNP interface. Its ionizable head group is crucial in these processes \parencite{hajjToolsTranslationNonviral2017,kowalskiDeliveringMessengerAdvances2019,mengNanoplatformsMRNATherapeutics2021}. At low pH, below their pK$_a$, ILs are charged, promoting mRNA encapsulation and aiding nanoparticle assembly. However, this charge can destabilize the LNP interface. As pH increases to physiological levels, ILs deprotonate, becoming neutral, which enhances interface stability and reduces interactions with anionic blood cell membranes, lowering toxicity. Upon cellular uptake, the endosomes which have an acidic environment causes ILs to re-protonate, thus destabilizing the LNP and facilitating cargo release. The ionizable nature of ILs is fundamental to LNP function and must be carefully considered in mechanistic modeling, as their pK$_a$ and structures are key design parameters influencing delivery efficiency and stability.}

However, the IL alone is not capable of forming a stable interface. As such, phospholipids, cholesterol, and PEGylated lipids are added to enhance the formation and stability of the interface. While this system may appear quite complex, the essential building blocks of each component share some commonalities in terms of their molecular interactions. The ionizable, phospho- and PEGylated lipids all include hydrophobic tails which interact favorably to form the interface. The ionizable and helper lipid both include cationic (typically amine) groups, while the mRNA and helper lipid include anionic (typically phosphate) groups. The electrostatic interactions between mRNA and the ionizable lipid are responsible for their initial complexation. Additionally, the PEG tail of the PEGylated lipid will experience hydrophilic interactions, and the nucleic acids can form hydrogen bonds due to the presence of the nucleobase groups. A more in-depth summary of the role of each component can be found in \parencite{eygeris_chemistry_2022,hald_albertsen_role_2022}.

To predict any thermodynamic property of interest for LNPs, a thermodynamic model is needed that is representative of the LNP system. As of the writing of this perspective, no such model has been developed. The building blocks mentioned previously, however, are not unique to LNPs. Systems such as polyelectrolytes, for which numerous thermodynamic models have been developed, often involve some or all of these building blocks. For polyelectrolytes, activity coefficient models such as extensions of the Non-Random Two-Liquid (NRTL) \parencite{yu_nonrandom_2019,li_modeling_2021} model have been used to predict activity coefficients of aqueous polyelectrolytes. Unfortunately, as highlighted by \textcite{zhang_research_2023}, such approaches are likely to perform poorly when considering denser systems (such as the LNP core) due to the treatment of the long-range electrostatic interactions. Other activity coefficient model approaches such as UNIFAC and COSMO-based approaches, despite their improved predictive capabilities, are likely to have the same limitations.

Thermodynamic perturbation theories are an alternative to activity coefficient models. In the case of polyelectrolytes, Zhang and co-workers \parencite{zhang_salting-out_2016,zhang_interfacial_2021} have developed a simple liquid-state (LS) theory capable of predicting complexation between chains of opposite and asymmetric charge, similar to the expected behavior between the ionizable lipid and mRNA, and has been shown to provide accurate representation of experimental systems. In principle, this approach could be used to estimate the partition coefficient of mRNA between the two phases, with the added benefit of being capable of modeling the interfacial properties of the system \parencite{zhang_interfacial_2021}.

One limitation of LS approach is the neglection of hydrophobic and hydrophilic interactions, which are likely to play a vital role in the formation and stabilization of the LNP interface. { In the case of the IL, its hydrophobicity is expected to vary depending on its deprotonation state, becoming more hydrophobic when the IL is neutral at physiological pH.} Approaches such as the Statistical Association Fluid Theory (SAFT) equation of state \parencite{chapman_saft_1989,chapman_new_1990} are more-suited to modeling such interactions, as demonstrated by their ability to accurately model bulk and equilibrium properties of hydrophobic+hydrophilic systems \parencite{gross_perturbed-chain_2001,lafitte_accurate_2013,papaioannou_group_2014}. 

Given both the SAFT and LS approaches are expressed as perturbation-free-energy expansions, in principle, it is possible to combine them, leading to a thermodynamic models, which accurately characterises the LNP system:
\begin{equation}
    A(V,T,\mathbf{n}) = A_\mathrm{id}+A_\mathrm{disp}+A_\mathrm{chain}+A_\mathrm{assoc}+A_\mathrm{ele}\,,
\end{equation}
where $A$, $V$, and $T$ are the Helmholtz free energy, volume and temperature of the system, respectively, and $\mathbf{n}$ is a vector containing the molar amounts of each component. The subscripts denote the ideal (id), dispersive/hydrophobic (disp), chain formation (chain), association/hydrophilic (assoc) and electrostatic (ele) contributions. With a Helmholtz free energy expression, following from the Gibbs-phase rule, it is possible to obtain any equilibrium property. For example, the activity coefficients ($\gamma_i$) of each species can be obtained from their chemical potential ($\mu_i$), which itself can be obtained as derivatives of the Helmholtz free energy:
\begin{equation}
    \mu_i = \!\left(\frac{\partial A}{\partial n_i}\right)_{\!\!V,T}\rightarrow \ \gamma_i = \frac{1}{x_i}\exp\!{\left(\frac{\mu_i-\mu_i^*}{RT}\right)}\,,
\end{equation}
where $x_i$ is the mole fraction of species $i$ and the superscript * denotes a property relating to a pure system of species $i$. 

{ In the case of the ionizable lipid, an additional constraint needs to be included in the formulation of the thermodynamic model to account for its deprotonatable site. While an association theory like SAFT could be used to model this behavior \parencite{perdomoPredictiveGroupcontributionFramework2023}, a more accurate approach would be to explicitly account for the reversible reaction:
\begin{center}
    \ce{IL + H$_3$O+ <=> HIL+ + H$_2$O}
\end{center}
This approach explicitly models the two charge states of the IL, as well as the hydronium ion. The equilibrium concentrations of each species in the presence of this reaction can be determined by including a modified version of the Henderson--Hasselbach equation:
\begin{equation}
    \mathrm{pH} = \mathrm{pK}_a + \log_{10} \frac{a_\mathrm{IL}^*}{a_{\mathrm{HIL}^+}^*}\,,
\end{equation}
where $a_i^*$ is the activity of species $i$ in which the reference system is identical to that used to obtain the pK$_a$. The inclusion of this equation accounts for the effects of system pH and the pK$_a$ of the IL on any of the thermodynamic properties of interest for the system.}

One of the more-vital thermodynamic properties, when considering the LNP system, is the relative solubility of mRNA between the LNP core and the bulk phase. This partition coefficient ($K_i$) can be obtained as
\begin{equation}
    K_\mathrm{mRNA} = \frac{x_\mathrm{mRNA}^\mathrm{LNP}}{x_\mathrm{mRNA}^\mathrm{bulk}}=\frac{\gamma_\mathrm{mRNA}^\mathrm{bulk}}{\gamma_\mathrm{mRNA}^\mathrm{LNP}}\,.
\end{equation}
Ideally, $K_\mathrm{mRNA}$ should be as large as possible to maximize the solubility of mRNA within the LNP. To even estimate the partition coefficient, the equilibrium compositions must be obtained in both the LNP and bulk phases at a given set of conditions such as pressure ($p_0$), temperature ($T_0$), and initial composition ($\mathbf{n}_0$). This information is obtained by minimizing the Gibbs free energy of the system:
\begin{equation}
    \min G(p_0,T_0,n_0)\,,
\end{equation}
This optimization presents itself as deceptively simple. In reality, multiple sophisticated algorithms have been proposed to solve this optimization \parencite{michelsen_isothermal_1982,pereira_held_2012}, made more challenging through the introduction of charged species \parencite{nikolaidis_rigorous_2022} { and reversible reactions \parencite{perezcisnerosReactiveSeparationSystems1997}}. The optimization can be simplified to the solution of two nonlinear algebraic equations: 
\begin{align}
    \sum_i \frac{n_i(1-K_i\exp(Z_i\psi))}{1+\phi(K_i\exp(Z_i\psi)-1)} &= 0\,, \\
    \sum_i \frac{n_iZ_i}{1+\phi(K_i\exp(Z_i\psi)-1)} &= 0\,,
\end{align}
which ensure that the mass balance and charge neutrality is satisfied, respectively, where $Z_i$ is the charge of species $i$, $\phi$ is the phase fraction of the LNP phase, and $\psi$ is the electrochemical potential difference between the two phases. The set of $K_i$ which satisfy these equations will correspond to the partition coefficients at equilibrium. 

Performing the above calculations can pose a significant challenge. However, the implementation of the free energy expressions and the methods needed to use them has been abstracted away in projects such as Clapeyron.jl \parencite{walker_clapeyronjl_2022} (and derivative packages such as cDFT.jl) which provide all the necessary tools to obtain the relevant properties. 

Nevertheless, one challenge remains whenever trying to use such thermodynamic models: parameterization. While the approach described above would be suitable for modeling LNP systems, the true limitation lies in obtaining the parameters which represent the system. These parameters are typically obtained through regression using experimental data. In the case of SAFT-type equations, the Perturbed-Chain SAFT (PC-SAFT) equation \parencite{gross_perturbed-chain_2001} has been used to model similar systems with some success \parencite{reschke_modeling_2014,reschke_modeling_2015,wysoczanska_solubility_2021}, where solubility data were used to fit the parameters. Unfortunately, due to the nature of the PC-SAFT equations, these parameters are not transferrable to LNP systems and acquiring such solubility data would be quite challenging. Alternative predictive approaches have been developed for both pure and mixture systems, including \textit{ab initio} 
\parencite{umer_pc-saft_2014,kaminski_sepp_2020,walker_ab_2022} and machine learning \parencite{winter_spt-nrtl_2022,habicht_predicting_2023,felton_ml-saft_2024,winter_understanding_2023} methods, the most promising of which are group-contribution-based approaches \parencite{sauer_comparison_2014} such as the SAFT-$\gamma$ Mie equation \parencite{papaioannou_group_2014,wehbe_phase_2022,bernet_modeling_2024,valsecchi_modelling_2024}, where, much like the illustration in Figure \ref{fig:lnp_simple}, molecules can be assembled from common moieties (or groups). Here, parameters are specific to the groups, which can be fitted to systems with more-abundant data and then extrapolated to the desired systems. In the case of SAFT-$\gamma$ Mie, most of the required groups have already been fitted, with the exception of the phosphate groups involved in mRNA and the phospholipids, which should be possible to parameterize using experimental data for systems involving phosphate groups (such as ionic liquids). The tools required for parameterization of model parameters are already available within the Clapeyron.jl framework and will be the topic of future study.

\section{Process Modeling Approaches}

\label{sec:4_Process_Modeling_Approaches}
\begin{figure}[htbp]
    \centering
    \includegraphics[width= 0.6\linewidth]{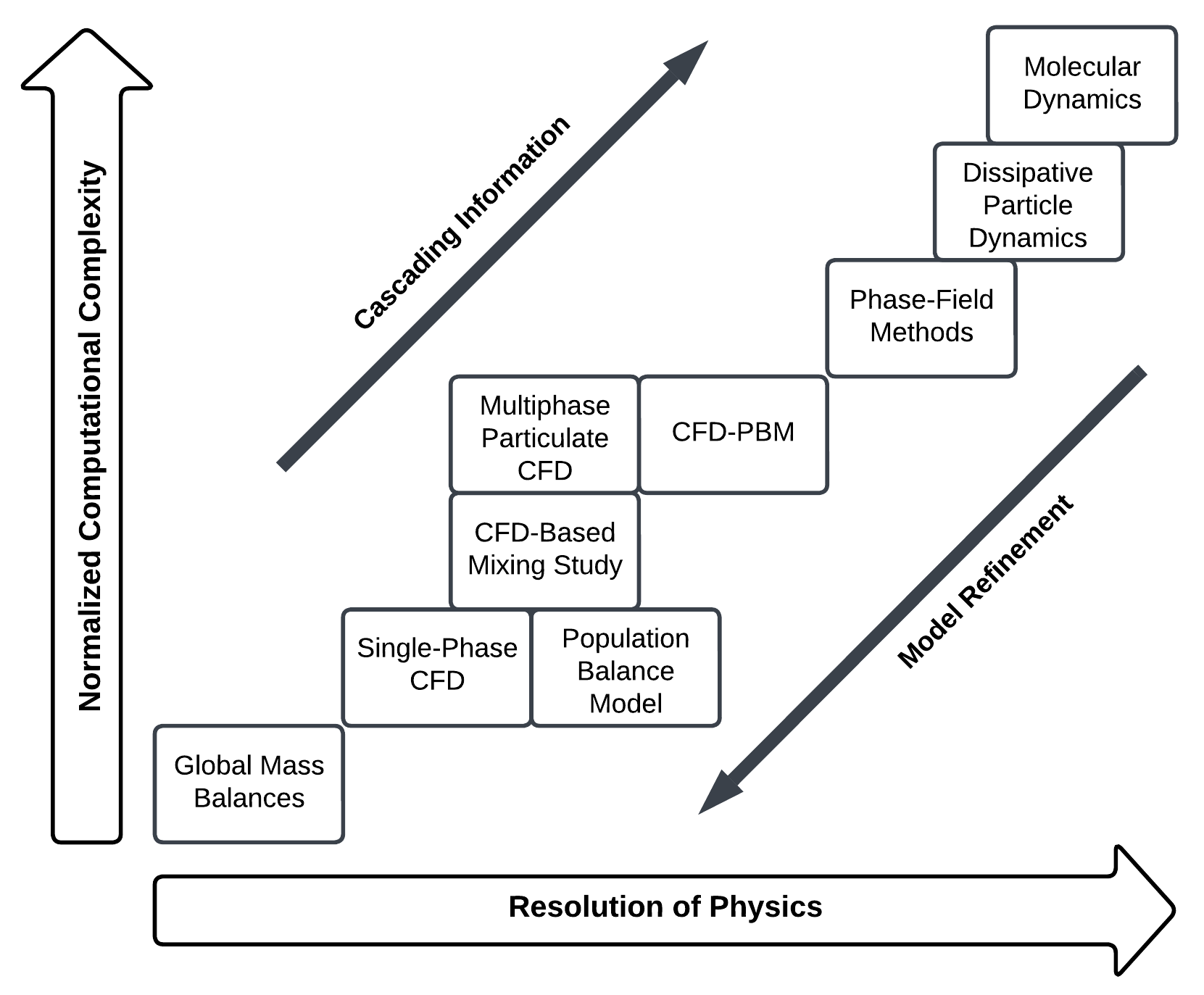}

    \vspace{-0.3cm}
    
    \caption{Summary of modeling strategies available to characterize LNP manufacturing sorted based on depth of physical/process insights gained and normalized computational complexity (i.e., computational cost incurred to resolve the same length- and time-scale).}
    \label{fig:model_strategy}
\end{figure}

This section outlines some of the modeling strategies that can be employed to study various aspects of the LNP production process. These methods can be categorized along two axes: computational and model complexity, and process and physical insights. A schematic of the various methods outlined is presented in Figure~\ref{fig:model_strategy}. The LNP formation is inherently a multi-scale and multi-physics process. A rigorous modeling approach needs to employ multiple methods at different length and time scales where results from simpler approaches e.g., mass and energy balances / single-phase CFD are used to inform more detailed methods while insights from more detailed methods can be used to refine simpler models.

\subsection{Mass and Energy Balances}

\begin{figure}[htbp]
    \centering
    \includegraphics[width= 0.5\linewidth]{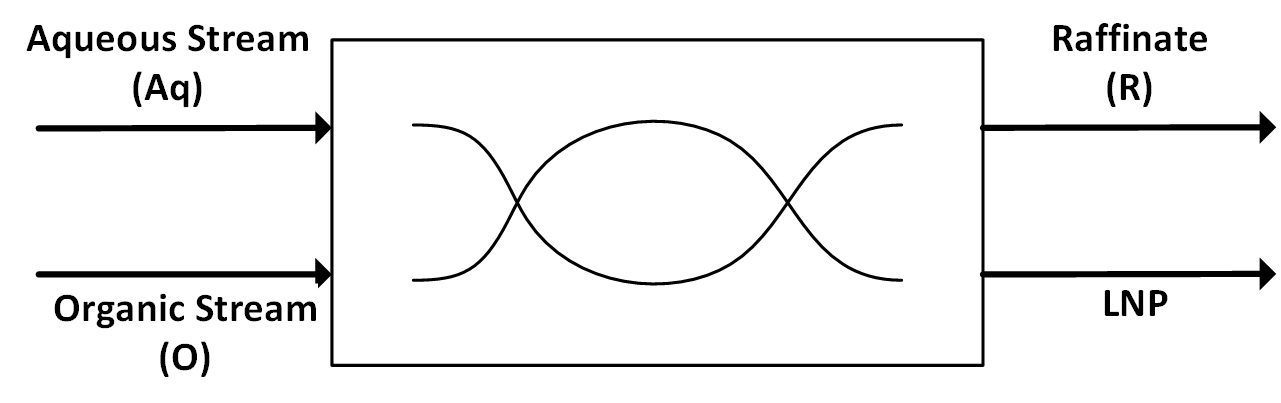}

\vspace{-0.3cm}
    
    \caption{Schematic of LNP mixer with two feed streams and two product streams as a result of LNP precipitation.}
    \label{fig:lle_schematic}
\end{figure}

Global mass and energy balances can be constructed by considering the LNP manufacturing process as a mixing process with two input streams (aqueous and organic) and two output streams (LNPs and the raffinate which consists of the solvents and residual lipids / nucleic acids). The model equations are derived by drawing a control volume around the mixer (see Fig.~\ref{fig:lle_schematic}) and tracking the material and energy flows in and out of the system. This model assumes that the LNP formation process is at equilibrium i.e., the individual species concentrations in the respective outlet phases are the equilibrium concentrations for a given set of material inputs and process conditions. Formulating these balances for the entire process is useful for ``closing" the mass balance of the system and characterizing the location of various species which could be either in the LNP or in the raffinate stream. It is also straightforward to extend the control volume for the mass balance to include the buffer exchange step. The energy balance is also useful for estimating the temperature rise as a result of mixing.   

The overall mass balance is given by
\begin{equation}
    \frac{dM}{dt} = F^{\textrm{Aq}} + F^{\textrm{O}} - F^{\textrm{R}} - F^{\textrm{LNP}},
    \label{eq:balance1}
\end{equation}
where $M$ is the total mass holdup in the mixer, $F^{i}$ is the mass flowrate of stream $i$ with the superscripts $\textrm{Aq}$, $\textrm{O}$, $\textrm{R}$, and $\textrm{LNP}$ corresponding to the aqueous, organics, raffinate, and LNP streams. Similarly, mass balances for each individual species can be written by introducing a mass fraction variable for species $j$ in stream $j$, $x_{j}^{i}$,
\begin{equation}
    \frac{dM_{j}}{dt} = F^{\textrm{Aq}}x_{j}^{\textrm{Aq}} + F^{\textrm{O}}x_{j}^{\textrm{O}} - F^{\textrm{R}}x_{j}^{\textrm{R}} - F^{\textrm{LNP}}x_{j}^{\textrm{LNP}},
    \label{eq:balance2}
\end{equation}
where $M_{j}$ is the mass holdup of species $j$ in the mixer. For an $N$ component system, $N-1$ species mass balance equations would be formulated with the mass fraction of the last species being inferred from the fact that, for any stream, the sum of mass fractions of all species is unity, i.e., $\sum_{j} x_{j}^{i} = 1$. Equation~\ref{eq:balance2} can be simplified by introducing the partition coefficient $k_{D,j} = \frac{x_{j}^{\textrm{LNP}}}{x_{j}^{\textrm{R}}}$. The advantage of introducing $k_{D}$ to the model is that it can be computed separately either experimentally or computationally using methods outlined in Sec.~\ref{sec:thermo}. Lastly, the energy balance is given by
\begin{equation}
    \frac{dH}{dt} = F^{\textrm{Aq}}h^{\mathrm{Aq}} + F^{\textrm{O}}h^{\mathrm{O}} - F^{\textrm{R}}h^{\mathrm{R}} - F^{\textrm{LNP}}h^{\mathrm{LNP}} - \dot{Q}_\mathrm{env},
    \label{eq:balance3}
\end{equation}
where $H$ is the total enthalpy holdup in the mixer, $h^{i}$ is the specific enthalpy of of stream $i$, and $\dot{Q}_{env}$ is the heat loss to the environment. Estimating the specific enthalpies of the respective streams can be quite challenging considering the complex species present. However, the typical LNP mixing process uses dilute aqueous buffer and ethanol streams as inputs (see Table~\ref{tbl:stream_composition}). Hence, a very reasonable approximation for the system is to simply treat the process as a two-component mixing process of only ethanol and water. In addition, considering the short residence time in the mixer, it is also reasonable to assume that heat transfer to the environment is negligible (i.e., $\dot{Q}_\mathrm{env} = 0$). Consequently, \eqref{eq:balance3} simplifies to
\begin{equation}
    \frac{dH}{dt} = F^{\textrm{Aq}} h^{\textrm{Water}} + F^{\textrm{O}}h^{\textrm{Ethanol}} - (F^{\textrm{R}} + F^{\textrm{LNP}})h^{\textrm{Water-Ethanol}}.
    \label{eq:balance4}
\end{equation}
Computing the mass and energy (simplified) balances can be performed easily with the help of a process simulator and accompanying thermodynamic model package (e.g., Aspen). The main source of temperature changes during the process is due to the mixing of ethanol with water which is known to be exothermic \parencite{peeters_endothermicity_1993}. However, the effect of heat of mixing has not been extensively discussed in the literature on LNP production and multiple computational fluid dynamics (CFD) studies investigating the mixing of water and ethanol have similarly neglected its effect \parencite{schikarski_direct_2017,orsi_waterethanol_2013}. While \eqref{eq:balance1}--\eqref{eq:balance4} are presented on a mass basis, the equations can be readily expressed on a molar basis instead. In addition, all the balance equations have an accumulation term (i.e., time derivative) for completeness, but it would typically be ignored as the mixing process normally operates at steady state. 

The relative simplicity of global mass and energy balances to characterize the process should be recognized as both a strength and weakness. On one hand, these balances are able to provide useful information on the process at low-cost. However, the balances are only able to provide coarse-grained information on the process and the quality of the results are a function of the quality of experimental data / thermodynamic models used to estimate the partition coefficients and enthalpies. 
Mass balances that span multiple unit operations (in particular the buffer exchange step) may be needed when key constituents of outlet streams are measured at the outlet of a downstream unit operation. For example, internal LNP properties are not measurable immediately downstream from the LNP formation, since their measurement first requires buffer exchange and then physical separation of LNPs from liquid. Lastly, non-equilibrium phenomena such as the trapping of water within the LNPs cannot be captured by global mass and energy balances.

\subsection{Computational Fluid Dynamics (CFD)}
\label{sec:cfd}
Computational fluid dynamics (CFD) is widely used to understand unit operations involving fluid mixing in pharmaceutical manufacturing \parencite{rantanen_future_2015}. While LNP manufacturing involves multiphase solid-liquid flow in the mixer, where the liquid phase is composed of a mixture of water and ethanol and the solid phase is composed of LNPs,  
experimentally observed mass fractions of LNP inside the mixer are comparatively low ($\leq5\%$) \parencite{he_size-controlled_2018,obrien_laramy_process_2023}. A single-phase liquid flow model is sometimes used to understand the main characteristics of the flow inside the mixer \parencite{buongiorno_convective_2006}. However, since the LNPs are a separate and distinct phase, multiphase flow modeling may be needed to enable more accurate process modeling by including the dynamic behaviour of the particulate phase. Single-phase models, while useful, cannot adequately explain solid-liquid interactions, the movement of particles, the agglomeration and dispersion processes, and changes in particle size. 

This section first reviews the governing equations for single-phase one- and two-component liquid flows and discusses the relative strengths and limitations of these models in characterizing the flow inside the mixer. Subsequently, we review the governing equations for solid-liquid flows based on a multiphase approach and outline why multiphase approaches can yield more realistic and reliable results compared to single-phase models.

\subsubsection{Single-Phase Models}
\label{sec:sin_com_hom_mod}

The incompressible Navier--Stokes equations, which are the governing equations for single-component liquid flow, can be used to understand the overall impact of the mixer geometry on flow. The associated total mass and linear momentum conservation equations, respectively, are 
\begin{alignat}{2}
\label{eq:1phom_mass_balance}     \nabla\cdot\boldsymbol{u} 
     &= 0, \\
\label{eq:1phom_momentum_balance}     \rho\frac{\partial\boldsymbol{u}}{\partial t}
     + \rho\boldsymbol{u}\cdot\nabla\boldsymbol{u}
     &= -\nabla p
     + \nabla\cdot\boldsymbol{\tau}
     + \rho \boldsymbol{b},
\end{alignat}
where $\boldsymbol{u}$, $p$, and $\boldsymbol{\tau}=\mu( \nabla\boldsymbol{u} + (\nabla\boldsymbol{u})^\top )$ are the velocity, pressure, and viscous stress, and $\rho$, $\mu$, and $\boldsymbol{b}$ are density, viscosity, and acceleration due to an external body force. Exothermic ethanol-water mixing, turbulence, and contrast of properties at different spatial locations inside the mixer are ignored in the above equations. While these equations provide an approximate idea of the fluid flow inside the mixer, the assumption of zero turbulence can give an incorrect estimate of shear rates in the flow. Estimating the shear stress is important, as it affects the structure of LNPs and the encapsulated genetic material \parencite{kim_engineering_2022}.

\subsubsection{Turbulence Models}
\label{sec:tur_mod}

While the Reynolds number (Re) typically used for microfluidic mixing in the mixers is not large enough to suggest transition to turbulence, the flow may become turbulent as a result of flow turning in some mixers with high local curvatures, such as the baffled mixture shown in Fig.~\ref{fig:mixer_geom}d. Such curvature can give rise to localized vortices in the flow \parencite{inguva_computer_2018}. Rapid momentum changes to fluid streams in a confined impinging jet mixer, as shown in Fig.~\ref{fig:mixer_geom}e, also can give rise to turbulent effects. In addition, ethanol-water mixing may also induce transition to turbulence at small Re \parencite{devos_impinging_2025}. Estimating these turbulent effects by directly simulating \eqref{eq:1phom_mass_balance}--\eqref{eq:1phom_momentum_balance} requires resolving the Kolmogorov length scale ($\lambda_K$), which is computationally very expensive. A more common approach is to use turbulence models to approximate the effects at small length scales, by extending \eqref{eq:1phom_mass_balance}--\eqref{eq:1phom_momentum_balance}.

\begin{figure}
    \centering
    \includegraphics[width=0.8\textwidth]{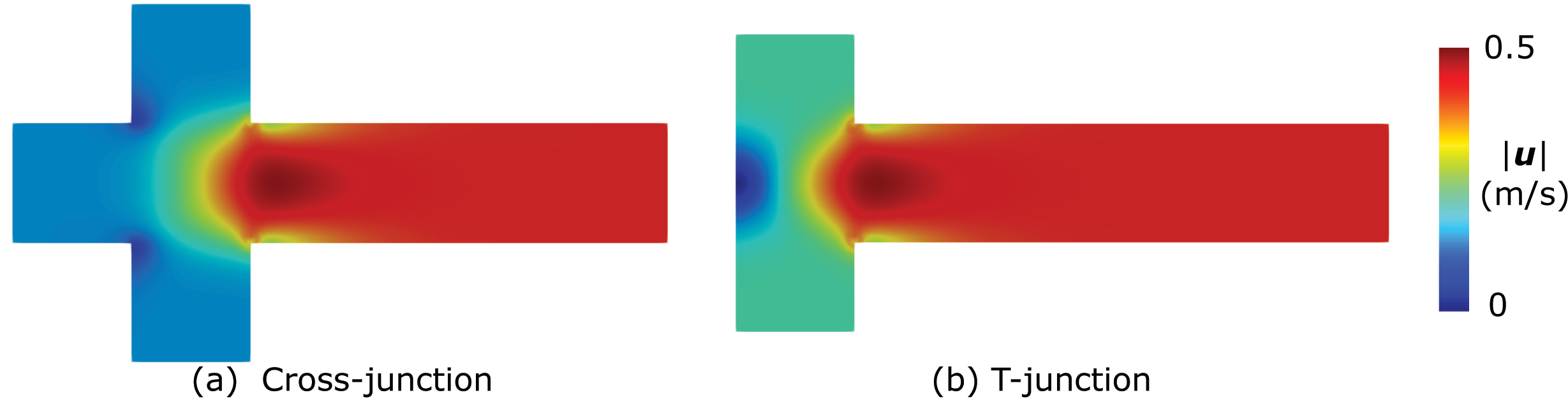}
    \caption{Distribution of magnitude of the velocity of water in (a) a T-junction and (b) a cross-junction mixer at steady state, solved using the $k$-$\varepsilon$ turbulence model. The flows in (a) are $F^{\text{left}}=$ \SI{125}{\micro \liter \per \minute} and $F^{\text{bottom}}=F^{\text{top}}=$ \SI{187.5}{\micro \liter \per \minute} and in (b) are $F^{\text{bottom}}=$ \SI{375}{\micro \liter \per \minute} and $F^{\text{top}}=$ \SI{125}{\micro \liter \per \minute}.}
    \label{fig:1componentGeometryVariation}
\end{figure}

A turbulence model uses computationally resolved physical quantities to approximate the effect of physical quantities that are not resolved computationally, such as sub-grid scale shear stress and viscosity. While the computational effort is reduced, properly choosing the turbulence model based on the target application is important. Commonly used turbulence models for \eqref{eq:1phom_mass_balance}--\eqref{eq:1phom_momentum_balance} can be broadly classified into Reynolds Averaged Navier Stokes (RANS) and Large Eddy Simulation (LES) models. Here, we provide a conceptual overview for each of these models.

The RANS model decomposes the flow variables into an ensemble-averaged and a fluctuating component: $\boldsymbol{u} = \langle\boldsymbol{u}\rangle + \boldsymbol{u}'$ and $p = \langle p \rangle + p'$. Substituting into \eqref{eq:1phom_mass_balance}--\eqref{eq:1phom_momentum_balance} gives the RANS equations, 
\begin{alignat}{2}
    \label{eq:1phom_ransAveraged}
    \nabla\cdot\langle\boldsymbol{u}\rangle 
     &= 0, \\
     \label{eq:1phom_momentum_balance_ransAvaregaed}
     \rho\frac{\partial\langle\boldsymbol{u}\rangle}{\partial t}
     + \rho\langle\boldsymbol{u}\rangle\cdot\nabla\langle\boldsymbol{u}\rangle
     &= -\nabla \langle p\rangle
     + \nabla\cdot\langle\boldsymbol{\tau}\rangle
     + \nabla\cdot\langle\boldsymbol{\tau'}\rangle
     + \rho \boldsymbol{b},
\end{alignat}
where $\langle\boldsymbol{\tau}\rangle=\mu( \nabla\langle\boldsymbol{u}\rangle + \nabla^\top\langle\boldsymbol{u}\rangle )$ and $\langle\boldsymbol{\tau'}\rangle$ are the ensemble-averaged viscous stress and Reynolds stress, respectively. Most engineering applications employ two-equation RANS models such as the $k$-$\varepsilon$ \parencite{launder_numerical_1983} and $k$-$\omega$ \parencite{wilcox_reassessment_1988} models, where the Reynolds stress is modeled using the Boussinesq eddy viscosity approximation,
\begin{equation}
    \label{eq:boussinesq_eddy_viscosity}
    \langle\boldsymbol{\tau'}\rangle = \mu_t( \nabla\langle\boldsymbol{u}\rangle + \nabla^\top\langle\boldsymbol{u}\rangle )
    - \frac{2}{3}\rho k \boldsymbol{I},
\end{equation}
where $\mu_t$ and $k$ are the turbulent viscosity and turbulent kinetic energy, respectively, and $\boldsymbol{I}$ is the identity matrix. As an illustration, in the $k$-$\varepsilon$ model, $k$ and $\mu_t$ are determined by solving
\begin{alignat}{2}
    \label{eq:turbkeps_k_advection_diffusion}
     \frac{\partial\left( \rho k \right)}{\partial t} 
     + \nabla\cdot\left( \rho k \langle\boldsymbol{u}\rangle \right)
     &= \nabla\cdot\left\{ \left( \mu 
     + \frac{\mu_t}{\sigma_k} \right)\!\nabla k \right\}
     + G_k
     - \rho\varepsilon
     + S_k,\\
     \label{eq:turbkeps_epsilon_advection_diffusion}
     \frac{\partial\left( \rho\varepsilon \right)}{\partial t} 
     + \nabla\cdot\left( \rho\varepsilon\langle\boldsymbol{u}\rangle \right)
     &= \nabla\cdot\left\{ \left(\mu 
     + \frac{\mu_t}{\sigma_\varepsilon} \right)\!\nabla\varepsilon\right\}
     + C_{1\varepsilon}\frac{\varepsilon}{k}G_k
     - C_{2\varepsilon}\rho\frac{\varepsilon^2}{k}
     + S_\varepsilon, \\
     \label{eq:turbkeps_turbulent_viscosity_source}
     \mu_t 
     &= \rho C_\mu \frac{k^2}{\varepsilon}, \;\;\;
     G_k = \frac{\mu_t}{2\mu^2} \boldsymbol{\tau}\colon\boldsymbol{\tau},
\end{alignat}
where $\varepsilon$, $S_k$, and $S_\varepsilon$ are the turbulent dissipation rate, the source of turbulent kinetic energy, and the source of turbulent dissipation, and $C_{1\varepsilon}=1.44$, $C_{2\varepsilon}=1.92$, $C_{\mu}=0.09$, $\sigma_k=1.0$, and $\sigma_\varepsilon=1.3$ are constants.

In the viscous sublayer near walls, the accuracy of the $k$-$\varepsilon$ model can significantly decrease. Standard wall functions developed by \textcite{launder_numerical_1974} simplify the modeling of near-wall turbulence by using empirical laws such as the logarithmic law of the wall discovered by \textcite{von_karman_mechanical_1931}. The wall function developed through these studies primarily use the logarithmic law of the turbulent boundary layer to calculate the velocity, turbulent kinetic energy, turbulent dissipation rate, temperature, and component distribution near the walls. The basic equations for predicting the wall velocity based on the wall function are
\begin{align}
U^{*} = \frac{1}{\kappa} \ln\!{(Ey^{*})} = U_{p}C_{\mu}^{1/4}k_{p}^{1/2}, \\
y^{*} = \frac{\rho C_{\mu}^{1/4}k_{p}^{1/2}}{\mu}y_{p},
\end{align}
where $\kappa$ is the von K\'arm\'an constant ($=0.4187$), $E$ is an empirical constant ($=9.793$), $U_{p}$ is the mean velocity of the fluid at the wall-adjacent cell centroid $P$, $k_{p}$ is the turbulent kinetic energy at the wall-adjacent cell centroid $P$, and $y_p$ is the distance from the centroid of the wall-adjacent cell to the wall.

After Launder and Spalding \parencite{launder_numerical_1974} proposed the use of wall functions, various wall function methods have been suggested, with each method modeling the flow near the wall under specific conditions. Scalable wall functions are effective when the grid near the wall becomes denser, resulting in a smaller $y^{+}$ value \parencite{marie_similarity_1997}. Non-equilibrium wall functions are designed to be applicable even when the flow near the wall is in a non-equilibrium state and is effective in cases where there are large pressure gradients or strong curvatures near the wall \parencite{kim_near-wall_1995}. Enhanced wall treatment is a hybrid approach that uses wall functions together with direction resolution in the low Re region to better capture the flow near the wall \parencite{fiuza_comparison_2018}.

Fig.~\ref{fig:1componentGeometryVariation} shows the steady-state profile of the magnitude of the velocity  for the flow of water in (a) a T-junction and (b) a cross-junction mixer using the $k$-$\varepsilon$ model.  Although the native $k$-$\varepsilon$ and $k$-$\omega$ models provide good quantification for turbulent effects in single-component, single-phase flows, their predictions are sometimes inadequate for regions near the walls and flows containing large pressure gradients and large curvatures. Modified forms of the native RANS models have been proposed, such as the RNG model \parencite{yakhot_development_1992} and the SST model \parencite{menter_two-equation_1994}, among others. For an extensive review of RANS models for T-mixers, see \textcite{frank_simulation_2010}.

Another commonly employed turbulence modeling approach is LES, which filters the governing equations into a large- and a small-scale part, and models the subgrid-scale (SGS) terms. For the sake of understanding LES modeling for the equations presented in the ensuing sections, we present the LES equations corresponding to \eqref{eq:1phom_mass_balance}--\eqref{eq:1phom_momentum_balance}, in addition to a scalar transport equation that governs the transport of a scalar $\upsilon$ by advection, diffusion, and reaction,
\begin{equation}
    \label{eq:scalar_transport}
    \frac{\partial\upsilon}{\partial t} 
    + \boldsymbol{u}\cdot\nabla\upsilon 
    = \nabla\cdot\left(D_\upsilon\nabla\upsilon\right)
    + S_\upsilon,
\end{equation}
where $D_\upsilon$ is the diffusion constant and $S_\upsilon$ is a source term for the production of $\upsilon$. Applying spatial filtering on \eqref{eq:1phom_mass_balance}--\eqref{eq:1phom_momentum_balance} and \eqref{eq:scalar_transport} gives
\begin{alignat}{2}
    \label{eq:les_conservation_mass}
    \nabla\cdot\boldsymbol{\overline{u}} &= 0, \\
    \label{eq:les_conservation_momentum}
    \rho\frac{\partial\boldsymbol{\overline{u}}}{\partial t}
    + \rho\boldsymbol{\overline{u}}\cdot\nabla\boldsymbol{\overline{u}} 
    &= -\nabla\overline{p}
    + \nabla\cdot\boldsymbol{\overline{\tau}}
    +\nabla\cdot\boldsymbol{\tilde{\tau}}
    +\rho\boldsymbol{b}, \\
    \label{eq:les_conservation_species}
    \frac{\partial\overline{\upsilon}}{\partial t}
    + \overline{\boldsymbol{u}}\cdot\nabla\overline{\upsilon} 
    &= \nabla\cdot\left( D_\upsilon\nabla\overline{\upsilon} +\boldsymbol{\Tilde{J}} \right)
    + \overline{S_\upsilon},
\end{alignat}
where the overline represents spatially filtered quantities, and the symbols $\boldsymbol{\tilde{\tau}}$ and $\boldsymbol{\Tilde{J}}$ are the SGS stress and SGS scalar flux respectively. SGS terms are typically modeled using a Dynamic Smagorinsky Model, 
\begin{alignat}{2}
    \label{eq:les_sgs_stress_trace}
    \text{tr}( \boldsymbol{\tilde{\tau}})
    = \frac{C_I\rho\Delta^2}{\mu^2}\boldsymbol{\overline{\tau}}\colon\boldsymbol{\overline{\tau}}, \;\;\;&\;\;\;
    \boldsymbol{\tilde{\tau}} =
    \frac{1}{3} \text{tr}( \boldsymbol{\tilde{\tau}})\boldsymbol{I}
    -\frac{C_s\rho\Delta^2}{2\mu^2} \sqrt{2\boldsymbol{\overline{\tau}}\colon\boldsymbol{\overline{\tau}}}
    \left( \boldsymbol{\overline{\tau}} - \frac{1}{3}\text{tr}(\boldsymbol{\overline{\tau}})\boldsymbol{I} \right),\\
    \label{eq:les_sgs_flux}
    \boldsymbol{\tilde{J}} 
    &= -\frac{C_s\rho\Delta^2}{2\mu \text{Sc}_t}\sqrt{2\boldsymbol{\overline{\tau}}\colon\boldsymbol{\overline{\tau}}}\,\nabla\overline{\upsilon},
\end{alignat}
where $\Delta$ is the computational mesh size, $\text{Sc}_t$ is the turbulent Schmidt number, and $C_s$ and $C_I$ are model coefficients determined using the Germano identity \parencite{germano_turbulence_1992}. For a more extensive discussion on LES models, see \textcite{tkatchenko_performances_2007}.

While incompressible turbulence models can be used to predict shear stresses inside the mixer at a reasonable computational cost, they cannot account for the spatial variation in properties in an ethanol-water mixture. The mixing of ethanol and water changes the density of the solution and is exothermic. Such a mixing process can be treated in the form of compressible fluid flow equations. Moreover, the turbulent models to simulate the mixing phenomena needs to also include the species and energy transport models \parencite{gatski_compressibility_2009}. 

\subsubsection{Species Transport Models}
\label{sec:spec_tra}

\begin{figure}
    \centering
    \includegraphics[width=0.8\textwidth]{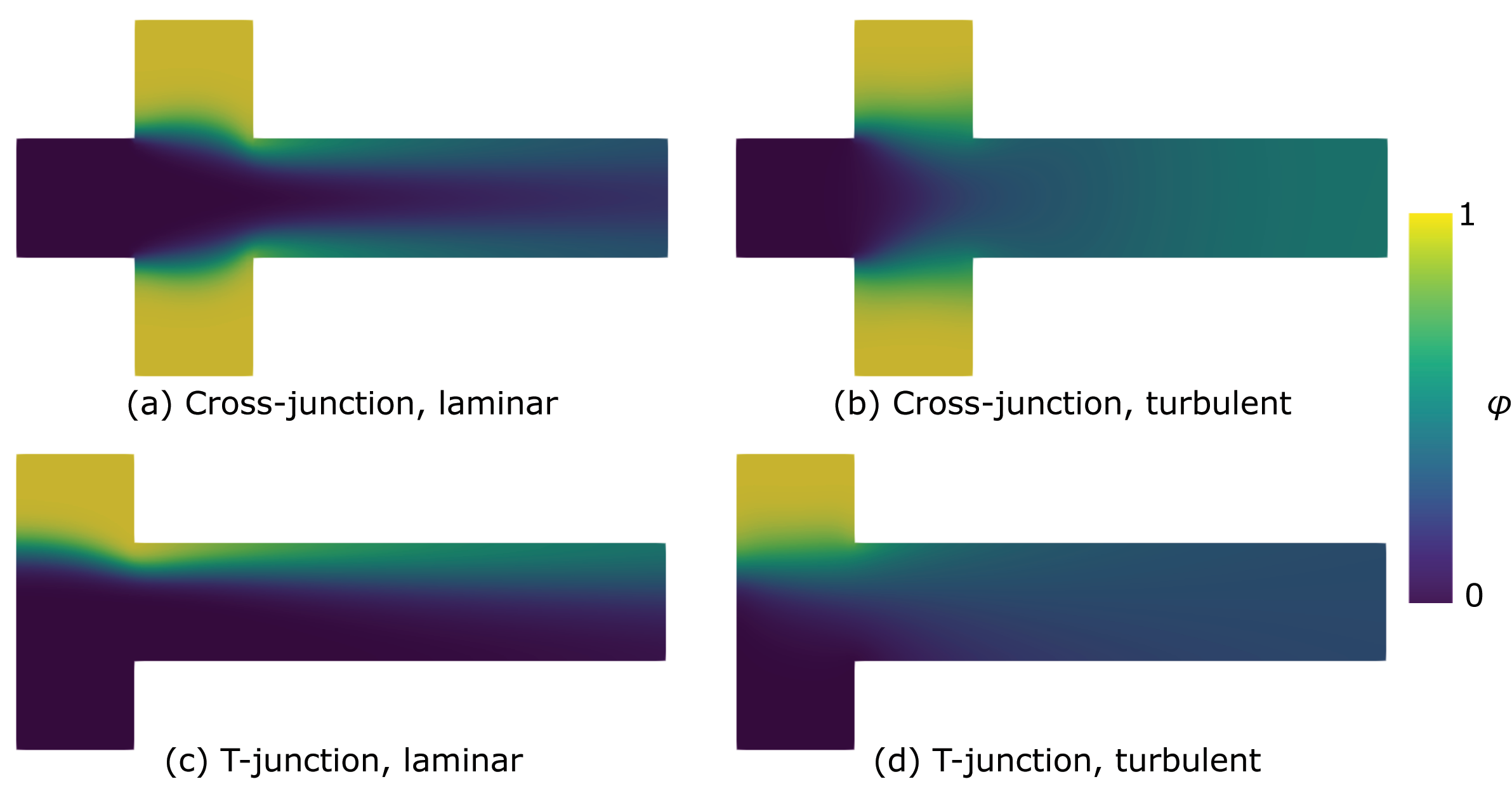}

    \vspace{-0.3cm}
    
    \caption{Distribution of the volume fraction of ethanol at steady-state in (a,b) cross-junction  and (c,d) T-junction mixers for (a,c) laminar and (b,d) turbulent flows, computed using the $k$-$\varepsilon$ turbulence model. Ethanol is injected from the top and bottom in (a,b), and from the top in (c,d) using the same flow rates as in Fig.~\ref{fig:1componentGeometryVariation}.}
    \label{fig:2componentMixing}
\end{figure}

Water and ethanol are two miscible species, and the extent of their mixing is key to determining LNP precipitation in the mixer. The species transport model for the various components can be represented by conservation equations of volume / mass fraction. Using the notation of volume fraction can be useful to consider changes in the composition in the mixer. The governing equations for two-component miscible flows are \parencite{ghorbani_computational_2021}
\begin{alignat}{2}
     \label{eq:2pvof_mass_balance}
     \frac{\partial\rho_m}{\partial t} 
     + \nabla\cdot\left( \rho_m\boldsymbol{u} \right)
     &= 0, \\
     \label{eq:2pvof_momentum_balance}
     \frac{\partial\left( \rho_m\boldsymbol{u} \right)}{\partial t} 
     + \rho_m\boldsymbol{u}\cdot\nabla\boldsymbol{u}
     &= -\nabla p
     + \nabla\cdot\boldsymbol{\tau}
     + \rho_m\boldsymbol{b}, \\
    \label{eq:2pvof_advection_vf}
     \frac{\partial\varphi}{\partial t}  
     + \boldsymbol{u}\cdot\nabla\varphi
     &= \nabla\cdot\left(D_{we}\nabla\varphi\right),
\end{alignat}
where $\rho_m(\varphi)$, $\mu_m(\varphi)$, $D_{we}$,  and $\varphi$ are the density, viscosity, and mass diffusivity of the mixture, and volume fraction of ethanol respectively. While linear interpolation functions are generally used for predicting density and viscosity of a liquid mixture \parencite{parke_solution_1999}, the viscosity of an ethanol-water mixture is $\sim3$ times higher than pure water \parencite{orsi_waterethanol_2013}. This viscous ethanol-water interfacial region could inhibit the formation of vortices, which reduces mixing efficiency at large Re. The density of the mixture varies linearly, and the difference between $\rho_w$ and $\rho_e$ is not large enough so that Rayleigh-Taylor instabilities become the primary source of transition to turbulence. However, shear at the ethanol-water interface could give rise to Kelvin-Helmholtz (KH) instabilities,
which are the primary instabilities that cause transition to turbulence during ethanol-water mixing \parencite{schikarski_direct_2017}. In addition, accurately quantifying the diffusion of water and ethanol requires resolving the Batchelor length scale, which is related to the Kolmogorov length scale by $\lambda_B = \lambda_K/\sqrt{\text{Sc}}$. Here, $\text{Sc}\approx600$ is the Schmidt number for water-ethanol mixing $(\text{Sc}=\mu_m/(\rho_m D_{we}))$. This makes the direct numerical simulation (DNS) of water-ethanol mixing computationally expensive even for moderate Re. Limited DNS of mixing in T-mixers \parencite{orsi_waterethanol_2013,schikarski_direct_2017,inguva_computer_2018} have shown that, for $\text{Re}\leq250$, the flow is stratified, the mixing is purely diffusive, and the vortices remain confined to individual liquids. As $\text{Re}\geq500$, strong KH instabilities are observed at the interface, and mixing is greatly enhanced due to turbulence. In Fig.~\ref{fig:2componentMixing}, we show the variation of the volume fraction of ethanol at steady state for water-ethanol mixing in cross-junction, and T-junction mixers under laminar and turbulent flow conditions.

The distribution of individual chemical species such as lipids and mRNA which are transported by the water-ethanol liquid mixture is neglected in simple CFD models. Since LNPs are formed from the self-assembly of these species, modeling the distribution of these species is critical to characterizing LNP formation. This section enumerates the transport equations for individual species of the mixture. Consider an organic stream composed of $I$ ionizable lipids (ILs), whose concentrations and valences are represented as $[\text{L}_i]$ and $z_i$ respectively, $i=1,\dots,I$; and $N$ neutral lipids (NLs) whose concentrations are represented as $[\text{L}_n]$, $n=1,\dots,N$. The transport of species due to the velocity of the underlying flow, molecular diffusion, electromigration (transport driven by electric potential for charged ILs and mRNA), and production/dissipation due to self assembly can be modeled by
\begin{alignat}{2}
    \label{eq:species_transport_macromixing_mRNA}
\rho\frac{\partial[\text{mRNA}]}{\partial t} + \rho\langle\boldsymbol{u}\rangle\cdot\nabla[\text{mRNA}]
    &= \nabla\cdot\left( \rho D_{t,\text{mRNA}}\!\left(\nabla[\text{mRNA}] 
    + z_{\text{mRNA}}[\text{mRNA}]\frac{e}{k_B T}\nabla V \right)\right)\!
    -\rho Y, \\
    \label{eq:species_transport_macromixing_IL}
    \rho\frac{\partial[\text{L}_i]}{\partial t}
    + \rho\langle\boldsymbol{u}\rangle\cdot\nabla[\text{L}_i]
    &= \nabla\cdot\left( \rho D_{t,\text{L}_i}\!\left(\nabla[\text{L}_i] 
    + z_i[\text{L}_i]\frac{e}{k_B T}\nabla V \right)\right)\!
    -\rho Y, \quad i=1,\dots,I, \\
    \label{eq:species_transport_macromixing_otherLipids}
    \rho\frac{\partial[\text{L}_n]}{\partial t}
    + \rho\langle\boldsymbol{u}\rangle\cdot\nabla[\text{L}_n]
    &= \nabla\cdot\left( \rho D_{t,\text{L}_n}\nabla[\text{L}_n] \right)
    -\rho Y, \quad n=1,\dots,N, \\
    \label{eq:species_transport_macromixing_LNP}
    \rho\frac{\partial[\text{LNP}]}{\partial t}
    + \rho\langle\boldsymbol{u}\rangle\cdot\nabla[\text{LNP}]
    &= \nabla\cdot\left( \rho D_{t,\text{LNP}}\nabla[\text{LNP}] \right)
    +\rho Y, \\
    \label{eq:species_transport_macromixing_potential}
    \nabla\cdot\left( \epsilon\epsilon_r\nabla V \right)
    &= -e\!\left( z_{\text{mRNA}}[\text{mRNA}] 
    + \sum_{i=1}^{I}z_{\text{L}_i}[\text{L}_i] \right),
\end{alignat}
where $[\text{mRNA}]$ and $z_{\text{mRNA}}$ are the concentration and valence of mRNA;  $\left[\text{LNP}\right]$ is the concentration of LNPs;  $e$ is the elementary charge; $k_B$ is the Boltzmann constant; $\epsilon\epsilon_r$ is the dielectric constant of the ethanol-water mixture; $D_{t,mRNA}$, $D_{t,L_i}$, $D_{t,L_n}$, and $D_{t,LNP}$ represent the total diffusivity of mRNA, ILs, NLs, and LNPs, respectively; $V$ is the induced electric potential; and $Y$ is a source term that describes the rate of self-assembly of lipids and mRNA to produce LNPs. These equations assume that the Nernst-Planck law is valid for the diffusion of charged species, namely ILs and mRNA in the liquid \parencite{del_rio_diffusion_2016}. 
Owing to the high dielectric constants of water and ethanol ($\epsilon\geq20$), a reasonable approximation is that the induced electric potential may not be large enough for Nernst--Planck diffusion to be significant. In that case, \eqref{eq:species_transport_macromixing_potential} becomes negligible, and \eqref{eq:species_transport_macromixing_mRNA}--\eqref{eq:species_transport_macromixing_IL} reduce to advection-diffusion-reaction equations. However, we include these terms for the sake of generality. The source term $Y$ is a complex nonlinear function of the species concentrations and valences, and is yet to be characterized experimentally for this system to the best of our knowledge. The total diffusivity is the sum of molecular and turbulent diffusivity, e.g., $D_{t,\text{mRNA}}=D_{\text{mRNA}}+D_t$. The turbulent diffusivity can be calculated using $D_t=\mu_t/(\rho \text{Sc}_t)$. For $\text{Re}\leq250$, molecular diffusivity is the predominant mechanism, whereas turbulent diffusivity dominates for $\text{Re}\geq500$ for the range of molecular diffusivities of the lipids. After the self-assembly, LNPs are precipitated out of solution and exist as small solid particles in the fluid. As such, \eqref{eq:species_transport_macromixing_LNP} describes the evolution equation of the solid phase in the fluid, assuming that the LNPs follow the streamlines of the flow. (This approximation is reasonable for nanometer-sized particles \parencite{di_carlo_continuous_2007}.)

DNS is accurate, but coupling to species balance equations previously outlined makes the problem intractable computationally \parencite{nguyen_solution_2016}. Additionally, the rate of self-assembly of lipids is not fast compared with the rate of mixing of water and ethanol at the SGS, which is referred to as micromixing \parencite{david_interpretation_1987}. These kind of flows, where the time scales for turbulent mixing and self--assembly are comparable and micromixing effects are important, are typically modeled using a Probability Distribution Function (PDF) approach for turbulent mixing \parencite{fox_computational_2003,meyer_micromixing_2009}. The PDF ($f_\phi$) is a function of the lipid concentrations, water and ethanol volume fractions, as well as the space and time coordinates; from which the individual species concentrations and the self--assembly source term may be obtained \parencite{fox_computational_2003}. Solution methods for the PDF include transported PDF models, where we explicitly solve transport equations for the PDF, for details see \parencite{pope_turbulent_2001}. Another approach is the finite-mode presumed PDF approach, where the PDF is presumed to be a composition of a finite number of Dirac delta functions \parencite{fox_computational_2003},
\begin{alignat}{2}
    \label{eq:presumed_pdf}
    f_\phi(\boldsymbol{\psi};\boldsymbol{\mathrm{x}},t) 
    &= \sum_{n=1}^{N_e} p_n(\boldsymbol{\mathrm{x}},t)
    \prod_{\alpha=1}^{N_s} \delta[\psi_\alpha-\langle\phi_\alpha\rangle_n(\boldsymbol{\mathrm{x}},t)]
\end{alignat}
where $N_s$ is the total number of species, $N_e$ is the total number of modes/environments, $p_n$ is the probability of mode $n$, and $\langle\phi_\alpha\rangle_n$ is the mean composition of scalar $\phi$ in mode $n$. For our system, we have $N_s=I+N+2$, corresponding to the ILs, NLs, mRNA, and LNP. A $3$-mode/environment model ($N_e=3$) has been previously used to describe micromixing of water-ethanol \parencite{da_rosa_multiscale_2018}, water-methanol \parencite{pirkle_computational_2015}, and water-BaCl$_2$ \parencite{marchisio_simulation_2001}, among others. For the $3$-environment model for binary liquid-liquid mixing,  environment $1$ describes pure water, environment $2$ describes pure ethanol, and environment $3$ describes the ethanol-water mixture. While other models, such as the $2$-and $4$-environment models are also available \parencite{wang_comparison_2004}, the 3-environment model aligns most naturally to this mixing process. Using $p_j$ to denote the probability of environment $j$ ($p_1+p_2+p_3=1$), and $\langle\xi\rangle_3$ to describe the concentration of ethanol in environment $3$, the governing equations for ethanol-water micromixing can be written as
\begin{alignat}{2}
\label{eq:micromixing}
\rho\frac{\partial p_j}{\partial t}
+\rho\langle\boldsymbol{u}\rangle\cdot\nabla\boldsymbol{p_j}
&= \nabla\cdot( \rho D_t\nabla p_j)
    + \rho G_j(p_1,p_2,p_3,\langle\xi\rangle_3)
    + \rho G_{s,j}\left(p_1,p_2,p_3,\langle\xi\rangle_3\right); \;\;\; j=1,2, \\
    \label{eq:component2_environemnt3}
    \rho\frac{\partial}{\partial t}( p_3\langle\xi\rangle_3)
    + \rho\langle\boldsymbol{u}\rangle\cdot\nabla\left( p_3\langle\xi\rangle_3 \right)
    &= \nabla\cdot\left[ \rho D_t\nabla\left( p_3\langle\xi\rangle_3 \right) \right]
    + \rho M^{(3)}(p_1,p_2,p_3,\langle\xi\rangle_3)
    + \rho M^{(3)}_{s}(p_1,p_2,p_3,\langle\xi\rangle_3),
\end{alignat}
where the functions $G_j$, $G_{s,j}$, $M^{(3)}$, and $M_s^{(3)}$ model micromixing effects \parencite{fox_computational_2003}. In particular, the functions $G_j$ and $M^{(3)}$ model micromixing in accordance with the interaction by exchange with the mean (IEM) theory, which states that the environment probabilities ($p_i$) and species concentrations homogenize to a mean value with the same rate constant ($\gamma$); whereas the functions $G_{s,j}$ and $M_s^{(3)}$ ensure that the mean variance of the scalar is correctly predicted using \eqref{eq:micromixing}--\eqref{eq:component2_environemnt3}. Their specific forms are enumerated in Table~\ref{tab:micromixing_constants}. The constants in Table~\ref{tab:micromixing_constants}, namely $\gamma$ and $\gamma_s$, and the mean variance of $\langle\xi\rangle_3$ (represented as $\langle\xi^{'2}\rangle$) are
\begin{alignat}{2}
    \label{eq:closure_functions_micromixing}
    \gamma&=\frac{2\langle\xi^{'2}\rangle\varepsilon}{k\left[p_1(1-p_1)(1-\langle\xi\rangle_3)^2+p_2(1-p_2)\langle\xi\rangle_3^2\right]}, \\
    \gamma_s&=\frac{2D_t\nabla\langle\xi\rangle_3\cdot\nabla\langle\xi\rangle_3}{(1-\langle\xi\rangle_3^2)^2+\langle\xi\rangle_3^2}, \\
    \langle\xi^{'2}\rangle&=p_1(1-p_1)-2p_1p_3\langle\xi\rangle_3+p_3(1-p_3)\langle\xi\rangle_3^2.
\end{alignat}

\begin{table}
\caption{Micromixing functions for the $3$--environment ethanol--water micromixing model}
    \centering
    \begin{tabular}{c c c c c c } 
 \toprule
 Function & Formula & Function & Formula & Function & Formula \\ 
 \hline
 $G_1$ & $-\gamma p_1(1-p_1)$ & $G_{s,1}$ & $\gamma_s p_3$ & $M^{(3)}$ & $\gamma \left(p_1(1-p_1)+p_2(1-p_2)\right)$ \\
 $G_2$ & $-\gamma p_2(1-p_2)$ & $G_{s,2}$ & $\gamma_s p_3$ & $M_s^{(3)}$ & $-2\gamma_s p_3$ \\
 \bottomrule
\end{tabular}
\label{tab:micromixing_constants}    
\end{table}

\subsubsection{Energy Transport Models}
\label{sec:non_iso_fl_mix}

Ethanol-water mixing is exothermic with an enthalpy of mixing of $\Delta H_{we}=-412$ J/mol for a mixture with $x_e=0.5$, where $x_e$ is the mole fraction of ethanol. This enthalpy of mixing causes a local increase in temperature near the ethanol-water interface. Quantifying this temperature rise may be important for certain flow rates and mixer geometries, as temperature changes the properties of mRNA and lipids \parencite{ball_achieving_2016}. The variation of temperature in the fluid is modeled by the energy balance
\begin{alignat}{2}
    \label{eq:2pvof_energy_balance}
     \frac{\partial( \rho_m C_{pm} T)}{\partial t}
     + \nabla\cdot\left( \rho_m\boldsymbol{u}C_{pm}T\right)
     &= \nabla\cdot\left( \kappa_m\nabla T \right)
     + S_h,
\end{alignat}
where $T$, $C_{pm}(\varphi)$, and $\kappa_m(\varphi)$ are the temperature, specific heat capacity, and thermal diffusivity of the mixture. For an extensive review of the dependence of specific heat capacity and thermal diffusivity of the mixture on the volume fraction of ethanol, see \parencite{parke_solution_1999}. 
The last term in the energy balance is the heat of mixing between ethanol and water, which can be represented as 
\begin{alignat}{2}
\label{eq:eq:heat_of_reaction_water_ethanol}
    S_h &= \left(\frac{\rho_m\varphi}{M_e}\right)\!\left(\frac{\rho_e F^{O} + \rho_w F^{Aq}}{\rho_m}\right)\!x_e (1-x_e) \sum_{n=0}^{4} B_n (1-2x_e)^n, \quad
    x_e &= \frac{\varphi/M_e}{\varphi/M_e + (1-\varphi)/M_w},
\end{alignat}
where $M_e$ and $M_w$ are the molar masses of ethanol and water, respectively, and $B_0=1580$, $B_1=1785$, $B_2=3487$, $B_3=3187$, and $B_4=1957$ J/mol are constants \parencite{boyne_enthalpies_1967}. When used alongside turbulence models, the velocity in \eqref{eq:2pvof_energy_balance} is replaced by the ensemble-averaged velocity for RANS models and spatially filtered velocity for LES models. For the range of Re relevant to LNP manufacturing ($\leq\mathcal{O}(10^4)$), the turbulent intensities are small and turbulent heating may be neglected \parencite{cafiero_heat_2014}, implying that the turbulence models reviewed in Sec.~\ref{sec:tur_mod} can be used directly.

\subsubsection{Multiphase Flow Models}
\label{sec:multiphase_model}

When LNPs precipitate from solution and form small particles in the fluid, it can be treated as a distinct solid phase dispersed in the liquid phase. Such flows, where different phases interact, are commonly treated mathematically as interpenetrating continua in CFD. The mixture model, which simultaneously considers the carrier phase and the dispersed phase, is efficient in the analysis of such multiphase flows. The model equations are
\begin{align}
&\frac{\partial \rho_{m}}{\partial t} + \nabla \cdot (\rho_{m}\mathbf{u}_{m}) = 0, \\
&\frac{\partial(\rho_{m}\mathbf{u}_{m})}{\partial t} + \nabla \cdot (\rho_{m}u_{m}u_{m}) = -\nabla p + \nabla \cdot (\mu_{m}(\nabla \mathbf{u}_{m} + (\nabla \mathbf{u}_{m})^{\top})) + \rho_{m}\mathbf{b} + \mathbf{F}_{\text{st}}, \\
&\frac{\partial (\alpha_{q}\rho_{q})}{\partial t} + \nabla \cdot (\alpha_{q}\rho_{q}\mathbf{u}_{q}) = 0,
\end{align}
where $\mathbf{F}_{\text{st}}$ represents the interfacial forces per unit volume; and $\alpha_{q}$, $\rho_{q}$, and $\mathbf{u}_{q}$ are the volume fraction, density, and velocity of phase $q$, respectively. The slip velocity (i.e., the relative velocity between phases), $\mathbf{u}_{\text{dr}}$ is defined as $\mathbf{u}_{\text{dr}} = \mathbf{u}_{p} - \mathbf{u}_{m}$, where $\mathbf{u}_{p}$ is the velocity of the dispersed phase. The mixture velocity $\mathbf{u}_{m}$ is related to the volume fractions and velocities of each phase by the relation,
\begin{equation}
\mathbf{u}_{m} = \sum_{q}\alpha_{q}\mathbf{u}_{q}.
\label{eq:multiphase_mixture}
\end{equation}

The mixture model is quite similar to the equations for single-phase flow previously described, but is expressed with the density and velocity of two or more phases. In addition, the momentum equation for the mixture includes addition terms due to the slip of the dispersed phase relative to the continuous phase. The slip velocity equation can be solved using empirical correlations or additional differential equations to account for the interactions between the phases due such as drag and drift forces \parencite{ansys_inc_ansys_2021}. The volume fraction of the dispersed phase is solved using the phase continuity equation. One way to simulate LNP formation and transport using the mixture model is to treat LNPs and the water-ethanol solution as two phases, with the species transport equation for ethanol and water added to represent the solution phase. A more accurate approach is to treat the system as one solid and multiple liquid phases.

Using only one momentum equation such as in the mixture model \eqref{eq:multiphase_mixture} makes it inherently difficult to accurately predict the multiphase flows. Specifically, if LNPs detach and disperse from the continuous phase flow (e.g., using a baffled mixer), the particles can have a completely different directional flow from the continuous phase within the cell. To better simulate such cases, an Euler--Euler multiphase model can be adopted which incorporates the mass and momentum conservation equations for each phase, thus calculating the velocity of each phase separately. The momentum equation is correspondingly expressed as
\begin{equation}
\frac{\partial (\alpha_{q} \rho_{q}\mathbf{u}_{q})}{\partial t} + \nabla \cdot (\alpha_{q}\rho_{q}\mathbf{u}_{q}\mathbf{u}_{q}) = -\alpha_{q}\nabla p + \nabla \cdot (\alpha_{q}\mathbf{\tau}_{q}) + \alpha_{q}\rho_{q}\mathbf{g} + \mathbf{M}_{q},
\end{equation}
where $p$ is assumed to be shared by all phases, $\mathbf{\tau}_{q}$ is the stress tensor for phase $q$, and $\mathbf{M}_{q}$ represents the rate of momentum exchange per unit volume between phases. 

When using the Euler--Euler model, interaction forces can be directly calculated using theoretical models based on the velocities computed for each phase \parencite{prosperetti_computational_2007}. In other words, the advantage of the Euler-Euler multiphase model is that it can more realistically represent interactions between phases by predicting the momentum of the phases separately. Specifically, it can accurately represent phenomena where vectors have different directions within a cell, making it well-suited for particulate processes frequently encountered in chemical engineering such as fluidized beds \parencite{van_wachem_methods_2003} and cyclones \parencite{narasimha_cfd_2012}. These features of the Euler--Euler model are also well-suited for predicting the behavior of nanoparticles in flow. 

To realistically predict the free movement of particles, the Discrete Element Method (DEM), which treats the particulate phase as individual particles and describes their trajectory using Newton's equations of motion, is an effective modeling approach. To predict the movement of the LNPs following the continuum, it is necessary to couple the DEM equations with the governing equations describing the flow of the continuous phase (i.e., the Navier--Stokes equations). This approach is called the Euler--Lagrangian multiphase model. The continuous phase of this model is treated in a similar way to a single-phase flow model, though additional interaction forces with the particulate phase can be incorporated depending on the extent of coupling deemed necessary. The motion of individual particles is governed by
\begin{equation}
    m_{p}\frac{d \mathbf{u}_{p}}{dt} = \mathbf{f}_{d} + \mathbf{f}_{g} + \mathbf{f}_{c},
\end{equation}
where $m_{p}$ is the particle mass, $\mathbf{u}_{p}$ is the velocity, $\mathbf{f}_{d}$ is the drag force \parencite{syamlal_hydrodynamics_1985,alobaid_progress_2022}, $\mathbf{f}_{g}$ is the gravitational force acting on the particle, and $\mathbf{f}_{c}$ is the force acting on the particle due to collisions. This modeling approach helps characterize the detailed transport phenomena of the LNPs in the mixer that otherwise cannot be resolved by simpler CFD models (i.e., single-phase models) \parencite{mahmoud_nanoparticle_2019}. For more details on the models used for $\mathbf{f}_{d}$ and on Eulerian--Lagrangian models in general, see \parencite{subramaniam_lagrangianeulerian_2013}. 

The CFD models in this section do not model particulate processes such as nucleation, growth, and agglomeration in the mixer. The evolution of the particulate size distribution can be modeling using the population balance framework described in the next section.


\subsection{Population Balance Models}
\label{sec:pbm}
Population balance models (PBMs) are a framework to study the spatiotemporal dynamics of a population that has a distribution over one or more intrinsic properties. Examples of intrinsic properties typically considered in PBMs include size (e.g., length, mass, volume), composition (e.g., concentrations and densities), and age. These descriptive capabilities of PBMs have resulted in their application in a wide range of physical, chemical, and biological systems such as crystallization and precipitation, multiphase flows, and cell cultures \parencite{ramkrishna_population_2014}. 

The PBM for a single species (e.g., LNPs) with $m$ intrinsic variables is \parencite{ramkrishna_population_2014,inguva_efficient_2022,inguva_efficient_2023},
\begin{equation}
    \frac{\partial n}{\partial t} + \sum_{i=1}^{m}\frac{\partial (G_{i}n)}{\partial a_{i}} + \nabla\cdot(\mathbf{u}n) = S + \nabla\cdot\left(D\nabla n\right), 
    \label{eq:pbm_general}
\end{equation}
where $n$ is the species number density, $G_{i}$ is the growth rate corresponding to the intrinsic variable $a_{i}$, $D$ is the particle diffusivity, and $S$ is a source term that describes various physical processes that change the number density of particles, e.g., breakage. To date, there are no published studies employing PBMs to study LNP production by rapid mixing. In this section, we summarize how PBMs from adjacent areas such as emulsification \parencite{hakansson_emulsion_2019,raikar_experimental_2009}, precipitation, and crystallization \parencite{woo_simulation_2006,schwarzer_predictive_2006} can guide model development for LNPs with a focus on modeling the particle size distribution.

\begin{figure}[htbp]
    \centering
    \includegraphics[width=0.7\textwidth]{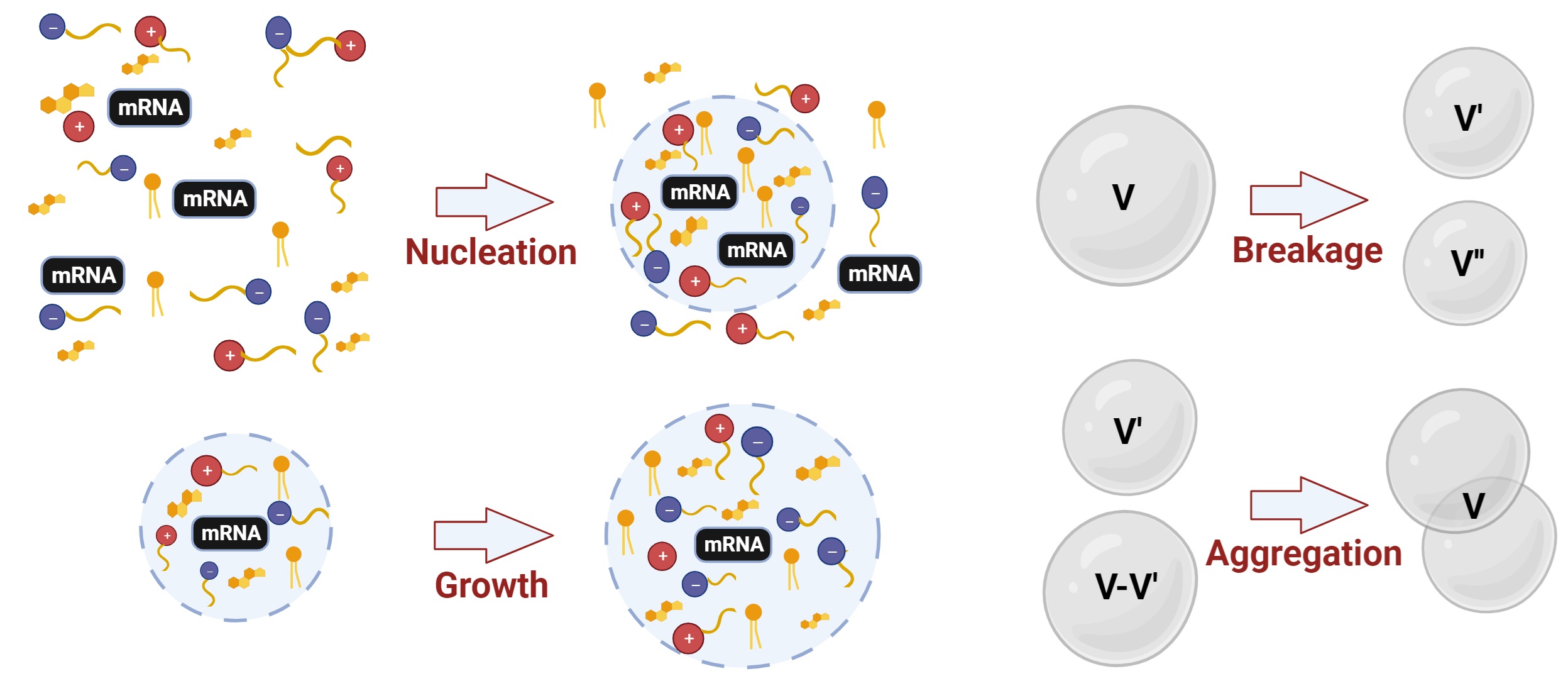}

\vspace{-0.2cm}
    
    \caption{ Schematic diagram of nucleation, growth, agglomeration, and breakage.}
    \label{fig:PBM_fig}
\end{figure}

Physically, LNP formation and dynamics can be modeled as having four predominant steps: 
\begin{itemize}
\item Nucleation, in which a small number of molecules of the various species associate to form an LNP,
\item Growth, in which the volume of the LNP increases as further quantities of various species are incorporated into the LNP from the bulk, 
\item Agglomeration/aggregation in which two or more LNPs collide and merge to form a larger LNP, and
\item Breakage, in which larger LNPs fragment into smaller LNPs. 
\end{itemize}
A schematic illustration of the four processes 
is shown in Fig.~\ref{fig:PBM_fig}. Tracking the LNP size $L$ as an intrinsic property, the PBM incorporating these processes is
\begin{equation}
    \frac{\partial n}{\partial t} + \frac{\partial (Gn)}{\partial L} + \nabla \cdot (\mathbf{u}n) = B_{n} + B_{a} - D_{a} + B_{b} - D_{b}
    + \nabla\cdot\left(D\nabla n\right),
    \label{eq:pbm_particlesize}
\end{equation}
where $B$ and $D$ refer to birth and death rates, and the subscripts $n$, $a$, and $b$ refer to nucleation, aggregation, and breakage, respectively. In the ensuing discussion, we outline how the underlying physics informs possible functional forms for the terms $G$, $B_n$, $B_a$, $D_a$, $B_b$, and $D_b$ for LNP formation. Exemplar functional forms for the various terms are presented in Table~\ref{tbl:pbm}.

\subsubsection{Nucleation}

Prior to formulating an expression for the nucleation rate, it is helpful to first consider the physics of the nucleation process. By creating a supersaturated environment, such as with the addition of an antisolvent, it becomes thermodynamically favorable for the solute(s) to form a new phase, such as the formation of a precipitate. Nucleation is the first step of this process and a nucleus can be understood as the smallest amount of the new phase that can exist independently \parencite{erdemir_nucleation_2009}. Nucleation mechanisms have been broadly categorized as either primary or secondary depending on whether nucleation occurs in the absence or presence of parent particles of the same kind respectively \parencite{xu_overview_2020}. For LNP manufacturing, which relies on rapid mixing of an antisolvent, high supersaturation is created locally at the mixing area, resulting in primary nucleation likely being the dominant mechanism \parencite{thorat_liquid_2012}, thus motivating the focus of the subsequent discussion on primary nucleation. 

Within the context of primary nucleation, the mechanism can be either homogeneous (nuclei form in the solution) or heterogeneous (nuclei form on structural inhomogeneities e.g., surfaces / foreign particles) \parencite{thanh_mechanisms_2014}. For some systems, the nuclei are formed by a two-step mechanism in which the fluid spontaneously forms highly concentrated liquid droplets within a dilute bulk fluid phase, which is then followed by the formation of nuclei \parencite{erdemir_nucleation_2009}. To our knowledge, nucleation rate expression based on two-step nucleation are rarely used, though bounds on the rate can be constructed \parencite{chen_identification_2012}. For precipitation processes, parsing the specific nucleation mechanism is challenging (e.g., see \parencite{roelands_analysis_2006,kugler_heterogeneous_2016}), and may not be necessary for developing a PBM as these complexities can be captured in the nucleation model with the use of fitted parameters. 

For primary nucleation, the birth rate is usually expressed in the PBM literature as 
\begin{equation}
    B_n=B_0 \delta(L- L_0),
   \label{eq:PBE_nuc}
\end{equation}
where $B_0$ is the nucleation rate, $L_0\geq0$ is the length of the smallest particle, and $\delta$ is the Dirac delta function which treats the nuclei as being mono-disperse, and is often a good assumption for nuclei that form in a turbulent mixing zone. Some studies replace the Dirac delta function with a broader probability distribution function (e.g., Gaussian), which can be useful for modeling poly-disperse nucleation \parencite{kumar_efficient_2008} or for numerical reasons \parencite{lapidot_calcium_2019}.

The nucleation rate $B_0$ can be modeled using various approaches. The first class of models uses mass-action kinetics to express $B_0$ as a function of the product of the concentration of reactants raised to the power of their stoichiometric coefficients \parencite{sajjadi_population_2009,liu_modeling_2014,thanh_mechanisms_2014}. In precipitation/crystallization applications, it is also common to see power law expressions for $B_{0}$ in terms of the supersaturation \parencite{omar_crystal_2017}. For multicomponent systems, the expression for the supersaturation can be modified accordingly (e.g., see  \parencite{schwarzer_combined_2004,guldenpfennig_how_2019}). The second class of models employ Arrhenius-type expressions, capturing properties such as the supersaturation, activities, or free energy directly. These models are often closely related to classical nucleation theory (e.g., see \parencite{di_pasquale_model_2012,di_pasquale_identification_2013,zhu_crystallization_2016}). 

The use of species concentrations (and supersaturation defined in terms of concentrations) for modeling nucleation, while straightforward and sufficient for many applications, is a simplification of the physics of the process. In reality, the species activities need to be used which requires an accurate description of the thermodynamics of the system. By using the species activities, the use of the supersaturation or activities or free energies in the nucleation model are equivalent. Thus far, there has been limited deployment of detailed thermodynamic modeling in the PBM literature, with most studies adapting specific semi-empirical methods to model precipitation of inorganic systems (e.g., see \parencite{peng_gypsum_2015,zhu_crystallization_2016,guldenpfennig_how_2019}). We also note the work of \textcite{widenski_use_2011} which employed various activity-coefficient thermodynamic models predictively to model the crystallization of acetaminophen with limited success as the thermodynamic models considered were not sufficiently accurate. For LNP systems, it is likely that advanced thermodynamic models, such as those described in Sec.~\ref{sec:thermo}, are necessary and additional work to interface the thermodynamics with the PBM is needed.

\subsubsection{Growth}

The growth of a particle is the process by which the characteristic dimension of a particle (e.g., diameter) increases as solute(s) from the bulk solution are incorporated into the particle \parencite{myerson_handbook_2019}. As growth, like nucleation, involves the formation of a new phase, the driving force for the process is also the supersaturation of the solute(s). The previous discussion on how the supersaturation is defined and employed for describing nucleation is directly relevant to the growth rate as well. Mechanistically, growth takes place sequentially with the transport of the solute(s) from the bulk solution to the particle-solution interface followed by the incorporation of the solute(s) into the particle. Transport of the heat of precipitation and/or counter-diffusion of liberated species/solvent to the bulk may also need to be accounted for \parencite{abegg_crystal_1968,myerson_handbook_2019}.

The first class of expressions for the growth rate $G$ employ a mechanistic understanding of the growth process by considering whether the growth is diffusion or reaction limited. When the growth is limited by either step, that step becomes the rate-determining step for growth, enabling one to formulate an expression incorporating mechanistic insight. Examples in the PBM literature include  \parencite{schwarzer_combined_2004,di_pasquale_identification_2013} for diffusion-controlled growth and \parencite{perala_two-step_2014,handwerk_mechanism-enabled_2019,hong_droplet-based_2021,pico_silver_2023} for reaction-controlled growth. The second class of growth rate expressions use an empirical power-law type functional form with the supersaturation raised to  power that is treated as a fitted parameter. Power-law type expressions, while a simplification, are commonly used in the PBM literature and are able to adequately describe important trends and experimental data even for complex systems. It is also possible to account for temperature-dependent effects on the growth rate by incorporating an Arrhenius-type expression for the prefactor in the growth rate term \parencite{myerson_handbook_2019}.  

A variety of modifications that include size-dependency to the growth rate expression have been proposed (e.g., see \parencite{szilagyi_digital_2022} and citations therein) that have mostly enabled the PBM to better fit experimental data. Size-dependent effects on growth can physically arise in situations such as when size-dependent solubility (i.e., the Gibbs-Thomson effect) is significant (typically at nano-scales) which results in smaller particles growing slower or even dissolving at the expense of larger particles \parencite{madras_temperature_2004,iggland_population_2012, szilagyi_digital_2022} or with growth rate dispersion \parencite{srisanga_crystal_2015}. Another situation relevant to nanoparticles where size-dependent effects may need to be addressed is when growth is mass transfer-limited which can also result in slower growth rates for smaller particles \parencite{mullin_crystallization_2001}.

\subsubsection{Agglomeration/Aggregation}

{ Particle-particle interactions between smaller particles that lead to the formation of a larger particle (i.e., agglomeration/aggregation) are highly relevant in LNP formation both in the rapid mixing and buffer exchange steps. During the rapid mixing step, it is likely that in addition to nucleation and growth processes described previously, agglomeration will also be a relevant phenomena that drives increases in particle size. In the buffer exchange step where the pH of the system is increased to physiological pH, it has been reported that agglomeration (often referred to as ``fusion" in the context of the LNP literature) of smaller initial nanoparticles formed during the rapid mixing step into the larger final LNP \parencite{kamanzi_quantitative_2024,kulkarni_fusion-dependent_2019}. It is likely that two separate PBMs would need to be formulated for the rapid mixing and buffer exchange steps considering the different phenomena and physics involved (e.g., during the buffer exchange step, no nucleation and growth are occurring). However, these two separate PBMs can be easily coupled to provide a model for the whole formation process as the results from the PBM for the rapid mixing step can be used as the input for the PBM for the buffer exchange step.}

Much of the PBM literature defines agglomeration and aggregation as distinct phenomena, with one phenomenon being where the merged particle is strongly integrated and the other phenomenon being where the merged particle is less tightly integrated. The literature is inconsistent, however, in terms of which word is applied to which type of merged particle. Given that the mechanistic modeling equations are the same from the perspective of the PBM, this review does not distinguish between the two phenomena. 

We consider the case of binary aggregation where a larger particle is formed by the collision and subsequent merging of two smaller particles. For LNP production, considering that the system is both dilute and that the difference in particle size is not very large, the binary agglomeration assumption is reasonably justifiable \parencite{ramkrishna_population_2000,baba_dempbm_2021}. For conceptualizing how the aggregation and breakage terms are developed, it can be helpful to think in terms of the particle volume $V$, but they can be readily expressed in terms of the particle size $L$ through the assumption of a relationship between $V$ and $L$ (e.g., $V \propto L^{3}$ for simple particles or $V \propto L^{D_{f}}$, where $D_{f}$ is the mass-based fractal dimension, for ramified aggregates) \parencite{marchisio_quadrature_2003,jeldres_population_2018}. 

To illustrate how the birth, $B_a$, and death, $D_a$, terms due to agglomeration are formulated, consider a particle of volume $V$. Particles of volume $V$ (with corresponding length $L$) can be formed by the collision of two smaller particles of volumes $V'$ (with corresponding length $\lambda$) and $V-V'(L-\lambda)$. Simultaneously, particles of volume $V$ can form larger particles by colliding with particles of any size \parencite{ramkrishna_population_2000,hakansson_dynamic_2009}. Considering that LNPs have reasonably defined shapes (i.e., spherical to first approximation and not complex ramified aggregated structures), the assumption of $V \propto L^{3}$ is reasonable, enabling $B_{a}$ and $D_{a}$, expressed in terms of $L$, to be written as \parencite{marchisio_quadrature_2003}
\begin{equation}
     B_a = \frac{L^{2}}{2}\! \int_{L_0}^{L} \frac{\beta((L^{3}-\lambda^{3})^{1/3}, \lambda)}{(L^{3} - \lambda^{3})^{2/3}} n((L^{3}-\lambda^{3})^{1/3},t) n(\lambda,t) \mathrm{d}\lambda, 
     \quad D_a = n(L,t) \int_{L_0}^\infty \beta((L, \lambda) n(\lambda ,t) \mathrm{d}\lambda,
     \label{eq:PBE_Agglo}
\end{equation}
where $\beta$ is a proportionality constant and is referred to as the agglomeration kernel or collision rate. The agglomeration kernel captures important information about the mechanism(s) by which collisions between particles occur that subsequently lead to agglomeration. Examples of various aggregation kernels can be found in \parencite{vanni_approximate_2000,marchisio_quadrature_2003,pena_modeling_2017,myerson_handbook_2019}. For systems containing nanoparticles in turbulent flows, the collision mechanisms (and consequent functional forms for $\beta$) commonly considered are due to Brownian motion and/or turbulence \parencite{schwarzer_predictive_2006,marchisio_use_2009,raponi_population_2023}. The contribution from the various mechanisms can be captured additively in the expression for $\beta$ (e.g., see \parencite{marchisio_use_2009,jeldres_population_2018,raponi_population_2023}). { As changes to the external environment i.e., the solvent composition, pH, and ionic strength, have a significant influence on LNPs in both the rapid mixing and buffer exchange steps, the effect of these changes on agglomeration can be captured by introducing a collision efficiency factor to the agglomeration kernel e.g., see \parencite{ahmad_population_2008}.} For modeling flash nanoprecipitation, a process closely related to LNP production, \textcite{cheng_competitive_2010,cheng_kinetic_2010} developed an aggregation model that incorporates significant mechanistic insight of the agglomeration process specific to polymeric nanoparticles containing a diblock copolymer.

\subsubsection{Breakage}

Breakage is the process by which larger particles fragment into smaller particles. Physically, breakage occurs when the forces acting on the particle exceeds the forces holding it together (e.g., interfacial tension). External stress on the particle can be introduced through a variety of mechanisms such exposure to a turbulent flow field (which results in turbulent inertial and viscous stresses) \parencite{hakansson_emulsion_2019}, or impact on a surface (e.g., during milling). The interested reader is directed to the review by \textcite{liao_literature_2009} for a detailed discussion on the breakage mechanism. For LNP systems, external stresses on the particles are likely to arise through exposure to the turbulent flow field in the mixer. However, considering the small particle size, comparatively low turbulent intensities (even for fully turbulent mixers), and the short residence time in the mixer, breakage is unlikely to play a significant role during LNP formation. Its inclusion in this review is for completeness. 

To illustrate how the birth $B_{b}$ and death $D_{b}$ terms due to breakage are formulated, consider a particle of volume $V$ (with corresponding length $L$) breaking into two daughter particles, which is the most common assumption in PBMs \parencite{wang_novel_2003} and is supported by experimental evidence for the fragmentation of small particles ($<\SI{500}{\micro\meter}$) \parencite{maas_experimental_2007,rasche_mathematical_2018}. Particles of volume $V$ can fragment to yield smaller particles of volume $V'$ (with corresponding length $\lambda$) and $V-V'$. More generally, particles can break into multiple daughter particles, in which $B_{b}$ and $D_{b}$ are given by \parencite{marchisio_quadrature_2003}
\begin{equation}
B_b = \int_L^\infty b(\lambda) s(L,\lambda) n(\lambda ,t)  \mathrm{d}\lambda, \quad    D_b = b(L) n(L,t),   
\label{eq:PBE_break}
\end{equation}
where $b(L)$ is the breakage kernel/frequency and $s(L,\lambda)$ is daughter size distribution function that describes the size of the daughter particles that are formed during breakage. The physics of the breakage process is captured in both these functions and the interested reader is referred to \parencite{wang_novel_2003,liao_literature_2009} and citations therein for examples and additional information for both functions.

\subsubsection{Next Steps}

Developing a PBM for tracking the particle size distribution of LNPs will likely be an iterative exercise requiring testing of various aspects of the PBM such as the presence/absence of specific terms and their functional forms. Simplifications where possible should also be considered. For instance, the time derivative could be neglected to consider the steady-state population distribution (e.g., see \parencite{schall_nucleation_2018}). However, this approach may pose numerical issues due to the solution of the resultant boundary-value problem often requiring a good initial guess for convergence. For a PBM-only model, it is also possible to reformulate the advective and diffusive terms in terms of residence time/outflow term. A helpful rule-of-thumb is to employ the simplest possible model that sufficiently captures the physics of the LNP formation process and is able to qualitatively and quantitatively explain experimental data. A range of efficient computational methods are available for the solution of PBMs (e.g., moment methods \parencite{marchisio_quadrature_2003,marchisio_solution_2005} and finite differences \parencite{inguva_efficient_2022,inguva_efficient_2023}). 

Thus far, this section has extensively discussed the application of PBMs to model the LNP particle size distribution. While this use case is important, the descriptive capabilities of PBMs motivates further development. As elaborated in Sec.~\ref{sec:cfd_pbm}, the PBM can be coupled to CFD models, thus providing a framework to resolve the spatio-temporal dynamics of LNP formation. Extensions to multidimensional PBMs, incorporating additional relevant intrinsic variables such as mRNA loading can be considered, but require novel methods to formulate expressions for the growth rates and source/sink terms.

\begin{table}[htbp]
\tiny
\centering
\caption{Exemplar functional forms for the various terms in a PBM: Nucleation rate $B_0$, Growth rate $G$, Aggregation kernel $\beta(L,\lambda)$, breakage kernel $b(L)$, and selection function $s(L,\lambda)$. }\renewcommand{\arraystretch}{1.5}
\begin{tabular}{ccl}    
\toprule
Function & Expression & Comments\\
\hline
$B_{0}$  & $k_1 \prod_{i = 1}^n (C_i)^{\alpha_i}$ & \makecell[l]{Mass-action kinetics (e.g., see \parencite{sajjadi_population_2009,liu_modeling_2014,pico_silver_2023}) \\ $C_i= \mbox{concentration of reactant }i$, $\alpha_i= \mbox{stoichiometric coefficient}$} \\
& $k_1 S^{n_b}$ & \makecell[l]{Power law expression (e.g., see \parencite{omar_crystal_2017}) \\ $S= \mbox{supersaturation ratio}$, $n_b=\mbox{fitted parameter}$} \\
& $k_1 \exp\!{\left(\frac{-B}{\ln^2\!{S}}\right)}$ & \makecell[l]{Classical nucleation theory \\ (e.g., see \parencite{schwarzer_combined_2004,roelands_analysis_2006,myerson_handbook_2019}) \\ $B=\mbox{fitted parameter}$, $S=\mbox{supersaturation ratio}$} \\
   
\hline

$G$& $k_2 \prod_{i=1}^n (C_i)^{\alpha_i}$ & \makecell[l]{Mass-action kinetics (e.g., see \parencite{sajjadi_population_2009,liu_modeling_2014,thanh_mechanisms_2014})\\ $C_i= \mbox{concentration of reactant }i$, $\alpha_i=\mbox{stoichiometric coefficient}$} \\
& $k_2 \exp\!{(-\frac{E_g}{RT})} S^{n_{g}}$ &   \makecell[l]{Power law with temperature dependence \parencite{myerson_handbook_2019} \\ $R=\mbox{ideal gas constant}$, $E_g=\mbox{activation energy}$, $n_g=\mbox{fitted parameter}$} \\
& $2\frac{\textrm{Sh} D C^*}{\rho_{m}} \frac{S-1}{L}$ & \makecell[l]{Diffusion-controlled growth \\ (e.g., see \parencite{schwarzer_combined_2004,di_pasquale_model_2012}) \\ $\textrm{Sh}=\mbox{Sherwood number}$, $D=\mbox{diffusion coefficient}$, \\ $C^*=\mbox{equilibrium concentration}$,  $\rho_m=\mbox{particle molar density}$} \\
\hline
  $\beta(L,\lambda )$ & $\frac{2kT}{3\mu}\frac{(L+\lambda)^2}{L\lambda}$& \makecell[l]{Collision due to Brownian motion \\ (e.g., see \parencite{smoluchowski_versuch_1918,schwarzer_combined_2004,di_pasquale_model_2012})} \\
 & $1.294\left(\frac{\epsilon}{\nu}\right)^{\!1/2} (L + \lambda)^3$& \makecell[l]{ Collision due to turbulence (e.g., see \parencite{di_pasquale_model_2012})} \\ 
 \hline
    
$b(L)$ & $b_0$& \makecell[l]{Constant (e.g., see \parencite{marchisio_quadrature_2003})} \\
 & $ b_0 L^m$  & Power law, $m=\mbox{fitted parameter}$ (e.g., see \parencite{hounslow_generic_2005,jeldres_population_2018})  \\

 \hline
    
$s(L,\lambda)$ &  
$\begin{cases}
2, \quad \text{if } L = \frac{\lambda}{2^{1/3}} \\
0, \quad \text{otherwise}
\end{cases}$
& Symmetric fragmentation (e.g., see \parencite{marchisio_quadrature_2003,hounslow_generic_2005})   \\ 
      \bottomrule
    \end{tabular}
    \renewcommand{\arraystretch}{1}
    \label{tbl:pbm}
\normalsize
\end{table}

\subsection{Coupled CFD-PBM Models}
\label{sec:cfd_pbm}

Coupling the CFD model (which provides detailed information on the flow field and transport of the various species in the mixer) with PBMs (which provides a framework to model key particulate processes) yields a powerful approach that can be used to model the evolution of the particle size distribution with increased predictive capability. The governing equations of a CFD-PBM model can be represented by combining the one of the multphase models based on the Eulerian representation outlined in Sec.~\ref{sec:multiphase_model} with the PBM outlined in Sec.~\ref{sec:pbm}. 

The coupled CFD-PBM model necessitates numerical methods for its solution and it can be computationally costly (the model is often high dimensional, typically involving 2-3 spatial dimensions, 1 or more intrinsic coordinates such as particle size, and time). Advances in algorithms have been pursued and currently available methods enable the efficient solution of CFD-PBM problems. The CFD-PDF-PBM method uses the cost savings from the PDF approach outlined in Sec.~\ref{sec:spec_tra} to speed up computations \parencite{woo_simulation_2006}. The MP-PIC-PBM method, based on a Eulerian--Lagrangian using the parcel assumption, decouples the PBM in 3-dimensional space and has been demonstrated to significantly decrease the computational costs of CFD-PBM models \parencite{kim_multi-phase_2020,kim_multi-scale_2021,kim_investigation_2024}.

\subsection{Phase-Field Models}
Phase-field models (PFMs) are a powerful class of continuum-scale models for studying microstructure evolution in multiphase systems and interfacial phenomena \parencite{lamorgese_phase_2011,chen_phase-field_2002,anderson_diffuse-interface_1998}. PFMs introduce and track the evolution of one or more auxiliary fields (the phase field(s)) which specify which phase is in each point in space in the system. It should be noted that PFMs typically resolve the physics at the length- and time-scales of a handful of LNPs $\mathcal{O}(\SI{1}{\micro \metre}, \SI{1}{\micro \second})$. Phase-field variables (also known as order parameters) can be categorized as either conserved (e.g., composition / density) or non-conserved (e.g., grain orientation / structural variants) with the Cahn--Hilliard equation being used to describe the former and the Allen--Cahn equation being used for the latter \parencite{chen_classical_2022,hohenberg_theory_1977}. The driving force for evolution in the system is the reduction in the total free energy with the PFMs themselves capturing contributions from factors such as bulk chemical free energy and interfacial energy \parencite{chen_phase-field_2002}. PFMs are readily extensible to account for additional factors contributing to the total free energy of the system such as electrostatics and also for external influences such as shear stresses and temperature effects \parencite{chen_phase-field_2002}. This would typically involve modifying the PFM and/or coupling it to other equations such as species, charge, momentum, and energy conservation equations (e.g., see \parencite{takaki_phase-field_2014,chiu_conservative_2011,guyer_phase_2004}), resulting in a set of nonlinear coupled partial differential equations that require numerical methods for solution. Considering the types of species present in typical LNP formulations and the rapid mixing manufacturing process, PFMs are positioned as an excellent technique to study the formation of LNPs during the precipitation step and subsequent evolution of internal structures during downstream processes.

To illustrate how PFMs are constructed and some of the nuances in their use, we start with the classic Cahn--Hilliard equation for a binary mixture. The starting point for any PFM is an expression for the total free energy of the system \parencite{rowlinson_translation_1979,cahn_free_1958},
\begin{equation}
    \label{eq:CH_Functional}
    \frac{\mathcal{F}}{k_{\text{B}}T} = \int_{V} \left( f(c_1) + \frac{\kappa}{\!2}( \nabla c_1) \right)^{\!2} dV,
\end{equation}
where $f(c_1)$ is the homogeneous free energy, $\kappa$ is the gradient energy parameter, $c_1$ is the concentration of species 1 in the system, and $V$ is the volume of the system. Note that concentration of species 2 is inferred by a mass balance e.g., $c_1 + c_2 = 1$. Equation~\ref{eq:CH_Functional} is often referred to as the Landau--Ginzburg free energy functional \parencite{nauman_nonlinear_2001,gurtin_generalized_1996}. The chemical potential, $\mu$, is the variational derivative of \eqref{eq:CH_Functional},
\begin{equation}
\label{eq:mu_CH}
    \mu = \frac{\delta \mathcal{F}}{\delta c_{1}} = \frac{\partial \mathcal{F}}{\partial c_{1}} - \nabla \cdot 
    \frac{\partial \mathcal{F}}{\partial \nabla c_{1}} = \frac{\partial f}{\partial c_{1}} - \kappa \nabla^{2}c_{1}.
\end{equation}
The Cahn--Hilliard equation is obtained by incorporating a constitutive relation for the flux of species 1, $\mathbf{J}_{1}$, based on linear irreversible thermodynamics \parencite{chen_classical_2022}, 
\begin{equation}
    \label{eq:CH_Binary}
    \frac{\partial c_{1}}{\partial t} = - \nabla \cdot \mathbf{J}_{1} = \nabla \cdot (L \nabla \mu) = \nabla \cdot \left[L \nabla\! \left( \frac{\partial f}{\partial c_{1}} - \kappa \nabla^{2}c_{1} \right) \right],
\end{equation}
where $L$ is the Onsanger coefficient relating the diffusive flux to the gradient in chemical potential and is given by 
\begin{equation}
L = \frac{D(c_{1})c_{1}}{k_{\text{b}}T},
\label{eq:mobility}
\end{equation}
where $D(c_{1})$ is the diffusivity. The inclusion of the concentration $c_{1}$ in \eqref{eq:mobility} is necessary to recover classical Fickian diffusion in the limit of an ideal mixture \parencite{nauman_nonlinear_2001}. Many PFMs consider the difference in chemical potential (i.e., $\mu_{ij} = \mu_{i} - \mu_{j})$ when formulating the constitutive equation and PFM (e.g., see \parencite{petrishcheva_exsolution_2012,nauman_morphology_1994,cahn_free_1958}). This has no substantial impact in the PFM formulation due to the requirements from the Onsanger reciprocal relationships on $L$ \parencite{petrishcheva_exsolution_2012}, though it can be convenient for model development. 

An inspection of \eqref{eq:CH_Binary} indicates there are three terms that need to be specified for a complete model: the diffusivity $D$, the homogeneous free energy $f$, and the gradient energy parameter $\kappa$. Note that $\kappa$ is intimately related to $f$ \parencite{inguva_continuum-scale_2021,nauman_nonlinear_2001}, which emphasizes the need for supplying a suitable thermodynamic model of the system to accurately model the physics of the process. A good thermodynamic model of the system should be able to capture the phase behavior (e.g., liquid-liquid equilibria and critical points). For the purposes of brevity, we direct the interested reader to the following references \parencite{inguva_continuum-scale_2021,manzanarez_modeling_2017,teichert_comparison_2017,nauman_nonlinear_2001,ariyapadi_gradient_1990} and citations therein for a detailed discussion on how to incorporate physically-appropriate methods for the various terms in the Cahn--Hilliard equation. 

To extend \eqref{eq:CH_Binary} to account for multiple species ($n \geq 3$), we first extend the free energy functional to incorporate additional species,
\begin{equation}
\label{eq:CH_Functional_ncomp}
\frac{\mathcal{F}}{k_{\text{b}}T} = \int_{V} \left( f(c_{1}, c_{2}, \dots, c_{n}) + \sum_{i}^{N-1}\frac{\kappa_{i}}{2}(\nabla c_{i})^{2} + \sum_{j>i} \sum_{i}^{N-1}\kappa_{ij} (\nabla c_{i}) (\nabla c_{j}) \right) dV,
\end{equation}
where $\kappa_{i}$ and $\kappa_{ij}$ are the self- and cross-gradient energy parameters. Note that with the presence of three or more species, cross-gradient terms appear in the free energy functional. $\kappa_{ij}$ can be understood as a tensor capturing a description of the interfacial tension between phases $i$ and $j$. For a ternary system, the chemical potentials $\mu_{ij}$ are \parencite{inguva_numerical_2020,petrishcheva_exsolution_2012,nauman_morphology_1994},
\begin{align}
\mu_{12} &= \mu_{1} - \mu_{2} = \frac{\partial f}{\partial c_{1}} - \frac{\partial f}{\partial c_{2}} - (\kappa_{1} - \kappa_{12}) \nabla^{2}c_{1} + (\kappa_{2} - \kappa_{12})\nabla^{2}c_{2},\nonumber \\
\mu_{13} &= \mu_{1} - \mu_{3} = \frac{\partial f}{\partial c_{1}} - \frac{\partial f}{\partial c_{3}} - \kappa_{1}\nabla^{2}c_{1} - \kappa_{12}\nabla^{2} c_{2}, \nonumber \\
\mu_{23} &= \mu_{2} - \mu_{3} = \frac{\partial f}{\partial c_{2}} - \frac{\partial f}{\partial c_{3}} - \kappa_{2}\nabla^{2}c_{2} - \kappa_{12}\nabla^{2}c_{1},
\end{align}
which when combined with a flux expression $\mathbf{J}_{i} = \sum_{j}L_{ij}\nabla\mu_{ij}$, gives the transport equations
\begin{align}
\label{eq:CH_Ternary}
\frac{\partial c_{1}}{\partial t} &= \nabla \cdot \left( L_{12}\nabla \mu_{12} + L_{13}\nabla \mu_{13} \right), \nonumber \\
\frac{\partial c_{2}}{\partial t} &= \nabla \cdot \left( -L_{12}\nabla \mu_{12} + L_{23}\nabla \mu_{23} \right),
\end{align}
with $c_{3}$ being inferred from a mass balance constraint. Further extensions to the PFM can be pursued either by accounting for additional species following a similar procedure outlined above or by incorporating additional physics. The incorporation of additional physics can be achieved by either modifying the free energy functional in  \eqref{eq:CH_Functional} (as is the case with classical Density Functional Theory\parencite{zhang_interfacial_2021}) or modifying the transport equations. In the case of the latter, consider the formulation of the Cahn--Hilliard Navier--Stokes equation system for a binary system. The Cahn--Hilliard equation is modified with the addition of a convective term,
\begin{equation}
\frac{\partial c_{1}}{\partial t} + \mathbf{u}\cdot \nabla c_{1} = \nabla \cdot \left[ L \nabla\! \left( \frac{\partial f}{\partial c_{1}} - \kappa \nabla^{2}c_{1} \right) \right],
\end{equation}
where $\mathbf{u}$ is the velocity vector and is obtained from the solution of the corresponding modified incompressible Navier--Stokes equations,
\begin{align}
\nabla \cdot \mathbf{u} &= 0, \\
\rho \frac{\partial \mathbf{u}}{\partial t} + \rho \mathbf{u}\cdot \nabla \mathbf{u} &= -\nabla p + \nabla \cdot \left[ \eta (\nabla \mathbf{u} + \nabla \mathbf{u}^\top) \right] \!+ \mathbf{F}_{b},
\end{align}
where $\rho$ is the fluid density, $\eta$ is the fluid viscosity, and $\mathbf{F}_{b}$ is the coupling body force which is described as a diffuse surface tension force and is given by $\mathbf{F}_{b} = -c_{1}\nabla \mu$ \parencite{nauman_nonlinear_2001,jacqmin_calculation_1999}. The modification of $\mathbf{F}_{b}$ to account for multiple species is straightforward \parencite{zhou_phase_2006} and can be further expressed in terms of differences in chemical potential via a Gibbs--Duhem relationship. For LNP systems, the viscosity of the various species can be significantly different. This can be accounted for by modifying $\eta$ to introduce a composition dependence e.g., $\eta = c_{1} \eta_{1} + (1-c_{1}) \eta_{2}$ where $\eta_{1}$ and $\eta_{2}$ are the viscosities of species 1 and 2 respectively, following common methods in multiphase CFD e.g., see \parencite{deshpande_evaluating_2012}.Variations in the interfacial tension arising due to temperature and composition effects that may occur for example when one or more species behave as a surfactant can be captured by modifying $\kappa$ to incorporate such dependencies \parencite{lamorgese_phase-field_2016}.

\begin{figure}[htb]
    \centering
    \begin{subfigure}{0.48\textwidth}
        \centering
        \includegraphics[width=0.75\textwidth]{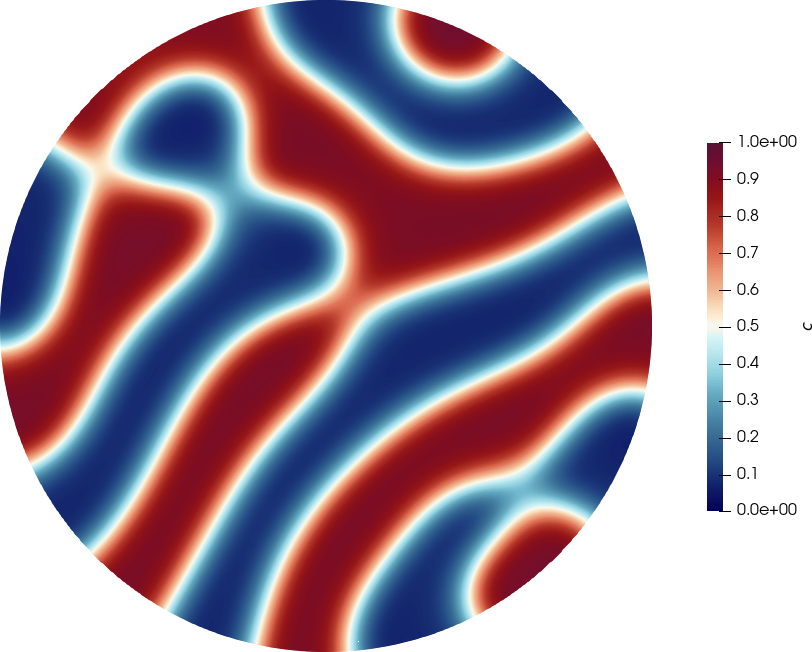} 
        \caption{Binary polymer blend on a 2D circular domain}
        \label{fig:pfm_circle}
    \end{subfigure}
    \hfill
    \centering
    \begin{subfigure}{0.48\textwidth}
        \centering
        \includegraphics[width=0.9\textwidth]{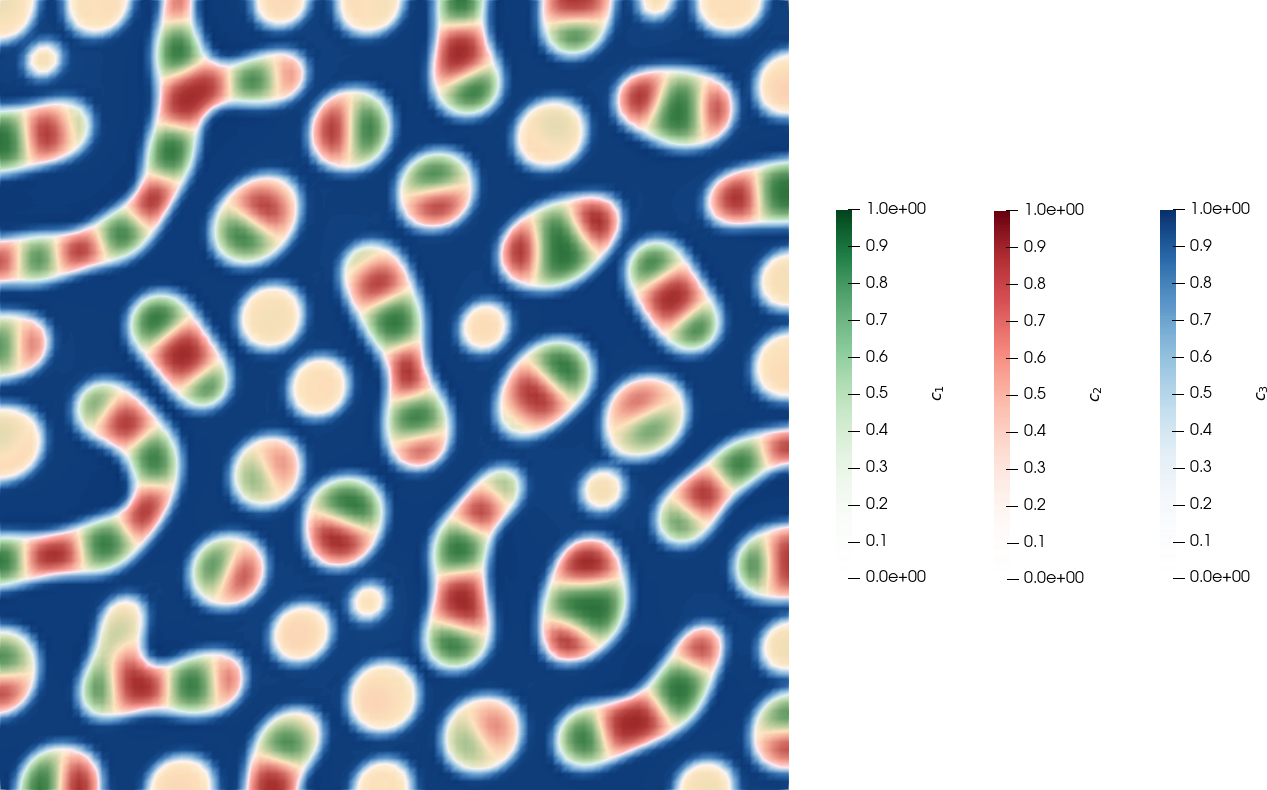} 
        \caption{Ternary polymer blend on a 2D rectangular domain}
        \label{fig:pfm_square}
    \end{subfigure}
    \caption{Exemplar phase-field model simulations of binary and ternary polymer blends using a Cahn--Hilliard model with degenerate mobility and logarithmic free potential on different geometries. Details of the simulation conditions can be found in the code repository. The color bars indicate the fraction of a species at a given point in space.}
    \label{fig:exemplar_pfm}
\end{figure}

Both the binary \eqref{eq:CH_Binary} and ternary \eqref{eq:CH_Ternary} PFMs with/without convection can form the basis of meaningful modeling of LNP formation with careful use. For example, the binary PFM can be used to approximate a ternary system and analyze the precipitation of a single lipid/polymer in the presence of an nonsolvent (e.g., see \parencite{hopp-hirschler_modeling_2018,kesler_modeling_2016}). Similarly, the ternary PFM, which is capable of demonstrating a diverse range of morphologies/patterns \parencite{inguva_numerical_2020,nauman_morphology_1994}, can characterize the precipitation of two species with the addition of a nonsolvent. Strategic choices of the initial and boundary conditions can also be considered to better describe the actual process. Exemplar simulations of the binary and ternary Cahn--Hilliard equations are shown in Fig.~\ref{fig:exemplar_pfm}. 

To fully describe the physics of a modern LNP formulation, the PFM would need to account for 4 or more species (i.e., lipids + RNA + Water + Ethanol), the complex thermodynamic and electrostatic interactions between some of the species, and shearing effects from the mixing process among other effects. It may also be useful to incorporate reactive terms into the PFM which can be used to describe the formation of colloidal aggregates \parencite{petersen_phase_2018,bazant_theory_2013}. The process is also inherently multi-scale as reactor-scale conditions such as solvent composition from mixing and fluid flow properties have an impact on the LNP structure formation. To our knowledge, the development and use of such a complex PFM in a multi-scale manner has yet to be undertaken in the literature. This is not surprising as requisite developments e.g., extensions to multiple components, coupling to other conservation equations, and numerical methods are still active areas of research. Consequently, a progressive approach where complexity is sequentially added to the PFM, with corresponding model benchmarking and validation at each step, will help in model development. Advances in adjacent fields can be leveraged to guide model formulation and solution. Particularly noteworthy areas include polymer precipitation \parencite{inguva_continuum-scale_2021,kesler_modeling_2016,vonka_modelling_2012}, polymeric membrane formation \parencite{hopp-hirschler_modeling_2018,zhou_phase_2006}, protein-RNA complex formation \parencite{natarajan_model_2023,grasselli_phase_2023,gasior_modeling_2020,}, and lipid membrane structure formation \parencite{arnold_active_2023,zhiliakov_experimental_2021}.

\subsection{Meso- and Molecular-Scale Methods}
\label{sec:md_sims}
Most of the approaches discussed above might be considered under the umbrella of continuum approaches where many of the degrees of freedom involved in the LNP system, such as bond vibrations and molecular orientation, have either been neglected or averaged over. When studying certain phenomena, such as particle diffusion or macro-scale structures, these degrees of freedom can have a significant impact. Solving the equations which represent the LNP system analytically is intractable. This motivates the use of meso- or molecular scale methods such as Molecular Dynamics (MD) simulations to explicitly account for all these degrees of freedom. 

\begin{figure}[ht!]
    \centering
    \includegraphics[width=0.6\textwidth]{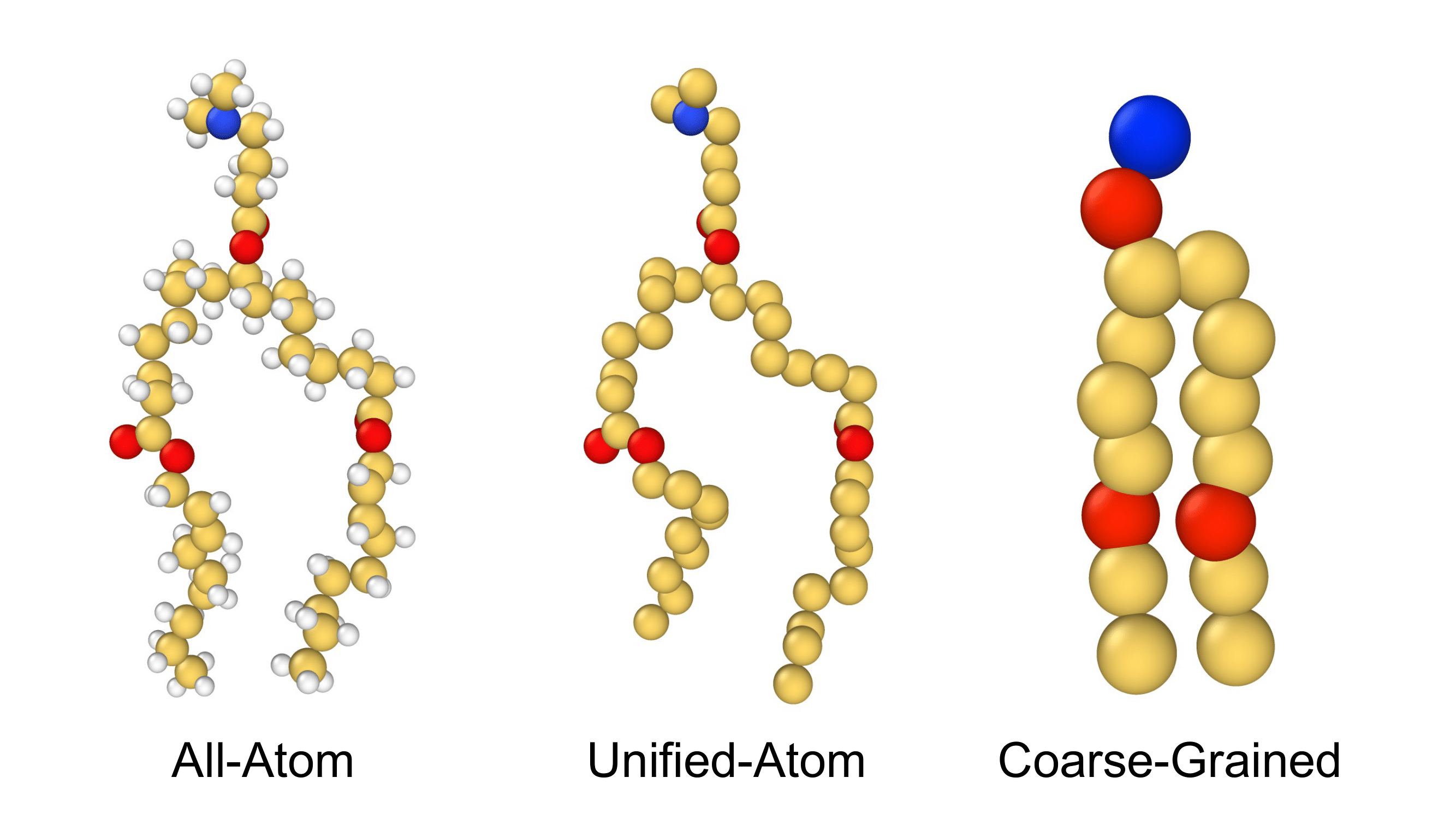}
    \vspace{-0.2cm}
    \caption{Different levels of representations of lipid-319 within molecular dynamics simulations: all-atom (left), unified-atom (center) and coarse-grained (right). Orange beads represent carbon atoms / alkyl groups, red beads represent oxygen atoms / carboxylic groups, blue beads represent nitrogen atoms / amine groups and white beads represent hydrogen atoms.}
    \label{fig:MD_rep}
\end{figure}

The most rigorous approach (with the exception of hybrid \textit{ab initio} methods) would use an all-atom (AA) representation of the species (left-most representation in Fig.\ \ref{fig:MD_rep}) where all atoms are accounted for explicitly. The parameters required to represent these species can be obtained from various standardized force fields (OPLS \parencite{jorgensen_development_1996}, AMBER \parencite{case_ambertools_2023}, CHARMM \parencite{brooks_charmm_2009}, etc.), which are compatible with most of the popular MD simulation packages (LAMMPS \parencite{thompson_lammps_2022}, GROMACS \parencite{abraham_gromacs_2015}, OpenMM \parencite{eastman_openmm_2017}, etc.). The challenge with this level of representation is the computational cost in simulating large systems. The smallest length ($\geq 100$ nm) and time ($\geq 1$ ms) scales of relevance for simulating LNP production is infeasible with any reasonably sized high-performance computing resources. As such, while all-atom resolution is desired, it is often more practical to simulate subsystems of the larger LNP system. For example, \textcite{trollmann_mrna_2022} used the CHARMM36 forcefield within GROMACS to study the pH-driven phase transition of the Comirnaty vaccine LNP by examining a lipid bilayer (19.1 nm thick) over a 4 ms simulation composed of the same components present in the full LNP. Indeed, this is the typical scale for performing all-atom simulations of lipid membranes \parencite{feller_molecular_2000,gumbart_molecular_2005,lindahl_membrane_2008,settanni_ph-dependent_2022} which is still capable of extracting information relevant to the full system. Attempting to simulate larger systems would be computationally challenging, not only due to the increased number of atoms being simulated, but the time required to equilibrate the system before even performing production simulations would dramatically increase as the magnitude of the slowest relaxation time will surely increase. \textcite{trollmann_mrna_2022} managed to simulate a 35-nm diameter LNP by restricting their system to maintain a spherical shape, at the cost of no longer being able to study the dynamics of the system.

The more common approach to simulate large-scale systems is to simply remove degrees of freedom within the system. For example, hydrogen atoms, due to their small size and rapid vibrations, can often be grouped with the heavy atom with which they are bonded into one larger group, resulting in a united-atom (UA) representation of the molecule (center representation in Fig.\ \ref{fig:MD_rep}). As hydrogen atoms can sometimes represent half of all atoms present in a system, this treatment can augment the length and time scales accessible within MD simulations. For such representations, force fields such as CHARMM \parencite{lee_charmm36_2014} and OPLS \parencite{jorgensen_optimized_1984} both provide UA variants, compatible with the various MD simulation packages. In the case of the CHARMM forcefield, a GUI has been developed where users can easily generate lipid bilayers to perform the simulations, bypassing much of the difficulty associated with initialising the simulations. While such approaches have not been used to study LNPs specifically as of yet, they have been used to study lipid bilayers \parencite{lee_charmm36_2014,das_molecular_2019} and small micelles \parencite{jorge_molecular_2008,roussel_multiscale_2014}, with structural properties proving to be very similar to those obtained from all-atom simulations. Unfortunately, even with this simpler representation, the computational cost is still too great to simulate large-scale systems.

Reaching the desired length scales requires dramatically simplifying the representation of species within the system such that multiple functional groups are represented by a single bead. Such a representation is often referred to as coarse-grained (CG); an example of such a representation is shown on the right in Fig.\ \ref{fig:MD_rep}. The objective of this representation is to maintain some structural information about the molecule, as well as some information regarding the types of interactions between functional groups. In principle, the force field parameter is adjusted to maintain some accurate representation of true system. In the case of the MARTINI force fields \parencite{souza_martini_2021}, numerous studies have been conducted to study LNPs (and similar systems) \parencite{lindahl_membrane_2008,bochicchio_interaction_2017} where, through much benchmarking \parencite{roussel_multiscale_2014,das_molecular_2019}, they have demonstrated being capable of reproducing results from all-atom representations. Another example of a CG approach would be dissipative particle dynamics (DPD) simulations where molecules are now represented by a few (sometimes a single) bead, practically losing the chemical identity of the molecule. In both of these approaches, the simplified approach allows researchers to study larger systems over larger time scales. In the case of DPD specifically, the improved representation of hydrodynamic interactions allows for a more accurate prediction of dynamic properties, as has been demonstrated in the case of lipid membranes \parencite{yang_computer_2010,angioletti-uberti_theory_2017,yong_effects_2023}, which would be almost inaccessible for all-atom approaches.

Regardless of which resolution is chosen when performing MD simulations, the subsequent analysis of trajectories and evaluation of specific properties require the use of tools such as MDAnalysis \parencite{michaud-agrawal_mdanalysis_2011}, mdcraft\cite{yeMDCraftPythonAssistant2024}, or TRAVIS \parencite{brehm_travisfree_2020}. In the case of the former, the package has been developed to enable user customization, as many of the studies cited in this section have done. However, an additional aspect to be aware of when performing MD simulations is that only one particular mode of interaction (for example, binding between ionizable lipids and mRNA, or the interaction between a lipid and the LNP interface) may be of interest. It is possible to use sampling methods to modify the simulations to study these interactions specifically. Such techniques can also be used to study interactions whose relaxation times are particularly slow and would require lengthy simulations to properly sample (an example of these techniques includes Replica-Exchange Molecular Dynamics). In recent years, these sampling techniques have been supplemented with Machine Learning, such as OPES, to significantly accelerate the sampling process. Packages such as PLUMED \parencite{bonomi_plumed_2009} and PySAGES \parencite{rico_pysages_2023} allow users to apply such techniques to their simulations and are compatible with multiple standard MD packages.

\section{Model-based Systems Engineering}
\label{sec:control}

The Quality-by-Design (QbD) framework, as articulated in the relevant regulatory guidance \parencite{us_food_and_drug_administration_q8r2_2009,yu_understanding_2014} highlights the importance of establishing relationships (i.e., models) between the critical process parameters (CPPs) in pharmaceutical manufacturing processes and the critical quality attributes (CQAs). It is highly desired that the established models can provide insight into the underlying physicochemical phenomena occurring within the process while enabling the interpretation of conditions beyond experimentation. These high-quality models, along with quality risk assessment, can further support the design of effective controllers for process regulation.

Section~\ref{sec:4_Process_Modeling_Approaches} describes modeling strategies for LNP production. Facilitated by the proposed process models, experiments can be optimally designed and planned to refine and validate the models, particularly focusing on identifying unspecified parameters in the proposed models. With the accessible CPPs and CQAs of interest in Figure~\ref{fig:cqa_schematic}, the models in Sec.~\ref{sec:4_Process_Modeling_Approaches} can be employed for the design of controls and soft sensors.

\subsection{Model-based Experiment Design}
Statistical experiment design is a widely adopted approach for strategically planning experiments. Employing first-principles models in the experimental design can yield the most informative data in terms of minimizing uncertainty in the estimated model parameters \parencite{franceschini_model-based_2008, abt_model-based_2018, shahmohammadi_using_2020}.

As discussed in Sec.~\ref{sec:4_Process_Modeling_Approaches}, the LNP formation process can be modeled by a set of integropartial differential-algebraic equations. These models can be represented as
\begin{equation}
\begin{aligned}
    & \boldsymbol{f}(\boldsymbol{x}(\boldsymbol{r}, t), {\boldsymbol{x}}_t(\boldsymbol{r}, t), {\boldsymbol{x}}_{\boldsymbol{r}}(\boldsymbol{r}, t), {\boldsymbol{x}}_{\boldsymbol{rr}}(\boldsymbol{r}, t), \boldsymbol{u}^c(\boldsymbol{r},t);   \boldsymbol{\theta}_{1}, \boldsymbol{\theta}_{2}, t) = 0 \\
    & {\boldsymbol{y}}(t) = \boldsymbol{h}(\boldsymbol{x}(\boldsymbol{r},t))
\end{aligned}
\label{eq:DAE}
\end{equation}
where $\boldsymbol{f}(\cdot)$ is the function comprising state variables $\boldsymbol{x}(\boldsymbol{r}, t)$, their derivatives with respect to position vector $\boldsymbol{r}$ and time $t$, and manipulated variables $\boldsymbol{u}^c(\boldsymbol{r},t)$; $\boldsymbol{\theta}_{1}$ and $\boldsymbol{\theta}_{2}$ are sets of specified and unspecified parameters, respectively; 
and $\boldsymbol{h}(\cdot)$ is the function relating the state and output variables ${\boldsymbol{y}}(t)$. The system properties of interest are selected as the state variables, such as the mass $M$ in (\ref{eq:balance1}), velocity $\boldsymbol{u}$ in (\ref{eq:1phom_momentum_balance}), and species density $n$ in (\ref{eq:pbm_general}). Based on these equations, the corresponding manipulated variables $\boldsymbol{u}^c(\boldsymbol{r},t)$ can be mass flowrates $F^{i}$ and species concentrations of the inlet flowrates. Usually the spatial derivatives and any integrals are approximated, which is known as the numerical method of lines \parencite{schiesser_numerical_1991}. This procedure results in a set of differential-algebraic equations of the same general form as \eqref{eq:DAE} but without having an explicit dependency on the spatial derivatives. To simplify the nomenclature, the rest of this section assumes that the numerical method of lines or an alternative method for removing the spatial derivatives and integrals has been applied. This assumption can be removed with small adjustments, although with more complex nomenclature.

Given the parametric model \eqref{eq:DAE} and experimental data, methods to estimate the unspecified parameters include maximum likelihood estimation \parencite{bogaerts_parameter_2004, canova_mechanistic_2023, destro_mechanistic_2023}, ordinary least squares \parencite{souza_procedure_2005}, and Bayesian estimation \parencite{hermanto_robust_2008,candy_bayesian_2016}. The aim of these approaches is to identify values of $\boldsymbol{\theta}_{2}$ that minimize the prediction error $\boldsymbol{\epsilon}(\boldsymbol{\theta}_{2}, t)$ between the predicted model outputs $\hat{\boldsymbol{y}}(t)$ and the observed data $\boldsymbol{y}_\mathrm{meas}(t)$ \parencite{ljung_system_1998}, i.e.,
\begin{equation}
 \hat{\boldsymbol{\theta}}_{2} = \arg \min_{\boldsymbol{\theta}_{2}} \boldsymbol{V_L}\big(\boldsymbol{\epsilon}(\boldsymbol{\theta}_{2}, t)\big)
\label{eq:ParamterEstimation}
\end{equation}
where $\boldsymbol{V_L}$ is a scalar-valued function of the prediction error, with a widely used choice being a quadratic function,  $\boldsymbol{V_L} = \frac{1}{2}\boldsymbol{\epsilon}^\top(\boldsymbol{\theta}_{2}, t) \boldsymbol{\Sigma}_{\boldsymbol{\nu}}^{-1} \boldsymbol{\epsilon}(\boldsymbol{\theta}_{2}, t)$, where $\boldsymbol{\Sigma_{\nu}}$ is the covariance matrix of the measurement errors. 

By evaluating the shape of the parameter likelihood function, the uncertainty of the parameter estimates can be quantified by a multivariate probability distribution function (pdf), which can be represented equivalently in terms of $100(1-\alpha)\%$ confidence regions with $\alpha\in (0,1)$. Integrals of the multivariate pdf can be taken to construct the single-parameter pdfs for each parameter, which can be represented equivalently in terms of a confidential interval \parencite{beck_parameter_1977}. Any estimated parameter for which the confidence interval is unbounded is not structurally or practically identifiable \parencite{raue_structural_2009,wieland_structural_2021,canova_mechanistic_2023}, in which case some of the predicted states may be inaccurate. Whether the parameter uncertainties result in poor accuracy of the predictions of interest can be assessed by propagating the model uncertainties through the process model to calculate prediction intervals for the model outputs of interest, e.g., the CQAs (e.g., \parencite{nagy_worst-case_2003,nagy_distributional_2007}). The process model and its uncertainties can be directly incorporated into the experimental design procedure, for a variety of objective functions, including minimization of the uncertainty in the model parameters or the prediction errors, and distinguishing between alternative hypothesized mechanisms \parencite{cho_experimental_2003, togkalidou_parameter_2004, bandara_optimal_2009,kutalik_optimal_2004, donckels_kernel-based_2009}.

\subsection{Model-based Control Design}

The CQAs for well-designed continuous processes are often controllable using
proportional-integral-derivative (PID) controllers \parencite{lakerveld_application_2015}. Model-based control can provide better product quality for processes that have strong multivariable interactions and/or often operate near constraints. The process models introduced in Sec.~\ref{sec:4_Process_Modeling_Approaches} can be incorporated into such control designs, either directly or indirectly through construction of a reduced-order model \parencite{paulson_fast_2018}. Model-based controllers can be desired to adapt to system changes \parencite{gutierrez_mpc-based_2014, oravec_robust_2018, hong_mechanistic_2021}. Model-based control strategies that have been used in manufacturing processes include feedforward control \parencite{ohkubo_hybrid_2023}, linear quadratic regulation \parencite{b_kanwar_modeling_2022}, model reference adaptive control \parencite{dochain_monitoring_1998, quo_adaptive_2011} and model predictive control (MPC) \parencite{mesbah_model_2017, hong_mechanistic_2021}. An optimal control formulation associated with the MPC strategy is 
\begin{equation}
\begin{aligned}
 &\min_{\boldsymbol{u}^c(t)} J = \int_{0}^{T_f} \Phi(\boldsymbol{x}(t), \boldsymbol{u}^c(t), t) dt \\
 &\text{subject to:}\\
 & \text{model equations in (\ref{eq:DAE}}) \\
 & \boldsymbol{u}^c_\mathrm{LB} \leq \boldsymbol{u}^c(t) \leq \boldsymbol{u}^c_\mathrm{UB} \\
 & \boldsymbol{x}_\mathrm{LB} \leq \boldsymbol{x}(t) \leq \boldsymbol{x}_\mathrm{UB} \\
\end{aligned}
\label{eq:MPC_formulation}
\end{equation}
In this optimization, the manipulated variables $\boldsymbol{u}^c(t)$ are determined to minimize the control objective $J$, representing the objective function $\Phi$ integrated over the time horizon $[0, T_f]$. The objective function for a continuous process is typically the mean-squared error of the difference between the measured and desired CQAs over a prediction horizon while penalizing sharp moves in the manipulated variables over a control horizon.\footnote{For batch and fed-batch processes, the objective function is typically defined as the closeness of the CQAs to their desired values at the end of the process \parencite{nagy_robust_2003}.} In this optimization formulation, the future manipulated variables are required to satisfy the model equation as well as input and state bounds, $\boldsymbol{u}^c(t) \in [\boldsymbol{u}^c_\mathrm{LB},~\boldsymbol{u}^c_\mathrm{UB}]$ and $\boldsymbol{x}(t) \in [\boldsymbol{x}_\mathrm{LB},~\boldsymbol{x}_\mathrm{UB}]$. 

As discussed in the previous section, the model equations, \eqref{eq:MPC_formulation}, are typically represented as ordinary differential equations (ODEs) or differential-algebraic equations (DAEs), which can be derived from the governing PDEs using such methods as polynomial approximation, the method of moments, the method of weighted residuals, and the finite difference, volume, or element methods applied to the spatial variables. Such methods have been applied for MPC design for the regulation of particle size distribution (e.g., \parencite{nagy_advances_2012}), which is one of the critical attributes in LNP production.

\subsection{Model-based Sensing Strategy}
When direct measurement of a physical property of interest becomes time-consuming, infeasible, or challenging, as is the case for many LNP properties and quality attributes (especially in a production setting), the integration of process models with analytical measurement data can be employed to implement a soft sensors \parencite{golabgir_observability_2015, mears_application_2017, y_jiang_review_2021}. Based on state estimation algorithms, these soft sensors can leverage the inherent relationships between accessible variables to estimate physical quantities that are either unmeasurable or challenging to measure. State estimators include Luenberger observers \parencite{duan_model_2020}, extended Kalman filters \parencite{de_assis_soft_2000}, and Bayesian estimators \parencite{mesbah_comparison_2011}. Most state estimators can be written in the form,
\begin{equation}
\begin{aligned}
    & \dot{\hat{\boldsymbol{x}}}(\boldsymbol{r},t) =  \boldsymbol{f}(\hat{\boldsymbol{x}}(\boldsymbol{r}, t), \boldsymbol{u}^c (\boldsymbol{r},t); \boldsymbol{\theta}_{1}, \boldsymbol{\theta}_{2}) + \boldsymbol{L} (\boldsymbol{y}(t)-\hat{\boldsymbol{y}}(t)), \\
    & \hat{\boldsymbol{y}}(t) = \boldsymbol{h}(\hat{\boldsymbol{x}}(\boldsymbol{r},t)),
\end{aligned}
\label{eq:Observer}
\end{equation}
where values of $\boldsymbol{\theta}_2$ have been credibly identified and specified, $\boldsymbol{L}$ is the observer gain, and ${\hat{\boldsymbol{x}}}(\boldsymbol{r},t)$ is the state estimate or the physical quantity of interest to be estimated.

The literature on soft sensors for bioprocesses is well established. For instance, consider the estimation of the specific growth rate of a recombinant Escherichia coli strain by using the measured heat flow produced by the cells and applying balance equations of biomass and heat \parencite{biener_calorimetric_2010}. An alternative state estimator based on measurable ammonia titration can achieve the same purpose \parencite{sundstrom_software_2008}. Given such state estimates, regulation of the specific growth rate becomes straightforward while ensuring the minimization of growth-inhibitory byproducts. State estimator designs of similar depth have been developed for the estimation of many other properties \parencite{komives_bioreactor_2003}. State estimators have also been demonstrated for bioprocess operations modeled by PDEs, including simulated moving bed chromatography \parencite{kupper_efficient_2009} and lyophilization \parencite{srisuma_thermal_2023}.

Given that rich literature on state estimation for bioprocessing, it is reasonable to expect that similar state estimation design strategies will be applicable to the continuous production of LNPs, which can be used in a real-time monitoring system for promptly identifying and addressing variations in the specified CQAs outlined in Fig.~\ref{fig:cqa_schematic} and Table~\ref{tbl:cqa}. Similar to other biotherapeutic manufacturing applications \parencite{kramer_hybrid_2019, golabgir_combining_2016, narayanan_hybrid-ekf_2020} -- by leveraging online and offline analytical techniques as highlighted in Table~\ref{tbl:cqa} and the models proposed in Sec.~\ref{sec:4_Process_Modeling_Approaches} -- real-time sensing strategies can be developed for the measurable and unmeasurable specifics of LNP production.

\section{Outlook}

LNPs are and will likely continue to remain an important delivery platform for nucleic acid therapeutics which are currently seeing intense levels of research and development, both from a drug discovery and manufacturing perspective. To that end, advances in the fundamental understanding of LNP formation will be essential to support both product and process development. In this article, we have outlined how various classes of mathematical modeling tools can be employed to study LNPs and improve their manufacturing. Such an endeavor will require adopting a multiscale approach where information from mixer-scale models (e.g., mass balances / CFD) need to be cascaded down to smaller-scale models (e.g., phase-field models) while insights from more detailed methods can be used to refine simpler models. We advocate for interested readers to adopt a progressive approach when developing models, whereby model development and experimental validation are first carried out on simpler systems (e.g., liposomal systems or LNPs with simpler formulations) prior to embarking on modeling a novel NAT-LNP system. 

While the development and use of such modeling strategies in the context of NAT-LNPs is nascent, this state of affairs should be viewed as an exciting opportunity for impactful fundamental and translational research. For some classes of models (e.g., CFD / population balance modeling), the basic know-how and tools are already well-established in the literature and adapting these methods to study LNP formation should be comparatively straightforward. Other classes of models (e.g., predictive thermodynamic modeling and phase-field modeling) likely require significant advances in the state of the art to be sufficient for describing NAT-LNP systems. However, simplified versions of these methods (e.g., phase-field models for binary/ternary mixtures) can still have utility if employed judiciously. For all classes of models, rigorous validation with experimental data is necessary, but can be challenging to do so considering the complexity and/or cost of measuring specific aspects of the LNPs such as its internal structure. Concomitant advances in analytical methods and sensor technology will be invaluable for supporting model development and deepening product/process understanding.

\section*{Availability of Code}

Example codes for implementing the various models and techniques discussed in the manuscript can be found at \url{https://github.com/pavaninguva/LNP_Models}.

\section*{Acknowledgments}

This research was supported by the U.S. Food and Drug Administration under the FDA BAA-22-00123 program, Award Number 75F40122C00200. Financial support is also acknowledged from the Agency for Science, Technology and Research (A*STAR), Singapore.

\section{Nomenclature}

\DefTblrTemplate{middlehead,lasthead}{default}{Continued from previous page}
\begin{longtblr}[caption = {Nomenclature},label = {tab:nomen}]{colspec={ccc},rowhead=1}
\toprule
Symbol & Description & Units \\
\midrule
& \textit{Transport Properties}  & \\
\midrule
$r$ & Radius of gyration & \SI{}{\meter}\\
$\mu$ & Viscosity & \SI{}{\pascal\second}\\
$k$ & Boltzmann constant & \SI{}{\joule\per\K}\\
$T$ & Temperature & \SI{}{\kelvin}\\
$N_n$ & Number of nucleotides & \SI{}{\meter} \\
$D_{AB}$ & Diffusivity coefficient &\SI{}{\meter\squared \per \second}\\
$D_{0}$ & Diffusion of solvent & \SI{}{\meter\squared \per \second}\\
$D_{r}$ & Rotational diffusivity & \SI{}{\radian \squared \per\second} \\
$D_{t}$ & Translational diffusivity & \SI{}{\meter\squared \per \second}\\
$\xi$ & Correlation length & \SI{}{\meter}\\
$k^\text{hydro}$ & Hydrodynamic interaction parameter & --\\
$N_p$ & Degree of polymerization &  -- \\
$M_r$ & Molecular weight &  \SI{}{\gram \per \mole}\\
$a$ & Length of a monomer &  \SI{}{\meter} \\
\midrule
& \textit{Thermodynamic Properties}  & \\
\midrule
$A$ & Helmholtz free energy & \SI{}{\joule}\\
$\mu$ & Chemical potential & \SI{}{\joule\per\mole}\\
$n$ & Moles & \SI{}{\mole} \\
$V$ & Volume &  \SI{}{\cubic \meter} \\
$T$ & Temperature &  \SI{}{\kelvin}\\
$\gamma$ & Activity coefficient & --\\
$x$ & Molar fraction & \SI{}{\mole\per\mole}\\
$R$ & Universal gas constant & \SI{}{\joule\per\kelvin\per\mole}\\
$K$ & Partition coefficient & --\\
$G$ & Gibbs free energy & \SI{}{\joule}\\
$p$ & Pressure & \SI{}{\pascal}\\
$Z$ & Charge & --\\
$\phi$ & Phase fraction & --\\
$\psi$ & Electrochemical potential difference & \SI{}{\joule\per\mole}\\
\midrule
& \textit{Mass and Energy Balances}  & \\
\midrule
$M$ & Total mass holdup in the mixer& \SI{}{\gram}\\
$M_{j}$ & Mass holdup of species $j$ in the mixer & \SI{}{\gram} \\
$F^{i}$ & Mass flowrate of stream $i$ &  \SI{}{\gram\per\second}\\
$x_{j}^{i}$ & Mass fraction of species $j$ in stream $i$ & -- \\
$K_{D,j}$ & Partition coefficient of species $j$ between LNP and raffinate phases & -- \\
$H$ & Total enthalpy holdup in the mixer & \SI{}{\joule}\\
$h^{i}$ & specific enthalpy of stream $i$ & \SI{}{\joule \per \gram} \\
\midrule
& \textit{Computational Fluid Dynamics}  & \\
\midrule
$\boldsymbol{u}$ & Velocity of the fluid & \SI{}{\meter\per\second}\\
$\rho$ & Density of fluid & \SI{}{\kilo \gram \per \cubic \meter} \\
$p$ & Pressure of fluid & \SI{}{\kilo\gram\per\meter\per\second\squared} \\
$T$ & Temperature of fluid &  \SI{}{\kelvin}\\
$\boldsymbol{\tau}$ & Viscous stress tensor & \SI{}{\kilo\gram\per\meter\per\second\squared} \\
$\boldsymbol{b}$ & Body force per unit mass & \SI{}{\meter\per\squared\second} \\
$\langle\boldsymbol{u}\rangle$ & Ensemble-averaged velocity of fluid & \SI{}{\meter\per\second}  \\
$\langle p \rangle$ & Ensemble-averaged pressure of fluid & \SI{}{\kilo\gram\per\meter\per\second\squared} \\
$\langle\boldsymbol{\tau}\rangle$ & Ensemble-averaged viscous stress tensor & \SI{}{\kilo\gram\per\meter\per\second\squared} \\
$\langle\boldsymbol{\tau'}\rangle$ & Reynolds stress & \SI{}{\kilo\gram\per\meter\per\second\squared} \\
$\overline{\boldsymbol{u}}$ & Filtered velocity of fluid & \SI{}{\meter\per\second} \\
$\overline{p}$ & Filtered pressure of fluid & \SI{}{\kilo\gram\per\meter\per\second\squared} \\
$\overline{\boldsymbol{\tau}}$ & Filtered viscous stress tensor & \SI{}{\kilo\gram\per\meter\per\second\squared} \\
$\Tilde{\boldsymbol{\tau}}$ & Sub-grid scale (SGS) stress & \SI{}{\kilo\gram\per\meter\per\second\squared} \\
$\Tilde{\boldsymbol{J}}$ & Sub-grid scale (SGS) scalar flux & -- \\
$\mu$ & Viscosity of fluid & \SI{}{\kilo\gram\per\meter\per\second} \\
$D_{we}$ & Mass diffusivity of water-ethanol mixture & \SI{}{\meter\squared\per\second} \\
$C_{pm}$ & Specific heat capacity of water-ethanol mixture & \SI{}{\joule\per\kilo\gram\per\kelvin} \\
$\kappa_m$ & Thermal diffusivity of water-ethanol mixture & \SI{}{\meter\squared\per\second} \\
$S_h$ & Heat of water-ethanol mixing & \SI{}{\joule\per\second} \\
$k$ & Turbulent kinetic energy & \SI{}{\meter\squared\per\second\squared} \\
$\mu_t$ & Turbulent viscosity & \SI{}{\kilo\gram\per\meter\per\second} \\
$S_k$ & Source of turbulent kinetic energy & \SI{}{\kilo\gram\per\meter\per\cubic\second} \\
$\varepsilon$ & Turbulent dissipation rate & \SI{}{\meter\squared\per\cubic\second} \\
$\kappa$ & von K\'arm\'an constant & \SI{}{-} \\
$S_\varepsilon$ & Source of turbulent dissipation & \SI{}{\kilo\gram\per\meter\per\second\tothe{4}} \\
$\varphi$  & Volume fraction of ethanol & -- \\
$\rho_w$ & Density of water & \SI{}{\kilo\gram\per\cubic\meter} \\
$\rho_e$ & Density of ethanol & \SI{}{\kilo\gram\per\cubic\meter} \\
$\rho_m$ & Density of mixture & \SI{}{\kilo\gram\per\cubic\meter} \\
$\mu_w$ & Dynamic viscosity of water & \SI{}{\kilo\gram\per\meter\per\second} \\
$\mu_e$ & Dynamic viscosity of ethanol & \SI{}{\kilo\gram\per\meter\per\second} \\
$\mu_m$ & Dynamic viscosity of mixture & \SI{}{\kilo\gram\per\meter\per\second} \\
$\lambda_K$ & Kolmogorov length scale & \SI{}{\meter}  \\
$\lambda_B$ & Batchelor length scale & \SI{}{\meter} \\
$[\text{mRNA}]$ & Concentration of mRNA & per convention \\
$[\text{L}_i]$ & Concentration of ionic lipid & per convention \\
$[\text{L}_n]$ & Concentration of neutral lipid & per convention \\
$[\text{LNP}]$ & Concentration of LNP & per convention \\
$V$ & Electric potential in the fluid & \SI{}{\kilo\gram\meter\squared\per\cubic\second\per\ampere} \\
$e$ & Elementary charge & \SI{}{\ampere\second} \\
$k_B$ & Boltzmann constant & \SI{}{\joule\per\kelvin} \\
$\epsilon\epsilon_r$ & Dielectric constant of fluid & \SI{}{\ampere\squared\second\tothe{4}\per\kilo\gram\per\cubic\meter} \\
$D_{\text{mRNA}}$ & Molecular diffusivity of mRNA & \SI{}{\meter\squared\per\second} \\
$D_{\text{L}_i}$ & Molecular diffusivity of ionic lipid & \SI{}{\meter\squared\per\second} \\
$D_{\text{L}_n}$ & Molecular diffusivity of neutral lipid & \SI{}{\meter\squared\per\second} \\
$ D_{\text{LNP}}$ & Molecular diffusivity of LNP & \SI{}{\meter\squared\per\second} \\
$D_{t,\text{mRNA}}$ & Total diffusivity of mRNA & \SI{}{\meter\squared\per\second} \\
$D_{t,\text{L}_i}$ & Total diffusivity of ionic lipid & \SI{}{\meter\squared\per\second} \\
$D_{t,\text{L}_n}$ & Total diffusivity of neutral lipid & \SI{}{\meter\squared\per\second} \\
$D_{t,\text{LNP}}$ & Total diffusivity of LNP & \SI{}{\meter\squared\per\second} \\
$p_1$ & Probability of environment containing pure water & -- \\
$p_2$ & Probability of environment containing pure ethanol & -- \\
$p_3$ & Probability of environment containing a mixture of water and ethanol & -- \\
$G_1,G_{s,1}$ & Micromixing function in environment $1$ & \SI{}{\per\second} \\
$G_2,G_{s,2}$ & Micromixing function in environment $2$ & \SI{}{\per\second} \\
$M^{(3)},M_s^{(3)}$ & Micromixing functions for scalar transport in environment $3$ & -- \\
$\gamma,\gamma_s$ & Micromixing constants & \SI{}{\per\second} \\
$\langle\xi\rangle_3$ & Fraction of ethanol in environment $3$ & -- \\
$\mathbf{F}_{\text{st}}$ & Interfacial force per unit volume $3$ & \SI{}{\kilo\gram\per\second\squared\per\meter\squared} \\
$\alpha_{q}$ & Volume fraction of phase $q$ & \SI{}{-} \\
$\rho_{q}$ & Density of phase $q$ & \SI{}{\kilo\gram\per\cubic\meter} \\
$\mathbf{u}_{q}$ & Velocity of phase $q$ & \SI{}{\meter\per\second} \\
$\mathbf{u}_{\text{dr}}$ & Slip velocity between phases & \SI{}{\meter\per\second} \\
$\mathbf{\tau}_{q}$ & Stress tensor for phase $q$ & \SI{}{\kilo\gram\per\meter\per\second\squared} \\
$\mathbf{M}_{q}$ & Rate of momentum exchange per unit volume between phases & \SI{}{\kilo\gram\per\meter\squared\per\second\squared} \\
$m_{p}$ & Mass of particle $p$ & \SI{}{\kilo\gram} \\
$\mathbf{u}_{p}$ & Velocity of particle $p$ & \SI{}{\meter\per\second} \\
$\mathbf{f}_{d}$ & Drag force on particle $p$ & \SI{}{\kilo\gram\meter\per\second\squared} \\
$\mathbf{f}_{g}$ & Gravitational force on particle $p$ & \SI{}{\kilo\gram\meter\per\second\squared} \\
$\mathbf{f}_{c}$ & Collision force on particle $p$ & \SI{}{\kilo\gram\meter\per\second\squared} \\
\midrule
& \textit{Population Balance Models} & \\
\midrule
$n(L,t)$ & Number density  & \SI{}{\meter \tothe{-4} }\\
$V$&Volume of LNP & \SI{}{\cubic \meter}\\
$L$ & Size of LNP & \SI{}{\meter}\\
$\lambda$ & Size of LNP & \SI{}{\meter}\\
$G$&Growth rate &  \SI{}{\meter\per\second}\\
$k_1$&Nucleation kinetic constants& \SI{}{\per\second}\\
$k_2$&Growth kinetic constants&  \SI{}{\cubic\meter\per\second}\\
$L_0$&Size of nucleated LNP  &   \SI{}{\meter }\\
$S$&Source term &   \SI{}{\meter\tothe{-4}\per\second}\\
$B_n$&Birth rate due to nucleation  &   \SI{}{\meter\tothe{-4}\per\second}\\
$B_a$& Birth rate due to agglomeration  &  \SI{}{\meter\tothe{-4}\per\second}\\
$D_a$& Death rate due to agglomeration &  \SI{}{\meter\tothe{-4}\per\second}\\
$B_b$&Birth rate due to breakage &   \SI{}{\meter\tothe{-4}\per\second}\\
$D_b$&Death rate due to breakage &   \SI{}{\meter\tothe{-4}\per\second}\\
$\beta$&Aggregation kernel&  \SI{}{\cubic \meter \per\second}\\
$b$& Breakage kernel &  \SI{}{\per\second} \\
$s$& Selection function &  --\\
$S$ & Supersaturation ratio & -- \\
$\alpha_{i}$& Stoichiometric coefficient & --  \\
$a,m,B,n_{b},n_{g}$  & Fitted parameter &   --\\
\midrule
& \textit{Phase-Field Models} & \\
\midrule
$\mathcal{F}$ & Total free energy of the system& \SI{}{\joule}\\
$f$ & Homogeneous free energy & \SI{}{\joule \per \cubic \meter} \\
$c_{i}$ & Concentration of species $i$ & -- \\
$\kappa_{i}$ & Self-gradient energy parameter &  \SI{}{\joule \per \meter}\\
$\kappa_{ij}$ & Cross-gradient energy parameter &  \SI{}{\joule \per \meter}\\
$V$ & Volume of system & \SI{}{\cubic \meter} \\
$\mu_{i}$ & Chemical potential of species $i$ & \SI{}{\joule \per \mole} \\
$\mu_{ij}$ & Difference in chemical potential between species $i$ and $j$ & \SI{}{\joule \per \mole} \\
$\mathbf{J}_{i}$ & Flux of species $i$ & \SI{}{\mole \per  \meter \squared \per \second} \\
$L$ & Mobility coefficient & \SI{}{\mole \squared \per \second \per \meter \per \joule } \\
$\mathbf{u}$ & Velocity & \SI{}{\meter \per \second} \\
$\rho$ & Fluid density & \SI{}{\kilo \gram \per \cubic \meter} \\
$p$ & Fluid pressure & \SI{}{\pascal}\\
$\eta$ & Fluid viscosity & \SI{}{\pascal \second} \\
$\mathbf{F}_{b}$ & Coupling surface tension body force & \SI{}{\joule \per \meter \squared} \\
\midrule
& \textit{Model-Based Systems Engineering} & \\
\midrule
$\boldsymbol{x}$ & State variable & -- \\
${\hat{\boldsymbol{x}}}$ & State estimate & --\\
$\boldsymbol{u}^c$ & Controllable variable & -- \\
$\boldsymbol{r}$ & Position vector in continuous phase& -- \\
$\boldsymbol{\theta}_{1}$ & Specified parameter& --\\
$\boldsymbol{\theta}_{2}$ & Unspecified parameter& -- \\
$\hat{\boldsymbol{\theta}}_{2}$ & Unspecified parameter estimate& -- \\
$\boldsymbol{f(\cdot)}$ & System function& -- \\
$\boldsymbol{h(\cdot)}$ & System output function& --\\
$\boldsymbol{\epsilon}$ & Prediction error& -- \\
${\boldsymbol{y}}$ & Model output& -- \\
$\hat{\boldsymbol{y}}$ & Model predicted output& -- \\
$\boldsymbol{y}_{meas}$ & Observed output& -- \\
$\boldsymbol{V_L}$ & Scalar-valued function & -- \\
$\boldsymbol{\Sigma_{\nu}}$ & Covariance matrix of measurement errors& -- \\
$J$ & Control objective& -- \\
$\Phi$ & Objective function& --\\ 
$T_f$ & Time horizon& \SI{}{\second}\\
$\boldsymbol{L}$ & Observer gain& --\\

\bottomrule
\end{longtblr}

\printbibliography

@article{fongIonCorrelationsTheir2021,
	title = {Ion Correlations and Their Impact on Transport in Polymer-Based Electrolytes},
	volume = {54},
	issn = {0024-9297, 1520-5835},
	url = {https://pubs.acs.org/doi/10.1021/acs.macromol.0c02545},
	doi = {10.1021/acs.macromol.0c02545},
	pages = {2575--2591},
	number = {6},
	journaltitle = {Macromolecules},
	shortjournal = {Macromolecules},
	author = {Fong, Kara D. and Self, Julian and {McCloskey}, Bryan D. and Persson, Kristin A.},
	urldate = {2022-01-22},
	date = {2021-03-23},
	langid = {english},
}

@article{onsagerZurTheorieElectrolyte1926,
	title = {Zur Theorie der Electrolyte. I},
	volume = {27},
	pages = {388--392},
	journaltitle = {Phys. Z},
	author = {Onsager, Lars},
	date = {1926},
}

@article{onsagerZurTheorieElectrolyte1927,
	title = {Zur Theorie der Electrolyte. {II}},
	volume = {28},
	pages = {277--298},
	journaltitle = {Phys. Z},
	author = {Onsager, Lars},
	date = {1927},
}

@article{muthukumarDynamicsPolyelectrolyteSolutions1997,
	title = {Dynamics of polyelectrolyte solutions},
	volume = {107},
	issn = {0021-9606},
	url = {https://doi.org/10.1063/1.474573},
	doi = {10.1063/1.474573},
	pages = {2619--2635},
	number = {7},
	journaltitle = {The Journal of Chemical Physics},
	shortjournal = {The Journal of Chemical Physics},
	author = {Muthukumar, M.},
	urldate = {2024-04-23},
	date = {1997-08-15},
}

@article{liuEffectElectrostaticInteractions1998,
	title = {Effect of electrostatic interactions on the structure and dynamics of a model polyelectrolyte. I. Diffusion},
	volume = {109},
	issn = {0021-9606},
	url = {https://doi.org/10.1063/1.477377},
	doi = {10.1063/1.477377},
	pages = {7556--7566},
	number = {17},
	journaltitle = {The Journal of Chemical Physics},
	shortjournal = {The Journal of Chemical Physics},
	author = {Liu, Hui and Skibinska, Lidia and Gapinski, Jacek and Patkowski, Adam and Fischer, Erhard W. and Pecora, R.},
	urldate = {2024-04-23},
	date = {1998-11-01},
}

@article{skibinskaEffectElectrostaticInteractions1999,
	title = {Effect of electrostatic interactions on the structure and dynamics of a model polyelectrolyte. {II}. Intermolecular correlations},
	volume = {110},
	issn = {0021-9606},
	url = {https://doi.org/10.1063/1.477887},
	doi = {10.1063/1.477887},
	pages = {1794--1800},
	number = {3},
	journaltitle = {The Journal of Chemical Physics},
	shortjournal = {The Journal of Chemical Physics},
	author = {Skibinska, Lidia and Gapinski, Jacek and Liu, Hui and Patkowski, Adam and Fischer, Erhard W. and Pecora, R.},
	urldate = {2024-04-23},
	date = {1999-01-15},
}

@article{fongOnsagerTransportCoefficients2020,
	title = {Onsager Transport Coefficients and Transference Numbers in Polyelectrolyte Solutions and Polymerized Ionic Liquids},
	volume = {53},
	issn = {0024-9297},
	url = {https://doi.org/10.1021/acs.macromol.0c02001},
	doi = {10.1021/acs.macromol.0c02001},
	pages = {9503--9512},
	number = {21},
	journaltitle = {Macromolecules},
	shortjournal = {Macromolecules},
	author = {Fong, Kara D. and Self, Julian and {McCloskey}, Bryan D. and Persson, Kristin A.},
	urldate = {2024-04-23},
	date = {2020-11-10},
	note = {Publisher: American Chemical Society},
}

@article{bernardElectrophoreticMobilityPolyelectrolyte1991,
	title = {Electrophoretic mobility in polyelectrolyte solutions},
	volume = {95},
	issn = {0022-3654},
	url = {https://doi.org/10.1021/j100176a086},
	doi = {10.1021/j100176a086},
	pages = {9508--9513},
	number = {23},
	journaltitle = {The Journal of Physical Chemistry},
	shortjournal = {J. Phys. Chem.},
	author = {Bernard, Olivier and Turq, Pierre and Blum, Lesser},
	urldate = {2023-05-07},
	date = {1991-11-01},
	note = {Publisher: American Chemical Society},
}

@article{lotgering-linPureSubstanceMixture2018,
	title = {Pure Substance and Mixture Viscosities Based on Entropy Scaling and an Analytic Equation of State},
	volume = {57},
	issn = {0888-5885},
	url = {https://doi.org/10.1021/acs.iecr.7b04871},
	doi = {10.1021/acs.iecr.7b04871},
	pages = {4095--4114},
	number = {11},
	journaltitle = {Industrial \& Engineering Chemistry Research},
	shortjournal = {Ind. Eng. Chem. Res.},
	author = {Lötgering-Lin, Oliver and Fischer, Matthias and Hopp, Madlen and Gross, Joachim},
	urldate = {2024-05-12},
	date = {2018-03-21},
}

@article{bellModifiedEntropyScaling2019,
	title = {Modified Entropy Scaling of the Transport Properties of the Lennard-Jones Fluid},
	volume = {123},
	issn = {1520-6106},
	url = {https://doi.org/10.1021/acs.jpcb.9b05808},
	doi = {10.1021/acs.jpcb.9b05808},
	pages = {6345--6363},
	number = {29},
	journaltitle = {The Journal of Physical Chemistry B},
	shortjournal = {J. Phys. Chem. B},
	author = {Bell, Ian H. and Messerly, Richard and Thol, Monika and Costigliola, Lorenzo and Dyre, Jeppe C.},
	urldate = {2024-05-12},
	date = {2019-07-25},
}

@article{melfiViscosityImidazoliumIonic2024,
	title = {Viscosity of imidazolium ionic liquids and mixtures of {ILs} from entropy scaling using the {PC}-{SAFT} and {ePC}-{SAFT} equations of state},
	volume = {401},
	issn = {0167-7322},
	url = {https://www.sciencedirect.com/science/article/pii/S0167732224005567},
	doi = {10.1016/j.molliq.2024.124500},
	pages = {124500},
	journaltitle = {Journal of Molecular Liquids},
	shortjournal = {Journal of Molecular Liquids},
	author = {Melfi, Diego Trevisan and Scurto, Aaron M.},
	urldate = {2024-05-12},
	date = {2024-05-01},
}

@article{rosenfeldRelationTransportCoefficients1977,
	title = {Relation between the transport coefficients and the internal entropy of simple systems},
	volume = {15},
	rights = {http://link.aps.org/licenses/aps-default-license},
	issn = {0556-2791},
	url = {https://link.aps.org/doi/10.1103/PhysRevA.15.2545},
	doi = {10.1103/PhysRevA.15.2545},
	pages = {2545--2549},
	number = {6},
	journaltitle = {Physical Review A},
	shortjournal = {Phys. Rev. A},
	author = {Rosenfeld, Yaakov},
	urldate = {2024-05-12},
	date = {1977-06-01},
	langid = {english},
}

@article{lopezDiffusionViscosityUnentangled2021,
	title = {Diffusion and Viscosity of Unentangled Polyelectrolytes},
	volume = {54},
	issn = {0024-9297},
	url = {https://doi.org/10.1021/acs.macromol.1c01169},
	doi = {10.1021/acs.macromol.1c01169},
	pages = {8088--8103},
	number = {17},
	journaltitle = {Macromolecules},
	shortjournal = {Macromolecules},
	author = {Lopez, Carlos G. and Linders, Jürgen and Mayer, Christian and Richtering, Walter},
	urldate = {2024-05-12},
	date = {2021-09-14},
}

@article{jagerResidualEntropyScaling2023,
	title = {Residual Entropy Scaling for Long-Chain Linear Alkanes and Isomers of Alkanes},
	volume = {62},
	issn = {0888-5885},
	url = {https://doi.org/10.1021/acs.iecr.2c04238},
	doi = {10.1021/acs.iecr.2c04238},
	pages = {3767--3791},
	number = {8},
	journaltitle = {Industrial \& Engineering Chemistry Research},
	shortjournal = {Ind. Eng. Chem. Res.},
	author = {Jäger, A. and Steinberg, L. and Mickoleit, E. and Thol, M.},
	urldate = {2024-05-12},
	date = {2023-03-01},
}

@article{hajjToolsTranslationNonviral2017,
	title = {Tools for translation: non-viral materials for therapeutic {mRNA} delivery},
	volume = {2},
	rights = {2017 Macmillan Publishers Limited},
	issn = {2058-8437},
	url = {https://www.nature.com/articles/natrevmats201756},
	doi = {10.1038/natrevmats.2017.56},
	shorttitle = {Tools for translation},
	pages = {1--17},
	number = {10},
	journaltitle = {Nature Reviews Materials},
	shortjournal = {Nat Rev Mater},
	author = {Hajj, Khalid A. and Whitehead, Kathryn A.},
	urldate = {2025-04-08},
	date = {2017-09-12},
	langid = {english},
	note = {Publisher: Nature Publishing Group},
}

@article{kowalskiDeliveringMessengerAdvances2019,
	title = {Delivering the Messenger: Advances in Technologies for Therapeutic {mRNA} Delivery},
	volume = {27},
	issn = {1525-0016, 1525-0024},
	url = {https://www.cell.com/molecular-therapy-family/molecular-therapy/abstract/S1525-0016(19)30053-X},
	doi = {10.1016/j.ymthe.2019.02.012},
	shorttitle = {Delivering the Messenger},
	pages = {710--728},
	number = {4},
	journaltitle = {Molecular Therapy},
	shortjournal = {Molecular Therapy},
	author = {Kowalski, Piotr S. and Rudra, Arnab and Miao, Lei and Anderson, Daniel G.},
	urldate = {2025-04-08},
	date = {2019-04-10},
	pmid = {30846391},
	note = {Publisher: Elsevier},
}

@article{mengNanoplatformsMRNATherapeutics2021,
	title = {Nanoplatforms for {mRNA} Therapeutics},
	volume = {4},
	rights = {© 2020 {WILEY}-{VCH} Verlag {GmbH} \& Co. {KGaA}, Weinheim},
	issn = {2366-3987},
	url = {https://onlinelibrary.wiley.com/doi/abs/10.1002/adtp.202000099},
	doi = {10.1002/adtp.202000099},
	pages = {2000099},
	number = {1},
	journaltitle = {Advanced Therapeutics},
	author = {Meng, Chaoyang and Chen, Zhe and Li, Gang and Welte, Thomas and Shen, Haifa},
	urldate = {2025-04-08},
	date = {2021},
	langid = {english},
	note = {\_eprint: https://onlinelibrary.wiley.com/doi/pdf/10.1002/adtp.202000099},
}

@article{perezcisnerosReactiveSeparationSystems1997,
	title = {Reactive separation systems—I. Computation of physical and chemical equilibrium},
	volume = {52},
	issn = {0009-2509},
	url = {https://www.sciencedirect.com/science/article/pii/S0009250996004241},
	doi = {10.1016/S0009-2509(96)00424-1},
	pages = {527--543},
	number = {4},
	journaltitle = {Chemical Engineering Science},
	shortjournal = {Chemical Engineering Science},
	author = {Pérez Cisneros, E. S. and Gani, R. and Michelsen, M. L.},
	urldate = {2025-04-08},
	date = {1997-02-01},
}

@article{perdomoPredictiveGroupcontributionFramework2023,
	title = {A predictive group-contribution framework for the thermodynamic modelling of {CO}2{\textless}math{\textgreater}{\textless}msub is="true"{\textgreater}{\textless}mrow is="true"{\textgreater}{\textless}/mrow{\textgreater}{\textless}mrow is="true"{\textgreater}{\textless}mn is="true"{\textgreater}2{\textless}/mn{\textgreater}{\textless}/mrow{\textgreater}{\textless}/msub{\textgreater}{\textless}/math{\textgreater} absorption in cyclic amines, alkyl polyamines, alkanolamines and phase-change amines: New data and {SAFT}-γ{\textless}math{\textgreater}{\textless}mi is="true"{\textgreater}γ{\textless}/mi{\textgreater}{\textless}/math{\textgreater} Mie parameters},
	volume = {566},
	issn = {0378-3812},
	url = {https://www.sciencedirect.com/science/article/pii/S0378381222002540},
	doi = {10.1016/j.fluid.2022.113635},
	shorttitle = {A predictive group-contribution framework for the thermodynamic modelling of {CO}2{\textless}math{\textgreater}{\textless}msub is="true"{\textgreater}{\textless}mrow is="true"{\textgreater}{\textless}/mrow{\textgreater}{\textless}mrow is="true"{\textgreater}{\textless}mn is="true"{\textgreater}2{\textless}/mn{\textgreater}{\textless}/mrow{\textgreater}{\textless}/msub{\textgreater}{\textless}/math{\textgreater} absorption in cyclic amines, alkyl polyamines, alkanolamines and phase-change amines},
	pages = {113635},
	journaltitle = {Fluid Phase Equilibria},
	shortjournal = {Fluid Phase Equilibria},
	author = {Perdomo, Felipe A. and Khalit, Siti H. and Graham, Edward J. and Tzirakis, Fragkiskos and Papadopoulos, Athanasios I. and Tsivintzelis, Ioannis and Seferlis, Panos and Adjiman, Claire S. and Jackson, George and Galindo, Amparo},
	urldate = {2025-04-08},
	date = {2023-03-01},
}

@article{yeMDCraftPythonAssistant2024,
	title = {{MDCraft}: A Python assistant for performing and analyzing molecular dynamics simulations of soft matter systems},
	volume = {9},
	rights = {All rights reserved},
	issn = {2475-9066},
	url = {https://joss.theoj.org/papers/10.21105/joss.07013},
	doi = {10.21105/joss.07013},
	shorttitle = {{MDCraft}},
	pages = {7013},
	number = {100},
	journaltitle = {Journal of Open Source Software},
	author = {Ye, Benjamin B. and Walker, Pierre J. and Wang, Zhen-Gang},
	urldate = {2024-08-16},
	date = {2024-08-15},
	langid = {english},
}

@techreport{von_karman_mechanical_1931,
	address = {Washington, USA},
	title = {Mechanical similitude and turbulence},
	number = {NACA-TM-611},
	institution = {National Advisory Committee on Aeronautics},
	author = {{von K\'arm\'an}, T.H.},
	year = {1931},
}

@article{thorat_liquid_2012,
	title = {Liquid antisolvent precipitation and stabilization of nanoparticles of poorly water soluble drugs in aqueous suspensions: {Recent} developments and future perspective},
	volume = {181-182},
	journal = {Chemical Engineering Journal},
	author = {Thorat, Alpana A. and Dalvi, Sameer V.},
	month = feb,
	year = {2012},
	pages = {1--34},
}

@article{ahmad_population_2008,
	title = {Population balance model ({PBM}) for flocculation process: {Simulation} and experimental studies of palm oil mill effluent ({POME}) pretreatment},
	volume = {140},
	doi = {10.1016/j.cej.2007.09.014},
	number = {1-3},
	journal = {Chemical Engineering Journal},
	author = {Ahmad, A.L. and Chong, M.F. and Bhatia, S.},
	month = jul,
	year = {2008},
	pages = {86--100},
}

@article{kamanzi_simultaneous_2021,
	title = {Simultaneous, {Single}-{Particle} {Measurements} of {Size} and {Loading} {Give} {Insights} into the {Structure} of {Drug}-{Delivery} {Nanoparticles}},
	volume = {15},
	language = {en},
	number = {12},
	journal = {ACS Nano},
	author = {Kamanzi, Albert and Gu, Yifei and Tahvildari, Radin and Friedenberger, Zachary and Zhu, Xingqi and Berti, Romain and Kurylowicz, Marty and Witzigmann, Dominik and Kulkarni, Jayesh A. and Leung, Jerry and Andersson, John and Dahlin, Andreas and Höök, Fredrik and Sutton, Mark and Cullis, Pieter R. and Leslie, Sabrina},
	month = dec,
	year = {2021},
	pages = {19244--19255},
}

@article{dragovic_sizing_2011,
	title = {Sizing and phenotyping of cellular vesicles using {Nanoparticle} {Tracking} {Analysis}},
	volume = {7},
	number = {6},
	journal = {Nanomedicine: Nanotechnology, Biology and Medicine},
	author = {Dragovic, Rebecca A. and Gardiner, Christopher and Brooks, Alexandra S. and Tannetta, Dionne S. and Ferguson, David J.P. and Hole, Patrick and Carr, Bob and Redman, Christopher W.G. and Harris, Adrian L. and Dobson, Peter J. and Harrison, Paul and Sargent, Ian L.},
	month = dec,
	year = {2011},
	pages = {780--788},
}

@article{nogueira_analytical_2024,
	title = {Analytical techniques for the characterization of nanoparticles for {mRNA} delivery},
	volume = {198},
	doi = {10.1016/j.ejpb.2024.114235},
	journal = {European Journal of Pharmaceutics and Biopharmaceutics},
	author = {Nogueira, Sara S. and Samaridou, Eleni and Simon, Johanna and Frank, Simon and Beck-Broichsitter, Moritz and Mehta, Aditi},
	month = may,
	year = {2024},
	pages = {114235},
}

@article{li_payload_2022,
	title = {Payload distribution and capacity of {mRNA} lipid nanoparticles},
	volume = {13},
	doi = {10.1038/s41467-022-33157-4},
	number = {1},
	journal = {Nature Communications},
	author = {Li, Sixuan and Hu, Yizong and Li, Andrew and Lin, Jinghan and Hsieh, Kuangwen and Schneiderman, Zachary and Zhang, Pengfei and Zhu, Yining and Qiu, Chenhu and Kokkoli, Efrosini and Wang, Tza-Huei and Mao, Hai-Quan},
	month = sep,
	year = {2022},
	pages = {5561},
}

@article{barriga_coupling_2022,
	title = {Coupling {Lipid} {Nanoparticle} {Structure} and {Automated} {Single}‐{Particle} {Composition} {Analysis} to {Design} {Phospholipase}‐{Responsive} {Nanocarriers}},
	volume = {34},
	issn = {0935-9648, 1521-4095},
	url = {https://onlinelibrary.wiley.com/doi/10.1002/adma.202200839},
	doi = {10.1002/adma.202200839},
	language = {en},
	number = {26},
	urldate = {2025-04-04},
	journal = {Advanced Materials},
	author = {Barriga, Hanna M. G. and Pence, Isaac J. and Holme, Margaret N. and Doutch, James J. and Penders, Jelle and Nele, Valeria and Thomas, Michael R. and Carroni, Marta and Stevens, Molly M.},
	month = jul,
	year = {2022},
	pages = {2200839},
}

@article{malburet_taylor_2023,
	title = {Taylor {Dispersion} {Analysis} to support lipid-nanoparticle formulations for {mRNA} vaccines},
	volume = {30},
	doi = {10.1038/s41434-022-00370-1},
	language = {en},
	number = {5},
	journal = {Gene Therapy},
	author = {Malburet, Camille and Leclercq, Laurent and Cotte, Jean-François and Thiebaud, Jérôme and Bazin, Emilie and Garinot, Marie and Cottet, Hervé},
	month = may,
	year = {2023},
	pages = {421--428},
}

@article{kinsey_determination_2022,
	title = {Determination of lipid content and stability in lipid nanoparticles using ultra high‐performance liquid chromatography in combination with a {Corona} {Charged} {Aerosol} {Detector}},
	volume = {43},
	doi = {10.1002/elps.202100244},
	language = {en},
	number = {9-10},
	urldate = {2025-04-04},
	journal = {ELECTROPHORESIS},
	author = {Kinsey, Caleb and Lu, Tian and Deiss, Alyssa and Vuolo, Kim and Klein, Lee and Rustandi, Richard R. and Loughney, John W.},
	month = may,
	year = {2022},
	pages = {1091--1100},
}

@article{parot_quality_2024,
	title = {Quality assessment of {LNP}-{RNA} therapeutics with orthogonal analytical techniques},
	volume = {367},
	issn = {01683659},
	url = {https://linkinghub.elsevier.com/retrieve/pii/S0168365924000506},
	doi = {10.1016/j.jconrel.2024.01.037},
	language = {en},
	journal = {Journal of Controlled Release},
	author = {Parot, Jeremie and Mehn, Dora and Jankevics, Hanna and Markova, Natalia and Carboni, Michele and Olaisen, Camilla and Hoel, Andrea D. and Sigfúsdóttir, Margrét S. and Meier, Florian and Drexel, Roland and Vella, Gabriele and McDonagh, Birgitte and Hansen, Terkel and Bui, Huong and Klinkenberg, Geir and Visnes, Torkild and Gioria, Sabrina and Urban-Lopez, Patricia and Prina-Mello, Adriele and Borgos, Sven Even and Caputo, Fanny and Calzolai, Luigi},
	month = mar,
	year = {2024},
	pages = {385--401},
}

@article{kulkarni_fusion-dependent_2019,
	title = {Fusion-dependent formation of lipid nanoparticles containing macromolecular payloads},
	volume = {11},
	issn = {2040-3364, 2040-3372},
	url = {https://xlink.rsc.org/?DOI=C9NR02004G},
	doi = {10.1039/C9NR02004G},
	language = {en},
	number = {18},
	journal = {Nanoscale},
	author = {Kulkarni, Jayesh A. and Witzigmann, Dominik and Leung, Jerry and Van Der Meel, Roy and Zaifman, Josh and Darjuan, Maria M. and Grisch-Chan, Hiu Man and Thöny, Beat and Tam, Yuen Yi C. and Cullis, Pieter R.},
	year = {2019},
	pages = {9023--9031},
}

@article{gilbert_evolution_2024,
	title = {Evolution of the structure of lipid nanoparticles for nucleic acid delivery: {From} in situ studies of formulation to colloidal stability},
	volume = {660},
	doi = {10.1016/j.jcis.2023.12.165},
	language = {en},
	journal = {Journal of Colloid and Interface Science},
	author = {Gilbert, Jennifer and Sebastiani, Federica and Arteta, Marianna Yanez and Terry, Ann and Fornell, Anna and Russell, Robert and Mahmoudi, Najet and Nylander, Tommy},
	month = apr,
	year = {2024},
	pages = {66--76},
}

@article{vargas_dialysis_2023,
	title = {Dialysis is a key factor modulating interactions between critical process parameters during the microfluidic preparation of lipid nanoparticles},
	volume = {54},
	doi = {10.1016/j.colcom.2023.100709},
	language = {en},
	journal = {Colloid and Interface Science Communications},
	author = {Vargas, Ronny and Romero, Miquel and Berasategui, Tomás and Narváez-Narváez, David A. and Ramirez, Patricia and Nardi-Ricart, Anna and García-Montoya, Encarna and Pérez-Lozano, Pilar and Suñe-Negre, Josep Mª and Moreno-Castro, Cristina and Hernández-Munain, Cristina and Suñe, Carlos and Suñe-Pou, Marc},
	month = may,
	year = {2023},
	pages = {100709},
}

@article{kamanzi_quantitative_2024,
	title = {Quantitative visualization of lipid nanoparticle fusion as a function of formulation and process parameters},
	volume = {18},
	doi = {10.1021/acsnano.3c12981},
	language = {en},
	number = {28},
	journal = {ACS Nano},
	author = {Kamanzi, Albert and Zhang, Yao and Gu, Yifei and Liu, Faith and Berti, Romain and Wang, Benjamin and Saadati, Fariba and Ciufolini, Marco A. and Kulkarni, Jayesh and Cullis, Pieter and Leslie, Sabrina},
	month = jul,
	year = {2024},
	pages = {18191--18201},
}

@article{wu_process_2025,
	title = {Process development of tangential flow filtration and sterile filtration for manufacturing of {mRNA}-lipid nanoparticles: {A} study on membrane performance and filtration modeling},
	volume = {675},
	journal = {International Journal of Pharmaceutics},
	author = {Wu, Wenjun and Oliveira, Liliam Teixeira and Jain, Aarushi and Karpov, Yury and Olsen, Kirstin and Wu, Yu and Gopalakrishna Panicker, Rajesh Krishnan},
	month = apr,
	year = {2025},
	pages = {125520},
}

@article{kimura_development_2020,
	title = {Development of a microfluidic-based post-treatment process for size-controlled lipid nanoparticles and application to {siRNA} delivery},
	volume = {12},
	doi = {10.1021/acsami.0c05489},
	language = {en},
	number = {30},
	journal = {ACS Applied Materials \& Interfaces},
	author = {Kimura, Niko and Maeki, Masatoshi and Sato, Yusuke and Ishida, Akihiko and Tani, Hirofumi and Harashima, Hideyoshi and Tokeshi, Manabu},
	month = jul,
	year = {2020},
	pages = {34011--34020},
}

@article{hardianto_effect_2023,
	title = {The effect of ethanol on lipid nanoparticle stabilization from a molecular dynamics simulation perspective},
	volume = {28},
	doi = {10.3390/molecules28124836},
	language = {en},
	number = {12},
	journal = {Molecules},
	author = {Hardianto, Ari and Muscifa, Zahra Silmi and Widayat, Wahyu and Yusuf, Muhammad and Subroto, Toto},
	month = jun,
	year = {2023},
	pages = {4836},
}

@article{devos_impinging_2025,
	title = {Impinging jet mixers: {A} review of their mixing characteristics, performance considerations, and applications},
	volume = {71},
	doi = {10.1002/aic.18595},
	number = {1},
	journal = {AIChE Journal},
	author = {Devos, Cedric and Mukherjee, Saikat and Inguva, Pavan and Singh, Shalini and Wei, Yi and Mondal, Sandip and Yu, Huiwen and Barbastathis, George and Stelzer, Torsten and Braatz, Richard D. and Myerson, Allan S.},
	month = jan,
	year = {2025},
	pages = {e18595},
}

@article{bazant_theory_2013,
	title = {Theory of chemical kinetics and charge transfer based on nonequilibrium thermodynamics},
	volume = {46},
	number = {5},
	journal = {Accounts of Chemical Research},
	author = {Bazant, Martin Z.},
	month = may,
	year = {2013},
	pages = {1144--1160},
}

@article{petersen_phase_2018,
	title = {Phase separation of stable colloidal clusters},
	volume = {2},
	number = {9},
	journal = {Physical Review Materials},
	author = {Petersen, Thomas and Bazant, Martin Z. and Pellenq, Roland J. M. and Ulm, Franz-Josef},
	month = sep,
	year = {2018},
	pages = {095602},
}

@article{lamorgese_phase-field_2016,
	title = {Phase-field modeling of interfacial dynamics in emulsion flows: {Nonequilibrium} surface tension},
	volume = {85},
	issn = {03019322},
	journal = {International Journal of Multiphase Flow},
	author = {Lamorgese, A. and Mauri, R.},
	month = oct,
	year = {2016},
	pages = {164--172},
}

@article{kim_multi-phase_2020,
	title = {Multi-phase particle-in-cell coupled with population balance equation ({MP}-{PIC}-{PBE}) method for multiscale computational fluid dynamics simulation},
	volume = {134},
	journal = {Computers \& Chemical Engineering},
	author = {Kim, Shin Hyuk and Lee, Jay H. and Braatz, Richard D.},
	month = mar,
	year = {2020},
	pages = {106686},
}

@article{kim_multi-scale_2021,
	title = {Multi-scale fluid dynamics simulation based on {MP}-{PIC}-{PBE} method for {PMMA} suspension polymerization},
	volume = {152},
	journal = {Computers \& Chemical Engineering},
	author = {Kim, Shin Hyuk and Lee, Jay H. and Braatz, Richard D.},
	month = sep,
	year = {2021},
	pages = {107391},
}

@article{kim_investigation_2024,
	title = {Investigation of particle flow effects in slug flow crystallization using the multiscale computational fluid dynamics simulation},
	volume = {297},
	journal = {Chemical Engineering Science},
	author = {Kim, Shin Hyuk and Hong, Moo Sun and Braatz, Richard D.},
	month = sep,
	year = {2024},
	pages = {120238},
}

@book{gatski_compressibility_2009,
	address = {Amsterdam},
	title = {Compressibility, {Turbulence} and {High} {Speed} {Flow}},
	publisher = {Elsevier},
	author = {Gatski, T. B. and Bonnet, Jean-Paul},
	year = {2009},
}

@incollection{mahmoud_nanoparticle_2019,
	address = {Amsterdam},
	title = {Nanoparticle behaviour in multiphase turbulent channel flow},
	volume = {46},
	isbn = {978-0-12-818634-3},
	language = {en},
	booktitle = {Computer {Aided} {Chemical} {Engineering}},
	publisher = {Elsevier},
	author = {Mahmoud, B.H. and Mortimer, L.F. and Fairweather, M. and Rice, H.P. and Peakall, J. and Harbottle, D.},
	year = {2019},
	pages = {607--612},
}

@article{narasimha_cfd_2012,
	title = {{CFD} modeling of hydrocyclones: {Prediction} of particle size segregation},
	volume = {39},
	doi = {10.1016/j.mineng.2012.05.010},
	journal = {Minerals Engineering},
	author = {Narasimha, M. and Brennan, M.S. and Holtham, P.N.},
	month = dec,
	year = {2012},
	pages = {173--183},
}

@article{van_wachem_methods_2003,
	title = {Methods for multiphase computational fluid dynamics},
	volume = {96},
	issn = {13858947},
	number = {1-3},
	journal = {Chemical Engineering Journal},
	author = {Van Wachem, B.G.M. and Almstedt, A.E.},
	year = {2003},
	pages = {81--98},
}

@book{prosperetti_computational_2007,
	address = {Cambridge, United Kingdom},
	title = {Computational {Methods} for {Multiphase} {Flow}},
	publisher = {Cambridge University Press},
	author = {Prosperetti, A. and Tryggvason, G.},
	year = {2007},
}

@book{ansys_inc_ansys_2021,
	address = {Canonsburg, PA},
	edition = {2021 R2},
	title = {Ansys {Fluent} {Theory} {Guide}},
	publisher = {ANSYS, Inc.},
	author = {{ANSYS, Inc.}},
	year = {2021},
}

@article{fiuza_comparison_2018,
	title = {Comparison of {K}-{E} turbulence model wall functions applied on a {T}-junction channel flow},
	volume = {4},
	number = {1},
	journal = {International Journal of Engineering Research and Science},
	author = {Fiuza, G.C.C. and Rezende, A.L.T.},
	year = {2018},
	pages = {60--70},
}

@inproceedings{kim_near-wall_1995,
	title = {A near-wall treatment using wall functions sensitized to pressure gradient},
	volume = {217},
	publisher = {ASME},
	author = {Kim, S.E. and Choudhury, D.},
	year = {1995},
	pages = {273--279},
}

@article{marie_similarity_1997,
	title = {Similarity law and turbulence intensity profiles in a bubbly boundary layer at low void fractions},
	volume = {23},
	number = {2},
	journal = {International Journal of Multiphase Flow},
	author = {Marié, J.L. and Moursali, E. and Tran-Cong, S.},
	year = {1997},
	pages = {227--247},
}

@article{launder_numerical_1974,
	title = {The numerical computation of turbulent flows},
	volume = {3},
	number = {2},
	journal = {Computer Methods in Applied Mechanics and Engineering},
	author = {Launder, B.E. and Spalding, D.B.},
	year = {1974},
	pages = {269--289},
}

@book{candy_bayesian_2016,
	address = {Hoboken, NJ},
	title = {Bayesian {Signal} {Processing}},
	isbn = {978-1-119-12549-5},
	publisher = {John Wiley \& Sons, Ltd},
	author = {Candy, James},
	year = {2016},
}

@article{hermanto_robust_2008,
	title = {Robust {Bayesian} estimation of kinetics for the polymorphic transformation of {L}‐glutamic acid crystals},
	volume = {54},
	issn = {0001-1541, 1547-5905},
	doi = {10.1002/aic.11623},
	number = {12},
	journal = {AIChE Journal},
	author = {Hermanto, Martin Wijaya and Kee, Nicholas C. and Tan, Reginald B. H. and Chiu, Min‐Sen and Braatz, Richard D.},
	month = dec,
	year = {2008},
	pages = {3248--3259},
}

@article{rowlinson_translation_1979,
	title = {Translation of {J}. {D}. van der {Waals}' ''{The} thermodynamik theory of capillarity under the hypothesis of a continuous variation of density"},
	volume = {20},
	issn = {0022-4715, 1572-9613},
	doi = {10.1007/BF01011513},
	language = {en},
	number = {2},
	journal = {Journal of Statistical Physics},
	author = {Rowlinson, J. S.},
	month = feb,
	year = {1979},
	pages = {197--200},
}

@book{ramkrishna_population_2000,
	address = {London},
	title = {Population {Balances}: {Theory} and {Applications} to {Particulate} {Systems} in {Engineering}},
	isbn = {978-0-12-576970-9},
	publisher = {Academic Press},
	author = {Ramkrishna, Doraiswami},
	year = {2000},
}

@article{boyne_enthalpies_1967,
	title = {Enthalpies of mixing of ethanol and water at 25°{C}},
	volume = {12},
	issn = {0021-9568, 1520-5134},
	doi = {10.1021/je60034a008},
	number = {3},
	journal = {Journal of Chemical \& Engineering Data},
	author = {Boyne, J. A. and Williamson, Arthur Gordon},
	month = jul,
	year = {1967},
	pages = {318--318},
}

@article{felton_ml-saft_2024,
	title = {{ML}-{SAFT}: {A} machine learning framework for {PCP}-{SAFT} parameter prediction},
	volume = {492},
	shorttitle = {{ML}-{SAFT}},
	language = {en},
	journal = {Chemical Engineering Journal},
	author = {Felton, Kobi and Rasßpe-Lange, Lukas and Rittig, Jan and Leonhard, Kai and Mitsos, Alexander and Meyer-Kirschner, Julian and Knösche, Carsten and Lapkin, Alexei},
	year = {2024},
	pages = {151999},
}

@incollection{barba_polymeric_2019,
	address = {Amsterdam},
	title = {Polymeric and lipid-based systems for controlled drug release: an engineering point of view},
	isbn = {978-0-12-816505-8},
	shorttitle = {Polymeric and lipid-based systems for controlled drug release},
	language = {en},
	booktitle = {Nanomaterials for {Drug} {Delivery} and {Therapy}},
	publisher = {Elsevier},
	author = {Barba, Anna Angela and Bochicchio, Sabrina and Dalmoro, Annalisa and Caccavo, Diego and Cascone, Sara and Lamberti, Gaetano},
	editor = {Grumezscu, Alexandru M.},
	year = {2019},
	doi = {10.1016/B978-0-12-816505-8.00013-8},
	pages = {267--304},
}

@article{winter_spt-nrtl_2022,
	title = {{SPT}-{NRTL}: {A} physics-guided machine learning model to predict thermodynamically consistent activity coefficients},
	volume = {568},
	shorttitle = {{SPT}-{NRTL}},
	journal = {Fluid Phase Equilibria},
	author = {Winter, Benedikt and Winter, Clemens and Esper, Timm and Schilling, Johannes and Bardow, André},
	month = sep,
	year = {2022},
	note = {arXiv:2209.04135 [physics]},
	pages = {113731},
}

@article{inguva_computer_2018,
	title = {Computer design of microfluidic mixers for protein/{RNA} folding studies},
	volume = {13},
	issn = {1932-6203},
	doi = {10.1371/journal.pone.0198534},
	language = {en},
	number = {6},
	journal = {PLoS ONE},
	author = {Inguva, Venkatesh and Kathuria, Sagar V. and Bilsel, Osman and Perot, Blair James},
	month = jun,
	year = {2018},
	pages = {e0198534},
}

@article{lamorgese_phase_2011,
	title = {Phase field approach to multiphase flow modeling},
	volume = {79},
	issn = {1424-9286, 1424-9294},
	doi = {10.1007/s00032-011-0171-6},
	number = {2},
	journal = {Milan Journal of Mathematics},
	author = {Lamorgese, Andrea G. and Molin, Dafne and Mauri, Roberto},
	month = dec,
	year = {2011},
	pages = {597--642},
}

@article{anderson_diffuse-interface_1998,
	title = {Diffuse-interface methods in fluid mechanics},
	volume = {30},
	issn = {0066-4189, 1545-4479},
	doi = {10.1146/annurev.fluid.30.1.139},
	number = {1},
	journal = {Annual Review of Fluid Mechanics},
	author = {Anderson, D. M. and McFadden, G. B. and Wheeler, A. A.},
	month = jan,
	year = {1998},
	pages = {139--165},
}

@article{destro_advanced_2024,
	title = {Advanced methodologies for model-based optimization and control of pharmaceutical processes},
	volume = {45},
	issn = {22113398},
	doi = {10.1016/j.coche.2024.101035},
	language = {en},
	journal = {Current Opinion in Chemical Engineering},
	author = {Destro, Francesco and Inguva, Pavan K and Srisuma, Prakitr and Braatz, Richard D},
	month = sep,
	year = {2024},
	pages = {101035},
}

@article{rasche_mathematical_2018,
	title = {Mathematical modelling of the evolution of the particle size distribution during ultrasound-induced breakage of aspirin crystals},
	volume = {132},
	issn = {02638762},
	doi = {10.1016/j.cherd.2018.01.014},
	journal = {Chemical Engineering Research and Design},
	author = {Rasche, Michael L. and Zeiger, Brad W. and Suslick, Kenneth S. and Braatz, Richard D.},
	year = {2018},
	pages = {170--177},
}

@article{maas_experimental_2007,
	title = {Experimental investigations and modelling of breakage phenomena in stirred liquid/liquid systems},
	volume = {85},
	issn = {02638762},
	doi = {10.1205/cherd06187},
	number = {5},
	journal = {Chemical Engineering Research and Design},
	author = {Maaß, S. and Gäbler, A. and Zaccone, A. and Paschedag, A.R. and Kraume, M.},
	year = {2007},
	pages = {703--709},
}

@article{wang_novel_2003,
	title = {A novel theoretical breakup kernel function for bubbles/droplets in a turbulent flow},
	volume = {58},
	issn = {00092509},
	doi = {10.1016/j.ces.2003.07.009},
	number = {20},
	journal = {Chemical Engineering Science},
	author = {Wang, Tiefeng and Wang, Jinfu and Jin, Yong},
	year = {2003},
	pages = {4629--4637},
}

@article{liao_literature_2009,
	title = {A literature review of theoretical models for drop and bubble breakup in turbulent dispersions},
	volume = {64},
	issn = {00092509},
	doi = {10.1016/j.ces.2009.04.026},
	language = {en},
	number = {15},
	journal = {Chemical Engineering Science},
	author = {Liao, Yixiang and Lucas, Dirk},
	year = {2009},
	pages = {3389--3406},
}

@article{hakansson_emulsion_2019,
	title = {Emulsion formation by homogenization: {Current} understanding and future perspectives},
	volume = {10},
	issn = {1941-1413, 1941-1421},
	shorttitle = {Emulsion {Formation} by {Homogenization}},
	doi = {10.1146/annurev-food-032818-121501},
	number = {1},
	journal = {Annual Review of Food Science and Technology},
	author = {Håkansson, Andreas},
	year = {2019},
	pages = {239--258},
}

@article{cheng_competitive_2010,
	title = {A competitive aggregation model for flash nanoprecipitation},
	volume = {351},
	issn = {00219797},
	doi = {10.1016/j.jcis.2010.07.066},
	language = {en},
	number = {2},
	journal = {Journal of Colloid and Interface Science},
	author = {Cheng, Janine Chungyin and Vigil, R.D. and Fox, R.O.},
	year = {2010},
	pages = {330--342},
}

@article{cheng_kinetic_2010,
	title = {Kinetic modeling of nanoprecipitation using {CFD} coupled with a population balance},
	volume = {49},
	issn = {0888-5885, 1520-5045},
	doi = {10.1021/ie100558n},
	language = {en},
	number = {21},
	journal = {Industrial \& Engineering Chemistry Research},
	author = {Cheng, Janine Chungyin and Fox, Rodney O.},
	month = nov,
	year = {2010},
	pages = {10651--10662},
}

@article{raponi_population_2023,
	title = {Population balance modelling of magnesium hydroxide precipitation: {Full} validation on different reactor configurations},
	volume = {477},
	doi = {10.1016/j.cej.2023.146540},
	language = {en},
	journal = {Chemical Engineering Journal},
	author = {Raponi, Antonello and Achermann, Ramona and Romano, Salvatore and Trespi, Silvio and Mazzotti, Marco and Cipollina, Andrea and Buffo, Antonio and Vanni, Marco and Marchisio, Daniele},
	year = {2023},
	pages = {146540},
}

@article{marchisio_use_2009,
	title = {On the use of bi-variate population balance equations for modelling barium titanate nanoparticle precipitation},
	volume = {64},
	issn = {00092509},
	doi = {10.1016/j.ces.2008.04.052},
	language = {en},
	number = {4},
	journal = {Chemical Engineering Science},
	author = {Marchisio, Daniele L.},
	year = {2009},
	pages = {697--708},
}

@article{vanni_approximate_2000,
	title = {Approximate population balance equations for aggregation–breakage processes},
	volume = {221},
	issn = {00219797},
	doi = {10.1006/jcis.1999.6571},
	number = {2},
	journal = {Journal of Colloid and Interface Science},
	author = {Vanni, Marco},
	month = jan,
	year = {2000},
	pages = {143--160},
}

@article{pena_modeling_2017,
	title = {Modeling and optimization of spherical agglomeration in suspension through a coupled population balance model},
	volume = {167},
	issn = {00092509},
	doi = {10.1016/j.ces.2017.03.055},
	journal = {Chemical Engineering Science},
	author = {Peña, Ramon and Burcham, Christopher L. and Jarmer, Daniel J. and Ramkrishna, Doraiswami and Nagy, Zoltan K.},
	year = {2017},
	pages = {66--77},
}

@article{marchisio_quadrature_2003,
	title = {Quadrature method of moments for aggregation–breakage processes},
	volume = {258},
	doi = {10.1016/S0021-9797(02)00054-1},
	number = {2},
	journal = {Journal of Colloid and Interface Science},
	author = {Marchisio, Daniele L. and Vigil, R.Dennis and Fox, Rodney O.},
	year = {2003},
	pages = {322--334},
}

@article{hakansson_dynamic_2009,
	title = {Dynamic simulation of emulsion formation in a high pressure homogenizer},
	volume = {64},
	doi = {10.1016/j.ces.2009.03.034},
	number = {12},
	journal = {Chemical Engineering Science},
	author = {Håkansson, Andreas and Trägårdh, Christian and Bergenståhl, Björn},
	year = {2009},
	pages = {2915--2925},
}

@article{baba_dempbm_2021,
	title = {{DEM}–{PBM} coupling method for the layering granulation of iron ore},
	volume = {378},
	journal = {Powder Technology},
	author = {Baba, Tomoya and Nakamura, Hideya and Takimoto, Hiroharu and Ohsaki, Shuji and Watano, Satoru and Takehara, Kenta and Higuchi, Takahide and Hirosawa, Toshiyuki and Yamamoto, Tetsuya},
	year = {2021},
	pages = {40--50},
}

@article{hong_droplet-based_2021,
	title = {Droplet-{Based} evaporative system for the estimation of protein crystallization kinetics},
	volume = {21},
	number = {11},
	journal = {Crystal Growth \& Design},
	author = {Hong, Moo Sun and Lu, Amos E. and Bae, Jaehan and Lee, Jong Min and Braatz, Richard D.},
	year = {2021},
	pages = {6064--6075},
}

@book{mullin_crystallization_2001,
	address = {Oxford},
	edition = {4},
	title = {Crystallization},
	publisher = {Butterworth-Heinemann},
	author = {Mullin, J.W.},
	year = {2001},
}

@article{srisanga_crystal_2015,
	title = {Crystal growth rate dispersion versus size-dependent crystal growth: {Appropriate} modeling for crystallization processes},
	volume = {15},
	doi = {10.1021/acs.cgd.5b00126},
	number = {5},
	journal = {Crystal Growth \& Design},
	author = {Srisanga, Sukanya and Flood, Adrian E. and Galbraith, Shaun C. and Rugmai, Supagorn and Soontaranon, Siriwat and Ulrich, Joachim},
	year = {2015},
	pages = {2330--2336},
}

@article{madras_temperature_2004,
	title = {Temperature effects on the transition from nucleation and growth to {Ostwald} ripening},
	volume = {59},
	issn = {00092509},
	doi = {10.1016/j.ces.2004.03.022},
	language = {en},
	number = {13},
	journal = {Chemical Engineering Science},
	author = {Madras, Giridhar and McCoy, Benjamin J.},
	year = {2004},
	pages = {2753--2765},
}

@article{iggland_population_2012,
	title = {Population balance modeling with size-dependent solubility: {Ostwald} ripening},
	volume = {12},
	issn = {1528-7483, 1528-7505},
	doi = {10.1021/cg201571n},
	number = {3},
	journal = {Crystal Growth \& Design},
	author = {Iggland, Martin and Mazzotti, Marco},
	month = mar,
	year = {2012},
	pages = {1489--1500},
}

@article{szilagyi_digital_2022,
	title = {Digital design of the crystallization of an active pharmaceutical ingredient using a population balance model with a novel size dependent growth rate expression. {From} development of a digital twin to \textit{in silico} optimization and experimental validation},
	volume = {22},
	doi = {10.1021/acs.cgd.1c01108},
	number = {1},
	journal = {Crystal Growth \& Design},
	author = {Szilágyi, Botond and Eren, Ayşe and Quon, Justin L. and Papageorgiou, Charles D. and Nagy, Zoltán K.},
	year = {2022},
	pages = {497--512},
}

@article{abegg_crystal_1968,
	title = {Crystal size distributions in continuous crystallizers when growth rate is size dependent},
	volume = {14},
	issn = {0001-1541, 1547-5905},
	doi = {10.1002/aic.690140121},
	number = {1},
	journal = {AIChE Journal},
	author = {Abegg, C. F. and Stevens, J. D. and Larson, M. A.},
	month = jan,
	year = {1968},
	pages = {118--122},
}

@article{komives_bioreactor_2003,
	title = {Bioreactor state estimation and control},
	volume = {14},
	issn = {09581669},
	doi = {10.1016/j.copbio.2003.09.001},
	language = {en},
	number = {5},
	journal = {Current Opinion in Biotechnology},
	author = {Komives, Claire and Parker, Robert S},
	month = oct,
	year = {2003},
	pages = {468--474},
}

@article{kupper_efficient_2009,
	title = {Efficient moving horizon state and parameter estimation for {SMB} processes},
	volume = {19},
	doi = {10.1016/j.jprocont.2008.10.004},
	language = {en},
	number = {5},
	journal = {Journal of Process Control},
	author = {Küpper, Achim and Diehl, Moritz and Schlöder, Johannes P. and Bock, Hans Georg and Engell, Sebastian},
	month = may,
	year = {2009},
	pages = {785--802},
}

@article{srisuma_thermal_2023,
	title = {Thermal imaging-based state estimation of a {Stefan} problem with application to cell thawing},
	volume = {173},
	issn = {00981354},
	doi = {10.1016/j.compchemeng.2023.108179},
	language = {en},
	journal = {Computers \& Chemical Engineering},
	author = {Srisuma, Prakitr and Pandit, Ajinkya and Zhang, Qihang and Hong, Moo Sun and Gamekkanda, Janaka and Fachin, Fabio and Moore, Nathan and Djordjevic, Dragan and Schwaerzler, Michael and Oyetunde, Tolutola and Tang, Wenlong and Myerson, Allan S. and Barbastathis, George and Braatz, Richard D.},
	month = may,
	year = {2023},
	pages = {108179},
}

@book{schiesser_numerical_1991,
	address = {San Diego, CA},
	title = {The {Numerical} {Method} of {Lines}: {Integration} of {Partial} {Differential} {Equations}},
	publisher = {Academic Press, Inc.},
	author = {Schiesser, W.E.},
	year = {1991},
}

@article{nagy_advances_2012,
	title = {Advances and new directions in crystallization control},
	volume = {3},
	issn = {1947-5438, 1947-5446},
	doi = {10.1146/annurev-chembioeng-062011-081043},
	language = {en},
	number = {1},
	journal = {Annual Review of Chemical and Biomolecular Engineering},
	author = {Nagy, Zoltan K. and Braatz, Richard D.},
	month = jul,
	year = {2012},
	pages = {55--75},
}

@article{golabgir_observability_2015,
	title = {Observability analysis of biochemical process models as a valuable tool for the development of mechanistic soft sensors},
	volume = {31},
	issn = {8756-7938},
	doi = {10.1002/btpr.2176},
	number = {6},
	journal = {Biotechnology Progress},
	author = {Golabgir, Aydin and Hoch, Thomas and Zhariy, Mariya and Herwig, Christoph},
	month = nov,
	year = {2015},
	pages = {1703--1715},
}

@article{cho_experimental_2003,
	title = {Experimental design in systems biology, based on parameter sensitivity analysis using a monte carlo method: {A} case study for the {TNFα}-mediated {NF}-κ {B} signal transduction pathway},
	volume = {79},
	issn = {0037-5497, 1741-3133},
	doi = {10.1177/0037549703040943},
	number = {12},
	journal = {Simulation},
	author = {Cho, Kwang-Hyun and Shin, Sung-Young and Kolch, Walter and Wolkenhauer, Olaf},
	month = dec,
	year = {2003},
	pages = {726--739},
}

@incollection{ljung_system_1998,
	address = {Boston, MA},
	title = {System {Identification}},
	isbn = {978-1-4612-1768-8},
	booktitle = {Signal {Analysis} and {Prediction}},
	publisher = {Birkhäuser},
	author = {Ljung, Lennart},
	editor = {Procházka, Ales and Uhlíř, Jan and Rayner, P. W. J. and Kingsbury, N. G.},
	year = {1998},
	doi = {10.1007/978-1-4612-1768-8_11},
	pages = {163--173},
}

@article{destro_mechanistic_2023,
	title = {Mechanistic modeling explains the production dynamics of recombinant adeno-associated virus with the baculovirus expression vector system},
	volume = {30},
	issn = {2329-0501},
	doi = {10.1016/j.omtm.2023.05.019},
	journal = {Molecular Therapy-Methods \& Clinical Development},
	author = {Destro, Francesco and Joseph, John and Srinivasan, Prasanna and Kanter, Joshua M. and Neufeld, Caleb and Wolfrum, Jacqueline M. and Barone, Paul W. and Springs, Stacy L. and Sinskey, Anthony J. and Cecchini, Sylvain and Kotin, Robert M. and Braatz, Richard D.},
	month = sep,
	year = {2023},
	pages = {122--146},
}

@techreport{us_food_and_drug_administration_liposome_2018,
	address = {Silver Spring, MD},
	title = {Liposome {Drug} {Products}: {Chemistry}, {Manufacturing}, and {Controls}; {Human} {Pharmacokinetics} and {Bioavailability}; and {Labeling} {Documentation}},
	author = {{U.S. Food and Drug Administration}},
	year = {2018},
}

@techreport{us_food_and_drug_administration_q8r2_2009,
	address = {Silver Spring, MD},
	title = {Q8({R2}) {Pharmaceutical} {Development}},
	institution = {U.S. Food and Drug Administration},
	author = {{U.S. Food and Drug Administration}},
	year = {2009},
}

@article{angioletti-uberti_theory_2017,
	title = {Theory, simulations and the design of functionalized nanoparticles for biomedical applications: {A} soft matter perspective},
	volume = {3},
	issn = {2057-3960},
	doi = {10.1038/s41524-017-0050-y},
	language = {en},
	number = {1},
	journal = {npj Computational Materials},
	author = {Angioletti-Uberti, Stefano},
	month = nov,
	year = {2017},
	pages = {1--15},
}

@article{bochicchio_interaction_2017,
	title = {Interaction of hydrophobic polymers with model lipid bilayers},
	volume = {7},
	copyright = {2017 The Author(s)},
	issn = {2045-2322},
	doi = {10.1038/s41598-017-06668-0},
	number = {1},
	journal = {Scientific Reports},
	author = {Bochicchio, D. and Panizon, E. and Monticelli, L. and Rossi, G.},
	month = jul,
	year = {2017},
	pages = {6357},
}

@article{das_molecular_2019,
	title = {Molecular dynamics simulation of interaction between functionalized nanoparticles with lipid membranes: {Analysis} of coarse-grained models},
	volume = {123},
	issn = {1520-6106},
	doi = {10.1021/acs.jpcb.9b08259},
	number = {49},
	journal = {The Journal of Physical Chemistry B},
	author = {Das, Mitradip and Dahal, Udaya and Mesele, Oluwaseun and Liang, Dongyue and Cui, Qiang},
	month = dec,
	year = {2019},
	pages = {10547--10561},
}

@article{zhang_research_2023,
	title = {Research progress of the ion activity coefficient of polyelectrolytes: {A} review},
	volume = {28},
	copyright = {http://creativecommons.org/licenses/by/3.0/},
	issn = {1420-3049},
	doi = {10.3390/molecules28052042},
	language = {en},
	number = {5},
	journal = {Molecules},
	author = {Zhang, Aokai and Yang, Xiuling and Yang, Feng and Zhang, Chunmei and Zhang, Qixiong and Duan, Gaigai and Jiang, Shaohua},
	month = jan,
	year = {2023},
	pages = {2042},
}

@article{yong_effects_2023,
	title = {Effects of shape on interaction dynamics of tetrahedral nanoplastics and the cell membrane},
	volume = {127},
	issn = {1520-6106},
	doi = {10.1021/acs.jpcb.2c07460},
	number = {7},
	journal = {The Journal of Physical Chemistry B},
	author = {Yong, Xin and Du, Ke},
	month = feb,
	year = {2023},
	pages = {1652--1663},
}

@article{yang_computer_2010,
	title = {Computer simulation of the translocation of nanoparticles with different shapes across a lipid bilayer},
	volume = {5},
	copyright = {2010 Springer Nature Limited},
	issn = {1748-3395},
	doi = {10.1038/nnano.2010.141},
	language = {en},
	number = {8},
	journal = {Nature Nanotechnology},
	author = {Yang, Kai and Ma, Yu-Qiang},
	month = aug,
	year = {2010},
	pages = {579--583},
}

@article{mears_application_2017,
	title = {Application of a mechanistic model as a tool for on-line monitoring of pilot scale filamentous fungal fermentation processes—{The} importance of evaporation effects},
	volume = {114},
	issn = {0006-3592},
	doi = {10.1002/bit.26187},
	number = {3},
	journal = {Biotechnology and Bioengineering},
	author = {Mears, Lisa and Stocks, Stuart M. and Albaek, Mads O. and Sin, Gürkan and Gernaey, Krist V.},
	month = mar,
	year = {2017},
	pages = {589--599},
}

@article{jorgensen_optimized_1984,
	title = {Optimized intermolecular potential functions for liquid hydrocarbons},
	volume = {106},
	issn = {0002-7863},
	doi = {10.1021/ja00334a030},
	number = {22},
	journal = {Journal of the American Chemical Society},
	author = {Jorgensen, William L. and Madura, Jeffry D. and Swenson, Carol J.},
	month = oct,
	year = {1984},
	pages = {6638--6646},
}

@article{jorge_molecular_2008,
	title = {Molecular dynamics simulation of self-assembly of n-decyltrimethylammonium bromide micelles},
	volume = {24},
	issn = {0743-7463},
	doi = {10.1021/la800291p},
	number = {11},
	journal = {Langmuir},
	author = {Jorge, Miguel},
	month = jun,
	year = {2008},
	pages = {5714--5725},
}

@article{lee_charmm36_2014,
	title = {{CHARMM36} united atom chain model for lipids and surfactants},
	volume = {118},
	issn = {1520-6106},
	doi = {10.1021/jp410344g},
	number = {2},
	journal = {The Journal of Physical Chemistry B},
	author = {Lee, Sarah and Tran, Alan and Allsopp, Matthew and Lim, Joseph B. and Hénin, Jérôme and Klauda, Jeffery B.},
	month = jan,
	year = {2014},
	pages = {547--556},
}

@article{lindahl_membrane_2008,
	title = {Membrane proteins: molecular dynamics simulations},
	volume = {18},
	issn = {0959-440X},
	shorttitle = {Membrane proteins},
	doi = {10.1016/j.sbi.2008.02.003},
	number = {4},
	journal = {Current Opinion in Structural Biology},
	author = {Lindahl, Erik and Sansom, Mark SP},
	month = aug,
	year = {2008},
	pages = {425--431},
}

@article{gumbart_molecular_2005,
	title = {Molecular dynamics simulations of proteins in lipid bilayers},
	volume = {15},
	issn = {0959-440X},
	doi = {10.1016/j.sbi.2005.07.007},
	number = {4},
	journal = {Current Opinion in Structural Biology},
	author = {Gumbart, James and Wang, Yi and Aksimentiev, Alekseij and Tajkhorshid, Emad and Schulten, Klaus},
	month = aug,
	year = {2005},
	pages = {423--431},
}

@article{eastman_openmm_2017,
	title = {{OpenMM} 7: {Rapid} development of high performance algorithms for molecular dynamics},
	volume = {13},
	issn = {1553-7358},
	shorttitle = {{OpenMM} 7},
	doi = {10.1371/journal.pcbi.1005659},
	language = {en},
	number = {7},
	journal = {PLOS Computational Biology},
	author = {Eastman, Peter and Swails, Jason and Chodera, John D. and McGibbon, Robert T. and Zhao, Yutong and Beauchamp, Kyle A. and Wang, Lee-Ping and Simmonett, Andrew C. and Harrigan, Matthew P. and Stern, Chaya D. and Wiewiora, Rafal P. and Brooks, Bernard R. and Pande, Vijay S.},
	month = jul,
	year = {2017},
	pages = {e1005659},
}

@article{case_ambertools_2023,
	title = {{AmberTools}},
	volume = {63},
	issn = {1549-9596},
	doi = {10.1021/acs.jcim.3c01153},
	number = {20},
	journal = {Journal of Chemical Information and Modeling},
	author = {Case, David A. and Aktulga, Hasan Metin and Belfon, Kellon and Cerutti, David S. and Cisneros, G. Andrés and Cruzeiro, Vinícius Wilian D. and Forouzesh, Negin and Giese, Timothy J. and Götz, Andreas W. and Gohlke, Holger and Izadi, Saeed and Kasavajhala, Koushik and Kaymak, Mehmet C. and King, Edward and Kurtzman, Tom and Lee, Tai-Sung and Li, Pengfei and Liu, Jian and Luchko, Tyler and Luo, Ray and Manathunga, Madushanka and Machado, Matias R. and Nguyen, Hai Minh and O’Hearn, Kurt A. and Onufriev, Alexey V. and Pan, Feng and Pantano, Sergio and Qi, Ruxi and Rahnamoun, Ali and Risheh, Ali and Schott-Verdugo, Stephan and Shajan, Akhil and Swails, Jason and Wang, Junmei and Wei, Haixin and Wu, Xiongwu and Wu, Yongxian and Zhang, Shi and Zhao, Shiji and Zhu, Qiang and Cheatham, Thomas E. III and Roe, Daniel R. and Roitberg, Adrian and Simmerling, Carlos and York, Darrin M. and Nagan, Maria C. and Merz, Kenneth M. Jr.},
	month = oct,
	year = {2023},
	pages = {6183--6191},
}

@article{ball_achieving_2016,
	title = {Achieving long-term stability of lipid nanoparticles: examining the effect of {pH}, temperature, and lyophilization},
	volume = {12},
	issn = {1178-2013},
	doi = {10.2147/IJN.S123062},
	language = {en},
	journal = {International Journal of Nanomedicine},
	author = {Ball, Rebecca and Bajaj, Palak and Whitehead, Kathryn},
	month = dec,
	year = {2016},
	pages = {305--315},
}

@incollection{launder_numerical_1983,
	address = {New York},
	title = {The {Numerical} {Computation} of {Turbulent} {Flows}},
	isbn = {978-0-08-030937-8},
	booktitle = {Numerical {Prediction} of {Flow}, {Heat} {Transfer}, {Turbulence} and {Combustion}},
	publisher = {Pergamon},
	author = {Launder, B.E. and Spalding, D.B.},
	year = {1983},
	pages = {96--116},
}

@article{bernet_modeling_2024,
	title = {Modeling the thermodynamic properties of saturated lactones in nonideal mixtures with the {SAFT}-γ {Mie} approach},
	volume = {69},
	issn = {0021-9568},
	doi = {10.1021/acs.jced.3c00358},
	journal = {Journal of Chemical \& Engineering Data},
	author = {Bernet, Thomas and Wehbe, Malak and Febra, Sara A. and Haslam, Andrew J. and Adjiman, Claire S. and Jackson, George and Galindo, Amparo},
	year = {2024},
	pages = {650--678},
}

@article{winter_understanding_2023,
	title = {Understanding the language of molecules: {Predicting} pure component parameters for the {PC}-{SAFT} equation of state from {SMILES}},
	journal = {arXiv},
	author = {Winter, Benedikt and Rehner, Philipp and Esper, Timm and Schilling, Johannes and Bardow, André},
	month = sep,
	year = {2023},
}

@article{umer_pc-saft_2014,
	title = {{PC}-{SAFT} parameters from ab initio calculations},
	volume = {362},
	issn = {03783812},
	doi = {10.1016/j.fluid.2013.08.037},
	journal = {Fluid Phase Equilib.},
	author = {Umer, Muhammad and Albers, Katja and Sadowski, Gabriele and Leonhard, Kai},
	year = {2014},
	pages = {41--50},
}

@article{swindells_absolute_1952,
	title = {Absolute viscosity of water at 20°{C}},
	volume = {48},
	issn = {0091-0635},
	doi = {10.6028/jres.048.001},
	language = {en},
	number = {1},
	journal = {Journal of Research of the National Bureau of Standards},
	author = {Swindells, J.F. and Coe, J.R. and Godfrey, T.B.},
	month = jan,
	year = {1952},
	pages = {1},
}

@book{kittel_introduction_2005,
	address = {Hoboken, NJ},
	edition = {8th},
	title = {Introduction to {Solid} {State} {Physics}},
	isbn = {978-0-471-41526-8},
	language = {eng},
	publisher = {Wiley},
	author = {Kittel, Charles},
	year = {2005},
}

@book{andreoli_membrane_1980,
	address = {Boston, MA},
	title = {Membrane {Physiology}},
	isbn = {978-1-4757-1720-4},
	language = {en},
	publisher = {Springer US},
	editor = {Andreoli, Thomas E. and Hoffman, Joseph F. and Fanestil, Darrell D.},
	year = {1980},
	doi = {10.1007/978-1-4757-1718-1},
}

@article{franzen_physicochemical_2011,
	title = {Physicochemical characterization of a {PEGylated} liposomal drug formulation using capillary electrophoresis},
	volume = {32},
	issn = {0173-0835, 1522-2683},
	doi = {10.1002/elps.201000552},
	language = {en},
	number = {6-7},
	journal = {Electrophoresis},
	author = {Franzen, Ulrik and Vermehren, Charlotte and Jensen, Henrik and Østergaard, Jesper},
	month = mar,
	year = {2011},
	pages = {738--748},
}

@article{cheng_induction_2023,
	title = {Induction of bleb structures in lipid nanoparticle formulations of {mRNA} leads to improved transfection potency},
	volume = {35},
	issn = {0935-9648, 1521-4095},
	doi = {10.1002/adma.202303370},
	number = {31},
	journal = {Advanced Materials},
	author = {Cheng, Miffy Hok Yan and Leung, Jerry and Zhang, Yao and Strong, Colton and Basha, Genc and Momeni, Arash and Chen, Yihang and Jan, Eric and Abdolahzadeh, Amir and Wang, Xinying and Kulkarni, Jayesh A. and Witzigmann, Dominik and Cullis, Pieter R.},
	month = aug,
	year = {2023},
	pages = {2303370},
}

@incollection{paulson_fast_2018,
	address = {Amsterdam},
	title = {Fast stochastic model  predictive control of end-to-end continuous pharmaceutical manufacturing},
	booktitle = {Process  {Systems} {Engineering} for {Pharmaceutical} {Manufacturing}},
	publisher = {Elsevier},
	author = {Paulson, Joel A. and Streif, S. and Findeisen, R. and Braatz, R.D.},
	editor = {Singh, Ravendra and Yuan, Zhihong},
	year = {2018},
}

@article{quo_adaptive_2011,
	title = {Adaptive control model reveals systematic feedback and key molecules in metabolic pathway regulation},
	volume = {18},
	doi = {10.1089/cmb.2010.0215},
	number = {2},
	journal = {Journal of Computational Biology},
	author = {Quo, Chang F. and Moffitt, Richard A. and Merrill, Alfred H. and Wang, May D.},
	year = {2011},
	pages = {169--182},
}

@article{eygeris_chemistry_2022,
	title = {Chemistry of lipid nanoparticles for {RNA} delivery},
	volume = {55},
	issn = {0001-4842},
	doi = {10.1021/acs.accounts.1c00544},
	number = {1},
	journal = {Accounts of Chemical Research},
	author = {Eygeris, Yulia and Gupta, Mohit and Kim, Jeonghwan and Sahay, Gaurav},
	month = jan,
	year = {2022},
	pages = {2--12},
}

@article{mesbah_model_2017,
	title = {Model predictive control of an integrated continuous pharmaceutical manufacturing pilot plant},
	volume = {21},
	issn = {1083-6160},
	doi = {10.1021/acs.oprd.7b00058},
	number = {6},
	journal = {Organic Process Research \& Development},
	author = {Mesbah, Ali and Paulson, Joel A. and Lakerveld, Richard and Braatz, Richard D.},
	month = jun,
	year = {2017},
	pages = {844--854},
}

@article{bandara_optimal_2009,
	title = {Optimal experimental design for parameter estimation of a cell signaling model},
	volume = {5},
	doi = {10.1371/journal.pcbi.1000558},
	number = {11},
	journal = {PLOS Computational Biology},
	author = {Bandara, Samuel and Schlöder, Johannes P. and Eils, Roland and Bock, Hans Georg and Meyer, Tobias},
	month = nov,
	year = {2009},
	pages = {1--12},
}

@article{gutierrez_mpc-based_2014,
	title = {An {MPC}-based control structure selection approach for simultaneous process and control design},
	volume = {70},
	issn = {0098-1354},
	doi = {10.1016/j.compchemeng.2013.08.014},
	journal = {Computers \& Chemical Engineering},
	author = {Gutierrez, G. and Ricardez-Sandoval, L.A. and Budman, H. and Prada, C.},
	month = nov,
	year = {2014},
	pages = {11--21},
}

@article{lakerveld_application_2015,
	title = {The application of an automated control strategy for an integrated continuous pharmaceutical pilot plant},
	volume = {19},
	issn = {1083-6160, 1520-586X},
	doi = {10.1021/op500104d},
	number = {9},
	journal = {Organic Process Research \& Development},
	author = {Lakerveld, Richard and Benyahia, Brahim and Heider, Patrick L. and Zhang, Haitao and Wolfe, Aaron and Testa, Christopher J. and Ogden, Sean and Hersey, Devin R. and Mascia, Salvatore and Evans, James M. B. and Braatz, Richard D. and Barton, Paul I.},
	month = sep,
	year = {2015},
	pages = {1088--1100},
}

@article{nagy_robust_2003,
	title = {Robust nonlinear model predictive control of batch processes},
	volume = {49},
	issn = {0001-1541, 1547-5905},
	doi = {10.1002/aic.690490715},
	number = {7},
	journal = {AIChE Journal},
	author = {Nagy, Zoltan K. and Braatz, Richard D.},
	month = jul,
	year = {2003},
	pages = {1776--1786},
}

@book{beck_parameter_1977,
	address = {New York},
	title = {Parameter {Estimation} in {Engineering} and {Science}},
	publisher = {John Wiley \& Sons, Inc},
	author = {Beck, James V. and Arnold, Kenneth J.},
	year = {1977},
}

@article{togkalidou_parameter_2004,
	title = {Parameter estimation and optimization of a loosely bound aggregating pharmaceutical crystallization using in situ infrared and laser backscattering measurements},
	volume = {43},
	issn = {0888-5885, 1520-5045},
	doi = {10.1021/ie0340847},
	number = {19},
	journal = {Industrial \& Engineering Chemistry Research},
	author = {Togkalidou, Timokleia and Tung, Hsien-Hsin and Sun, Yongkui and Andrews, Arthur T. and Braatz, Richard D.},
	month = sep,
	year = {2004},
	pages = {6168--6181},
}

@article{nagy_distributional_2007,
	title = {Distributional uncertainty analysis using power series and polynomial chaos expansions},
	volume = {17},
	issn = {09591524},
	doi = {10.1016/j.jprocont.2006.10.008},
	number = {3},
	journal = {Journal of Process Control},
	author = {Nagy, Z.K. and Braatz, R.D.},
	month = mar,
	year = {2007},
	pages = {229--240},
}

@article{yu_understanding_2014,
	title = {Understanding pharmaceutical quality by design},
	volume = {16},
	issn = {1550-7416},
	doi = {10.1208/s12248-014-9598-3},
	number = {4},
	journal = {The AAPS Journal},
	author = {Yu, Lawrence X. and Amidon, Gregory and Khan, Mansoor A. and Hoag, Stephen W. and Polli, James and Raju, G. K. and Woodcock, Janet},
	month = jul,
	year = {2014},
	pages = {771--783},
}

@article{widenski_use_2011,
	title = {Use of predictive solubility models for isothermal antisolvent crystallization modeling and optimization},
	volume = {50},
	issn = {0888-5885, 1520-5045},
	doi = {10.1021/ie102393y},
	number = {13},
	journal = {Industrial \& Engineering Chemistry Research},
	author = {Widenski, David J. and Abbas, Ali and Romagnoli, Jose A.},
	month = jul,
	year = {2011},
	pages = {8304--8313},
}

@article{chen_identification_2012,
	title = {Identification of nucleation rates in droplet-based microfluidic systems},
	volume = {77},
	issn = {00092509},
	doi = {10.1016/j.ces.2012.03.026},
	journal = {Chemical Engineering Science},
	author = {Chen, K. and Goh, L. and He, G. and Kenis, P.J.A. and Zukoski, C.F. and Braatz, R.D.},
	month = jul,
	year = {2012},
	pages = {235--241},
}

@article{kugler_heterogeneous_2016,
	title = {On heterogeneous nucleation during the precipitation of barium sulfate},
	volume = {114},
	issn = {02638762},
	doi = {10.1016/j.cherd.2016.07.024},
	journal = {Chemical Engineering Research and Design},
	author = {Kügler, Ricco T. and Beißert, Katharina and Kind, Matthias},
	month = oct,
	year = {2016},
	pages = {30--38},
}

@article{roelands_analysis_2006,
	title = {Analysis of nucleation rate measurements in precipitation processes},
	volume = {6},
	issn = {1528-7483, 1528-7505},
	doi = {10.1021/cg050678w},
	number = {6},
	journal = {Crystal Growth \& Design},
	author = {Roelands, C. P. Mark and Ter Horst, Joop H. and Kramer, Herman J. M. and Jansens, Pieter J.},
	month = jun,
	year = {2006},
	pages = {1380--1392},
}

@article{brooks_charmm_2009,
	title = {{CHARMM}: {The} biomolecular simulation program},
	volume = {30},
	copyright = {Copyright © 2009 Wiley Periodicals, Inc.},
	issn = {1096-987X},
	shorttitle = {{CHARMM}},
	doi = {10.1002/jcc.21287},
	language = {en},
	number = {10},
	journal = {Journal of Computational Chemistry},
	author = {Brooks, B. R. and Brooks III, C. L. and Mackerell Jr., A. D. and Nilsson, L. and Petrella, R. J. and Roux, B. and Won, Y. and Archontis, G. and Bartels, C. and Boresch, S. and Caflisch, A. and Caves, L. and Cui, Q. and Dinner, A. R. and Feig, M. and Fischer, S. and Gao, J. and Hodoscek, M. and Im, W. and Kuczera, K. and Lazaridis, T. and Ma, J. and Ovchinnikov, V. and Paci, E. and Pastor, R. W. and Post, C. B. and Pu, J. Z. and Schaefer, M. and Tidor, B. and Venable, R. M. and Woodcock, H. L. and Wu, X. and Yang, W. and York, D. M. and Karplus, M.},
	year = {2009},
	pages = {1545--1614},
}

@article{shahmohammadi_using_2020,
	title = {Using prior parameter knowledge in model-based design of experiments for pharmaceutical production},
	volume = {66},
	issn = {0001-1541},
	doi = {10.1002/aic.17021},
	number = {11},
	journal = {AIChE Journal},
	author = {Shahmohammadi, Ali and McAuley, Kimberley B.},
	month = nov,
	year = {2020},
	pages = {e17021},
}

@article{narayanan_hybrid-ekf_2020,
	title = {Hybrid-{EKF}: {Hybrid} model coupled with extended {Kalman} filter for real-time monitoring and control of mammalian cell culture},
	volume = {117},
	issn = {0006-3592},
	doi = {10.1002/bit.27437},
	number = {9},
	journal = {Biotechnology and Bioengineering},
	author = {Narayanan, Harini and Behle, Lars and Luna, Martin F. and Sokolov, Michael and Guillén-Gosálbez, Gonzalo and Morbidelli, Massimo and Butté, Alessandro},
	month = sep,
	year = {2020},
	pages = {2703--2714},
}

@article{golabgir_combining_2016,
	title = {Combining mechanistic modeling and {Raman} spectroscopy for real-time monitoring of fed-batch {Penicillin} production},
	volume = {88},
	issn = {0009-286X},
	doi = {10.1002/cite.201500101},
	number = {6},
	journal = {Chemie Ingenieur Technik},
	author = {Golabgir, Aydin and Herwig, Christoph},
	month = jun,
	year = {2016},
	pages = {764--776},
}

@article{xu_overview_2020,
	title = {Overview of secondary nucleation: {From} fundamentals to application},
	volume = {59},
	issn = {0888-5885, 1520-5045},
	doi = {10.1021/acs.iecr.0c03304},
	number = {41},
	journal = {Industrial \& Engineering Chemistry Research},
	author = {Xu, Shijie and Hou, Zhongbi and Chuai, Xiaoyu and Wang, Yanfei},
	month = oct,
	year = {2020},
	pages = {18335--18356},
}

@article{di_pasquale_identification_2013,
	title = {Identification of nucleation rate parameters with {MD} and validation of the {CFD} model for polymer particle precipitation},
	volume = {91},
	issn = {02638762},
	doi = {10.1016/j.cherd.2013.05.027},
	number = {11},
	journal = {Chemical Engineering Research and Design},
	author = {Di Pasquale, N. and Marchisio, D.L. and Carbone, P. and Barresi, A.A.},
	month = nov,
	year = {2013},
	pages = {2275--2290},
}

@article{di_pasquale_model_2012,
	title = {Model validation for precipitation in solvent-displacement processes},
	volume = {84},
	issn = {00092509},
	doi = {10.1016/j.ces.2012.08.043},
	journal = {Chemical Engineering Science},
	author = {Di Pasquale, N. and Marchisio, D.L. and Barresi, A.A.},
	month = dec,
	year = {2012},
	pages = {671--683},
}

@article{zhu_crystallization_2016,
	title = {Crystallization of calcium sulphate during phosphoric acid production: {Modeling} particle shape and size distribution},
	volume = {138},
	issn = {18777058},
	doi = {10.1016/j.proeng.2016.02.098},
	journal = {Procedia Engineering},
	author = {Zhu, Zhilong and Peng, You and Hatton, T. Alan and Samrane, Kamal and Myerson, Allan S. and Braatz, Richard D.},
	year = {2016},
	pages = {390--402},
}

@article{guldenpfennig_how_2019,
	title = {How to estimate material parameters for multiphase, multicomponent precipitation modeling},
	volume = {19},
	issn = {1528-7483, 1528-7505},
	doi = {10.1021/acs.cgd.9b00027},
	number = {5},
	journal = {Crystal Growth \& Design},
	author = {Güldenpfennig, Andreas and Pflug, Lukas and Peukert, Wolfgang},
	month = may,
	year = {2019},
	pages = {2785--2793},
}

@article{schwarzer_combined_2004,
	title = {Combined experimental/numerical study on the precipitation of nanoparticles},
	volume = {50},
	issn = {0001-1541, 1547-5905},
	doi = {10.1002/aic.10277},
	number = {12},
	journal = {AIChE Journal},
	author = {Schwarzer, Hans‐Christoph and Peukert, Wolfgang},
	month = dec,
	year = {2004},
	pages = {3234--3247},
}

@article{erdemir_nucleation_2009,
	title = {Nucleation of crystals from solution: {Classical} and two-step models},
	volume = {42},
	issn = {0001-4842, 1520-4898},
	doi = {10.1021/ar800217x},
	number = {5},
	journal = {Accounts of Chemical Research},
	author = {Erdemir, Deniz and Lee, Alfred Y. and Myerson, Allan S.},
	month = may,
	year = {2009},
	pages = {621--629},
}

@article{di_carlo_continuous_2007,
	title = {Continuous inertial focusing, ordering, and separation of particles in microchannels},
	volume = {104},
	issn = {0027-8424, 1091-6490},
	url = {https://pnas.org/doi/full/10.1073/pnas.0704958104},
	doi = {10.1073/pnas.0704958104},
	number = {48},
	journal = {Proceedings of the National Academy of Sciences},
	author = {Di Carlo, Dino and Irimia, Daniel and Tompkins, Ronald G. and Toner, Mehmet},
	month = nov,
	year = {2007},
	pages = {18892--18897},
}

@incollection{de_kee_general_2022,
	address = {Melville, New York},
	title = {General {Rigid} {Bead}-{Rod} {Macromolecular} {Theory}},
	isbn = {978-0-7354-2469-2},
	booktitle = {Recent {Advances} in {Rheology}},
	publisher = {AIP Publishing LLC},
	author = {Kanso, Mona and Giacomin, Alan Jeffrey},
	editor = {De Kee, Daniel and Ramachandran, Arun},
	month = jul,
	year = {2022},
	doi = {10.1063/9780735424715_002},
	pages = {2--1--2--32},
}

@article{retamal_marin_zeta_2017,
	title = {Zeta potential measurements for non-spherical colloidal particles – {Practical} issues of characterisation of interfacial properties of nanoparticles},
	volume = {532},
	issn = {09277757},
	doi = {10.1016/j.colsurfa.2017.04.010},
	language = {en},
	journal = {Colloids and Surfaces A: Physicochemical and Engineering Aspects},
	author = {Retamal Marín, R.R. and Babick, F. and Hillemann, L.},
	month = nov,
	year = {2017},
	pages = {516--521},
}

@article{bulcha_viral_2021,
	title = {Viral vector platforms within the gene therapy landscape},
	volume = {6},
	issn = {2059-3635},
	doi = {10.1038/s41392-021-00487-6},
	number = {1},
	journal = {Signal Transduction and Targeted Therapy},
	author = {Bulcha, Jote T. and Wang, Yi and Ma, Hong and Tai, Phillip W. L. and Gao, Guangping},
	month = feb,
	year = {2021},
	pages = {53},
}

@article{smoluchowski_versuch_1918,
	title = {Versuch einer mathematischen {Theorie} der {Koagulationskinetik} kolloider {Lösungen}},
	volume = {92U},
	issn = {2196-7156, 0942-9352},
	doi = {10.1515/zpch-1918-9209},
	number = {1},
	journal = {Zeitschrift für Physikalische Chemie},
	author = {Smoluchowski, M. V.},
	month = nov,
	year = {1918},
	pages = {129--168},
}

@article{menter_two-equation_1994,
	title = {Two-equation eddy-viscosity turbulence models for engineering applications},
	volume = {32},
	issn = {0001-1452, 1533-385X},
	doi = {10.2514/3.12149},
	language = {en},
	number = {8},
	journal = {AIAA Journal},
	author = {Menter, F. R.},
	month = aug,
	year = {1994},
	pages = {1598--1605},
}

@article{pope_turbulent_2001,
	title = {Turbulent flows},
	volume = {12},
	issn = {0957-0233, 1361-6501},
	doi = {10.1088/0957-0233/12/11/705},
	number = {11},
	journal = {Measurement Science and Technology},
	author = {Pope, Stephen B},
	month = nov,
	year = {2001},
	pages = {2020--2021},
}

@article{germano_turbulence_1992,
	title = {Turbulence: {The} filtering approach},
	volume = {238},
	issn = {0022-1120, 1469-7645},
	doi = {10.1017/S0022112092001733},
	language = {en},
	journal = {Journal of Fluid Mechanics},
	author = {Germano, M.},
	month = may,
	year = {1992},
	pages = {325--336},
}

@article{brehm_travisfree_2020,
	title = {{TRAVIS}—{A} free analyzer for trajectories from molecular simulation},
	volume = {152},
	issn = {0021-9606, 1089-7690},
	doi = {10.1063/5.0005078},
	number = {16},
	journal = {The Journal of Chemical Physics},
	author = {Brehm, M. and Thomas, M. and Gehrke, S. and Kirchner, B.},
	month = apr,
	year = {2020},
	pages = {164105},
}

@article{vaz_translational_1985,
	title = {Translational diffusion of lipids in liquid crystalline phase phosphatidylcholine multibilayers. {A} comparison of experiment with theory},
	volume = {24},
	issn = {0006-2960, 1520-4995},
	doi = {10.1021/bi00324a037},
	language = {en},
	number = {3},
	journal = {Biochemistry},
	author = {Vaz, Winchil L. C. and Clegg, Robert M. and Hallmann, Dieter},
	month = jan,
	year = {1985},
	pages = {781--786},
}

@article{king_translational_1973,
	title = {Translational and rotational diffusion of tobacco mosaic virus from polarized and depolarized light scattering},
	volume = {12},
	issn = {0006-3525, 1097-0282},
	doi = {10.1002/bip.1973.360120817},
	language = {en},
	number = {8},
	journal = {Biopolymers},
	author = {King, T. A. and Knox, A. and McAdam, J. D. G.},
	month = aug,
	year = {1973},
	pages = {1917--1926},
}

@techreport{gallud_time_2021,
	type = {preprint},
	title = {Time evolution of {PEG}-shedding and serum protein coronation determines the cell uptake kinetics and delivery of lipid nanoparticle formulated {mRNA}},
	language = {en},
	institution = {Biophysics},
	author = {Gallud, A. and Munson, M. J. and Liu, K. and Idström, A. and Barriga, H. M. G. and Tabaei, S. R. and Aliakbarinodehi, N. and Ojansivu, M. and Lubart, Q. and Doutch, J. J. and Holme, M. N. and Evenäs, L. and Lindfors, L. and Stevens, M. M. and Collén, A. and Sabirsh, A. and Höök, F. and Esbjörner, E. K.},
	month = aug,
	year = {2021},
	doi = {10.1101/2021.08.20.457104},
}

@article{cardellini_thermal_2016,
	title = {Thermal transport phenomena in nanoparticle suspensions},
	volume = {28},
	issn = {0953-8984, 1361-648X},
	doi = {10.1088/0953-8984/28/48/483003},
	number = {48},
	journal = {Journal of Physics: Condensed Matter},
	author = {Cardellini, Annalisa and Fasano, Matteo and Bozorg Bigdeli, Masoud and Chiavazzo, Eliodoro and Asinari, Pietro},
	month = dec,
	year = {2016},
	pages = {483003},
}

@article{hohenberg_theory_1977,
	title = {Theory of dynamic critical phenomena},
	volume = {49},
	issn = {0034-6861},
	doi = {10.1103/RevModPhys.49.435},
	language = {en},
	number = {3},
	journal = {Reviews of Modern Physics},
	author = {Hohenberg, P. C. and Halperin, B. I.},
	month = jul,
	year = {1977},
	pages = {435--479},
}

@article{hald_albertsen_role_2022,
	title = {The role of lipid components in lipid nanoparticles for vaccines and gene therapy},
	volume = {188},
	issn = {0169409X},
	doi = {10.1016/j.addr.2022.114416},
	language = {en},
	journal = {Advanced Drug Delivery Reviews},
	author = {Hald Albertsen, Camilla and Kulkarni, Jayesh A. and Witzigmann, Dominik and Lind, Marianne and Petersson, Karsten and Simonsen, Jens B.},
	month = sep,
	year = {2022},
	pages = {114416},
}

@article{cheng_role_2016,
	title = {The role of helper lipids in lipid nanoparticles ({LNPs}) designed for oligonucleotide delivery},
	volume = {99},
	issn = {0169409X},
	doi = {10.1016/j.addr.2016.01.022},
	language = {en},
	journal = {Advanced Drug Delivery Reviews},
	author = {Cheng, Xinwei and Lee, Robert J.},
	month = apr,
	year = {2016},
	pages = {129--137},
}

@article{boedtker_preparation_1958,
	title = {The preparation and characterization of essentially uniform tobacco mosaic virus particles},
	volume = {80},
	issn = {0002-7863, 1520-5126},
	doi = {10.1021/ja01543a049},
	language = {en},
	number = {10},
	journal = {Journal of the American Chemical Society},
	author = {Boedtker, Helga and Simmons, Norman S.},
	month = may,
	year = {1958},
	pages = {2550--2556},
}

@article{strelkova_petersen_mixing_2023,
	title = {The mixing method used to formulate lipid nanoparticles affects {mRNA} delivery efficacy and organ tropism},
	volume = {192},
	issn = {09396411},
	doi = {10.1016/j.ejpb.2023.10.006},
	language = {en},
	journal = {European Journal of Pharmaceutics and Biopharmaceutics},
	author = {Strelkova Petersen, Daria M. and Chaudhary, Namit and Arral, Mariah L. and Weiss, Ryan M. and Whitehead, Kathryn A.},
	month = nov,
	year = {2023},
	pages = {126--135},
}

@article{damase_limitless_2021,
	title = {The limitless future of {RNA} therapeutics},
	volume = {9},
	issn = {2296-4185},
	doi = {10.3389/fbioe.2021.628137},
	journal = {Frontiers in Bioengineering and Biotechnology},
	author = {Damase, Tulsi Ram and Sukhovershin, Roman and Boada, Christian and Taraballi, Francesca and Pettigrew, Roderic I. and Cooke, John P.},
	month = mar,
	year = {2021},
	pages = {628137},
}

@article{verma_landscape_2023,
	title = {The landscape for lipid-nanoparticle-based genomic medicines},
	volume = {22},
	issn = {1474-1776, 1474-1784},
	doi = {10.1038/d41573-023-00002-2},
	language = {en},
	number = {5},
	journal = {Nature Reviews Drug Discovery},
	author = {Verma, Malvika and Ozer, Imran and Xie, Wen and Gallagher, Ryan and Teixeira, Alexandra and Choy, Michael},
	month = may,
	year = {2023},
	pages = {349--350},
}

@article{thorn_journey_2022,
	title = {The journey of a lifetime — development of {Pfizer}’s {COVID}-19 vaccine},
	volume = {78},
	issn = {09581669},
	doi = {10.1016/j.copbio.2022.102803},
	language = {en},
	journal = {Current Opinion in Biotechnology},
	author = {Thorn, Chelsea R and Sharma, Divya and Combs, Rodney and Bhujbal, Sonal and Romine, Jennifer and Zheng, Xiaolu and Sunasara, Khurram and Badkar, Advait},
	month = dec,
	year = {2022},
	pages = {102803},
}

@article{michelsen_isothermal_1982,
	title = {The isothermal flash problem. {Part} {II}. {Phase}-split calculation},
	volume = {9},
	issn = {0378-3812},
	doi = {10.1016/0378-3812(82)85002-4},
	number = {1},
	journal = {Fluid Phase Equilib.},
	author = {Michelsen, Michael L.},
	month = dec,
	year = {1982},
	pages = {21--40},
}

@article{sharma_immunostimulatory_2024,
	title = {The immunostimulatory nature of {mRNA} lipid nanoparticles},
	volume = {205},
	issn = {0169409X},
	doi = {10.1016/j.addr.2023.115175},
	language = {en},
	journal = {Advanced Drug Delivery Reviews},
	author = {Sharma, Preeti and Hoorn, Daniek and Aitha, Anjaiah and Breier, Dor and Peer, Dan},
	month = feb,
	year = {2024},
	pages = {115175},
}

@article{pereira_held_2012,
	title = {The {HELD} algorithm for multicomponent, multiphase equilibrium calculations with generic equations of state},
	volume = {36},
	issn = {00981354},
	doi = {10.1016/j.compchemeng.2011.07.009},
	language = {en},
	journal = {Computers \& Chemical Engineering},
	author = {Pereira, Frances E. and Jackson, George and Galindo, Amparo and Adjiman, Claire S.},
	month = jan,
	year = {2012},
	pages = {99--118},
}

@article{rantanen_future_2015,
	title = {The future of pharmaceutical manufacturing sciences},
	volume = {104},
	issn = {00223549},
	doi = {10.1002/jps.24594},
	language = {en},
	number = {11},
	journal = {Journal of Pharmaceutical Sciences},
	author = {Rantanen, Jukka and Khinast, Johannes},
	month = nov,
	year = {2015},
	pages = {3612--3638},
}

@article{chen_formation_2019,
	title = {The formation and physicochemical properties of {PEGylated} deep eutectic solvents},
	volume = {43},
	issn = {1144-0546, 1369-9261},
	doi = {10.1039/C9NJ02196E},
	language = {en},
	number = {22},
	journal = {New Journal of Chemistry},
	author = {Chen, Wenjun and Bai, Xiyue and Xue, Zhimin and Mou, Hongyu and Chen, Jiangang and Liu, Zhaotie and Mu, Tiancheng},
	year = {2019},
	pages = {8804--8810},
}

@article{lovegrove_flow_2023,
	title = {The flow of anisotropic nanoparticles in solution and in blood},
	volume = {3},
	issn = {2766-8509, 2766-2098},
	doi = {10.1002/EXP.20220075},
	number = {6},
	journal = {Exploration},
	author = {Lovegrove, Jordan Thomas and Kent, Ben and Förster, Stephan and Garvey, Christopher J. and Stenzel, Martina H.},
	month = dec,
	year = {2023},
	pages = {20220075},
}

@article{kulkarni_current_2021,
	title = {The current landscape of nucleic acid therapeutics},
	volume = {16},
	issn = {1748-3387, 1748-3395},
	doi = {10.1038/s41565-021-00898-0},
	language = {en},
	number = {6},
	journal = {Nature Nanotechnology},
	author = {Kulkarni, Jayesh A. and Witzigmann, Dominik and Thomson, Sarah B. and Chen, Sam and Leavitt, Blair R. and Cullis, Pieter R. and van der Meel, Roy},
	month = jun,
	year = {2021},
	pages = {630--643},
}

@article{evans_surface_1980,
	title = {Surface viscosities of phospholipids alone and with cholesterol in monolayers at the air‐water interface},
	volume = {15},
	issn = {0024-4201, 1558-9307},
	doi = {10.1007/BF02534225},
	language = {en},
	number = {7},
	journal = {Lipids},
	author = {Evans, R. W. and Williams, M. A. and Tinoco, J.},
	month = jul,
	year = {1980},
	pages = {524--533},
}

@article{kise_submillisecond_2014,
	title = {Submillisecond mixing in a continuous-flow, microfluidic mixer utilizing mid-infrared hyperspectral imaging detection},
	volume = {14},
	issn = {1473-0197, 1473-0189},
	doi = {10.1039/C3LC51171E},
	language = {en},
	number = {3},
	journal = {Lab Chip},
	author = {Kise, Drew P. and Magana, Donny and Reddish, Michael J. and Dyer, R. Brian},
	year = {2014},
	pages = {584--591},
}

@article{pilz_studies_1972,
	title = {Studies on the conformation of {DNA}‐dependent {RNA} polymerase in solution by small‐angle {X}‐ray measurements},
	volume = {28},
	issn = {0014-2956, 1432-1033},
	doi = {10.1111/j.1432-1033.1972.tb01904.x},
	language = {en},
	number = {2},
	journal = {European Journal of Biochemistry},
	author = {Pilz, Ingrid and Kratky, Otto and Rabussay, Dietmar},
	month = jul,
	year = {1972},
	pages = {205--220},
}

@article{raue_structural_2009,
	title = {Structural and practical identifiability analysis of partially observed dynamical models by exploiting the profile likelihood},
	volume = {25},
	issn = {1367-4811, 1367-4803},
	doi = {10.1093/bioinformatics/btp358},
	language = {en},
	number = {15},
	journal = {Bioinformatics},
	author = {Raue, A. and Kreutz, C. and Maiwald, T. and Bachmann, J. and Schilling, M. and Klingmüller, U. and Timmer, J.},
	month = aug,
	year = {2009},
	pages = {1923--1929},
}

@article{evers_stateart_2018,
	title = {State‐of‐the‐art design and rapid‐mixing production techniques of lipid nanoparticles for nucleic acid delivery},
	volume = {2},
	issn = {2366-9608, 2366-9608},
	doi = {10.1002/smtd.201700375},
	language = {en},
	number = {9},
	journal = {Small Methods},
	author = {Evers, Martijn J. W. and Kulkarni, Jayesh A. and van der Meel, Roy and Cullis, Pieter R. and Vader, Pieter and Schiffelers, Raymond M.},
	month = sep,
	year = {2018},
	pages = {1700375},
}

@article{erfle_stabilized_2019,
	title = {Stabilized production of lipid nanoparticles of tunable size in {Taylor} flow glass devices with high-surface-quality {3D} microchannels},
	volume = {10},
	issn = {2072-666X},
	doi = {10.3390/mi10040220},
	language = {en},
	number = {4},
	journal = {Micromachines},
	author = {Erfle, Peer and Riewe, Juliane and Bunjes, Heike and Dietzel, Andreas},
	month = mar,
	year = {2019},
	pages = {220},
}

@article{perelman_spontaneous_2023,
	title = {Spontaneous confinement of {mRNA} molecules at biomolecular condensate boundaries},
	volume = {12},
	issn = {2073-4409},
	doi = {10.3390/cells12182250},
	language = {en},
	number = {18},
	journal = {Cells},
	author = {Perelman, Rebecca T. and Schmidt, Andreas and Khan, Umar and Walter, Nils G.},
	month = sep,
	year = {2023},
	pages = {2250},
}

@article{parke_solution_1999,
	title = {Solution properties of ethanol in water},
	volume = {67},
	issn = {03088146},
	doi = {10.1016/S0308-8146(99)00124-7},
	number = {3},
	journal = {Food Chemistry},
	author = {Parke, Sneha A. and Birch, Gordon G.},
	month = nov,
	year = {1999},
	pages = {241--246},
}

@article{marchisio_solution_2005,
	title = {Solution of population balance equations using the direct quadrature method of moments},
	volume = {36},
	issn = {00218502},
	doi = {10.1016/j.jaerosci.2004.07.009},
	language = {en},
	number = {1},
	journal = {Journal of Aerosol Science},
	author = {Marchisio, Daniele L. and Fox, Rodney O.},
	month = jan,
	year = {2005},
	pages = {43--73},
}

@article{nguyen_solution_2016,
	title = {Solution of population balance equations in applications with fine particles: {Mathematical} modeling and numerical schemes},
	volume = {325},
	issn = {00219991},
	doi = {10.1016/j.jcp.2016.08.017},
	language = {en},
	journal = {Journal of Computational Physics},
	author = {Nguyen, T.T. and Laurent, F. and Fox, R.O. and Massot, M.},
	month = nov,
	year = {2016},
	pages = {129--156},
}

@article{wysoczanska_solubility_2021,
	title = {Solubility of {DNP}-amino acids and their partitioning in biodegradable {ATPS}: {Experimental} and {ePC}-{SAFT} modeling},
	volume = {527},
	issn = {0378-3812},
	doi = {10.1016/j.fluid.2020.112830},
	journal = {Fluid Phase Equilibria},
	author = {Wysoczanska, Kamila and Nierhauve, Birte and Sadowski, Gabriele and Macedo, Eugénia A. and Held, Christoph},
	month = jan,
	year = {2021},
	pages = {112830},
}

@article{de_jesus_solid_2015,
	title = {Solid lipid nanoparticles as nucleic acid delivery system: {Properties} and molecular mechanisms},
	volume = {201},
	issn = {01683659},
	doi = {10.1016/j.jconrel.2015.01.010},
	language = {en},
	journal = {Journal of Controlled Release},
	author = {de Jesus, Marcelo B. and Zuhorn, Inge S.},
	month = mar,
	year = {2015},
	pages = {1--13},
}

@article{sundstrom_software_2008,
	title = {Software sensors for fermentation processes},
	volume = {31},
	issn = {1615-7605},
	doi = {10.1007/s00449-007-0157-5},
	number = {2},
	journal = {Bioprocess and Biosystems Engineering},
	author = {Sundström, Heléne and Enfors, Sven-Olof},
	month = feb,
	year = {2008},
	pages = {145--152},
}

@article{de_assis_soft_2000,
	title = {Soft sensors development for on-line bioreactor state estimation},
	volume = {24},
	issn = {0098-1354},
	doi = {10.1016/S0098-1354(00)00489-0},
	number = {2},
	journal = {Computers \& Chemical Engineering},
	author = {de Assis, Adilson José and Filho, Rubens Maciel},
	month = jul,
	year = {2000},
	pages = {1099--1103},
}

@article{he_size-controlled_2018,
	title = {Size-controlled lipid nanoparticle production using turbulent mixing to enhance oral {DNA} delivery},
	volume = {81},
	issn = {17427061},
	doi = {10.1016/j.actbio.2018.09.047},
	language = {en},
	journal = {Acta Biomaterialia},
	author = {He, Zhiyu and Hu, Yizong and Nie, Tianqi and Tang, Haoyu and Zhu, Jinchang and Chen, Kuntao and Liu, Lixin and Leong, Kam W. and Chen, Yongming and Mao, Hai-Quan},
	month = nov,
	year = {2018},
	pages = {195--207},
}

@article{malburet_size_2022,
	title = {Size and charge characterization of lipid nanoparticles for {mRNA} vaccines},
	volume = {94},
	issn = {0003-2700, 1520-6882},
	doi = {10.1021/acs.analchem.1c04778},
	language = {en},
	number = {11},
	journal = {Analytical Chemistry},
	author = {Malburet, Camille and Leclercq, Laurent and Cotte, Jean-François and Thiebaud, Jérôme and Bazin, Emilie and Garinot, Marie and Cottet, Hervé},
	month = mar,
	year = {2022},
	pages = {4677--4685},
}

@article{marchisio_simulation_2001,
	title = {Simulation of turbulent precipitation in a semi-batch {Taylor}-{Couette} reactor using {CFD}},
	volume = {47},
	issn = {00011541, 15475905},
	doi = {10.1002/aic.690470314},
	language = {en},
	number = {3},
	journal = {AIChE Journal},
	author = {Marchisio, D. L. and Barresi, A. A. and Fox, R. O.},
	month = mar,
	year = {2001},
	pages = {664--676},
}

@article{frank_simulation_2010,
	title = {Simulation of turbulent and thermal mixing in {T}-junctions using {URANS} and scale-resolving turbulence models in {ANSYS} {CFX}},
	volume = {240},
	issn = {00295493},
	doi = {10.1016/j.nucengdes.2009.11.008},
	language = {en},
	number = {9},
	journal = {Nuclear Engineering and Design},
	author = {Frank, Th. and Lifante, C. and Prasser, H.-M. and Menter, F.},
	month = sep,
	year = {2010},
	pages = {2313--2328},
}

@article{woo_simulation_2006,
	title = {Simulation of mixing effects in antisolvent crystallization using a coupled {CFD}-{PDF}-{PBE} approach},
	volume = {6},
	issn = {1528-7483, 1528-7505},
	doi = {10.1021/cg0503090},
	language = {en},
	number = {6},
	journal = {Crystal Growth \& Design},
	author = {Woo, Xing Yi and Tan, Reginald B. H. and Chow, Pui Shan and Braatz, Richard D.},
	month = jun,
	year = {2006},
	pages = {1291--1303},
}

@article{pico_silver_2023,
	title = {Silver nanoparticles synthesis in microfluidic and well-mixed reactors: {A} combined experimental and {PBM}-{CFD} study},
	volume = {474},
	issn = {13858947},
	shorttitle = {Silver nanoparticles synthesis in microfluidic and well-mixed reactors},
	doi = {10.1016/j.cej.2023.145692},
	language = {en},
	journal = {Chemical Engineering Journal},
	author = {Pico, Paula and Nathanael, Konstantia and Lavino, Alessio D. and Kovalchuk, Nina M. and Simmons, Mark J.H. and Matar, Omar K.},
	month = oct,
	year = {2023},
	pages = {145692},
}

@article{kaminski_sepp_2020,
	title = {{SEPP}: {Segment}-based equation of state parameter prediction},
	volume = {65},
	issn = {0021-9568},
	doi = {10.1021/acs.jced.0c00733},
	number = {12},
	journal = {J. Chem. Eng. Data},
	author = {Kaminski, Sebastian and Leonhard, Kai},
	month = dec,
	year = {2020},
	pages = {5830--5843},
}

@article{maldonado-camargo_scale-dependent_2017,
	title = {Scale-dependent rotational diffusion of nanoparticles in polymer solutions},
	volume = {9},
	issn = {2040-3364, 2040-3372},
	doi = {10.1039/C7NR01603D},
	language = {en},
	number = {33},
	journal = {Nanoscale},
	author = {Maldonado-Camargo, Lorena and Yang, Chuncheng and Rinaldi, Carlos},
	year = {2017},
	pages = {12039--12050},
}

@article{zhang_salting-out_2016,
	title = {Salting-out and salting-in of polyelectrolyte solutions: {A} liquid-state theory study},
	volume = {49},
	issn = {0024-9297, 1520-5835},
	doi = {10.1021/acs.macromol.6b02160},
	language = {en},
	number = {24},
	journal = {Macromolecules},
	author = {Zhang, Pengfei and Alsaifi, Nayef M. and Wu, Jianzhong and Wang, Zhen-Gang},
	month = dec,
	year = {2016},
	pages = {9720--9730},
}

@article{chapman_saft_1989,
	title = {{SAFT}: {Equation}-of-state solution model for associating fluids},
	volume = {52},
	issn = {03783812},
	shorttitle = {{SAFT}},
	doi = {10.1016/0378-3812(89)80308-5},
	language = {en},
	journal = {Fluid Phase Equilib.},
	author = {Chapman, W.G. and Gubbins, K.E. and Jackson, G. and Radosz, M.},
	month = dec,
	year = {1989},
	pages = {31--38},
}

@article{wada_rotational_1971,
	title = {Rotational diffusion of tobacco mosaic virus},
	volume = {55},
	issn = {0021-9606, 1089-7690},
	doi = {10.1063/1.1676311},
	number = {4},
	journal = {The Journal of Chemical Physics},
	author = {Wada, A. and Ford, N. C. and Karasz, F. E.},
	month = aug,
	year = {1971},
	pages = {1798--1802},
}

@article{oravec_robust_2018,
	title = {Robust model predictive control and {PID} control of shell-and-tube heat exchangers},
	volume = {159},
	issn = {0360-5442},
	doi = {10.1016/j.energy.2018.06.106},
	journal = {Energy},
	author = {Oravec, Juraj and Bakošová, Monika and Trafczynski, Marian and Vasičkaninová, Anna and Mészáros, Alajos and Markowski, Mariusz},
	month = sep,
	year = {2018},
	pages = {1--10},
}

@article{wang_rna_2020,
	title = {{RNA} therapeutics on the rise},
	volume = {19},
	issn = {1474-1776, 1474-1784},
	doi = {10.1038/d41573-020-00078-0},
	language = {en},
	number = {7},
	journal = {Nature Reviews Drug Discovery},
	author = {Wang, Feng and Zuroske, Travis and Watts, Jonathan K.},
	month = jul,
	year = {2020},
	pages = {441--442},
}

@article{nikolaidis_rigorous_2022,
	title = {Rigorous phase equilibrium calculation methods for strong electrolyte solutions: {The} isothermal flash},
	volume = {558},
	issn = {0378-3812},
	doi = {10.1016/j.fluid.2022.113441},
	language = {en},
	journal = {Fluid Phase Equilibria},
	author = {Nikolaidis, Ilias K. and Novak, Nefeli and Kontogeorgis, Georgios M. and Economou, Ioannis G.},
	month = jul,
	year = {2022},
	pages = {113441},
}

@article{cardenas_review_2023,
	title = {Review of structural design guiding the development of lipid nanoparticles for nucleic acid delivery},
	volume = {66},
	issn = {13590294},
	doi = {10.1016/j.cocis.2023.101705},
	language = {en},
	journal = {Current Opinion in Colloid \& Interface Science},
	author = {Cárdenas, Marité and Campbell, Richard A. and Yanez Arteta, Marianna and Lawrence, M. Jayne and Sebastiani, Federica},
	month = aug,
	year = {2023},
	pages = {101705},
}

@article{mckenzie_recent_2021,
	title = {Recent progress in non-native nucleic acid modifications},
	volume = {50},
	issn = {0306-0012, 1460-4744},
	doi = {10.1039/D0CS01430C},
	language = {en},
	number = {8},
	journal = {Chemical Society Reviews},
	author = {McKenzie, Luke K. and El-Khoury, Roberto and Thorpe, James D. and Damha, Masad J. and Hollenstein, Marcel},
	year = {2021},
	pages = {5126--5164},
}

@article{wilcox_reassessment_1988,
	title = {Reassessment of the scale-determining equation for advanced turbulence models},
	volume = {26},
	issn = {0001-1452, 1533-385X},
	doi = {10.2514/3.10041},
	language = {en},
	number = {11},
	journal = {AIAA Journal},
	author = {Wilcox, David C.},
	month = nov,
	year = {1988},
	pages = {1299--1310},
}

@article{chen_rapid_2012,
	title = {Rapid discovery of potent {siRNA}-containing lipid nanoparticles enabled by controlled microfluidic formulation},
	volume = {134},
	issn = {0002-7863, 1520-5126},
	doi = {10.1021/ja301621z},
	language = {en},
	number = {16},
	journal = {Journal of the American Chemical Society},
	author = {Chen, Delai and Love, Kevin T. and Chen, Yi and Eltoukhy, Ahmed A. and Kastrup, Christian and Sahay, Gaurav and Jeon, Alvin and Dong, Yizhou and Whitehead, Kathryn A. and Anderson, Daniel G.},
	month = apr,
	year = {2012},
	pages = {6948--6951},
}

@article{daniel_quality_2022,
	title = {Quality by {Design} for enabling {RNA} platform production processes},
	volume = {40},
	issn = {01677799},
	doi = {10.1016/j.tibtech.2022.03.012},
	language = {en},
	number = {10},
	journal = {Trends in Biotechnology},
	author = {Daniel, Simon and Kis, Zoltán and Kontoravdi, Cleo and Shah, Nilay},
	month = oct,
	year = {2022},
	pages = {1213--1228},
}

@misc{rico_pysages_2023,
	title = {{PySAGES}: flexible, advanced sampling methods accelerated with {GPUs}},
	shorttitle = {{PySAGES}},
	doi = {10.48550/arXiv.2301.04835},
	publisher = {arXiv},
	author = {Rico, Pablo F. Zubieta and Schneider, Ludwig and Pérez-Lemus, Gustavo R. and Alessandri, Riccardo and Dasetty, Siva and Menéndez, Cintia A. and Wu, Yiheng and Jin, Yezhi and Xu, Yinan and Nguyen, Trung D. and Parker, John A. and Ferguson, Andrew L. and Whitmer, Jonathan K. and de Pablo, Juan J.},
	month = apr,
	year = {2023},
	note = {arXiv:2301.04835 [physics]},
}

@article{goncalves_pvt_2010,
	title = {{PVT}, viscosity, and surface tension of ethanol: {New} measurements and literature data evaluation},
	volume = {42},
	issn = {00219614},
	doi = {10.1016/j.jct.2010.03.022},
	language = {en},
	number = {8},
	journal = {The Journal of Chemical Thermodynamics},
	author = {Gonçalves, F.A.M.M. and Trindade, A.R. and Costa, C.S.M.F. and Bernardo, J.C.S. and Johnson, I. and Fonseca, I.M.A. and Ferreira, A.G.M.},
	month = aug,
	year = {2010},
	pages = {1039--1049},
}

@article{alobaid_progress_2022,
	title = {Progress in {CFD} simulations of fluidized beds for chemical and energy process engineering},
	volume = {91},
	issn = {03601285},
	doi = {10.1016/j.pecs.2021.100930},
	language = {en},
	journal = {Progress in Energy and Combustion Science},
	author = {Alobaid, Falah and Almohammed, Naser and Massoudi Farid, Massoud and May, Jan and Rößger, Philip and Richter, Andreas and Epple, Bernd},
	month = jul,
	year = {2022},
	pages = {100930},
}

@article{obrien_laramy_process_2023,
	title = {Process robustness in lipid nanoparticle production: {A} comparison of microfluidic and turbulent jet mixing},
	volume = {20},
	issn = {1543-8384, 1543-8392},
	doi = {10.1021/acs.molpharmaceut.3c00390},
	language = {en},
	number = {8},
	journal = {Molecular Pharmaceutics},
	author = {O’Brien Laramy, Matthew N. and Costa, Antonio P. and Cebrero, Yareli Maciel and Joseph, Johnson and Sarode, Apoorva and Zang, Nanzhi and Kim, Lee Joon and Hofmann, Kate and Wang, Shirley and Goyon, Alexandre and Koenig, Stefan G. and Hammel, Michal and Hura, Greg L.},
	month = aug,
	year = {2023},
	pages = {4285--4296},
}

@article{kon_principles_2022,
	title = {Principles for designing an optimal {mRNA} lipid nanoparticle vaccine},
	volume = {73},
	issn = {09581669},
	doi = {10.1016/j.copbio.2021.09.016},
	language = {en},
	journal = {Current Opinion in Biotechnology},
	author = {Kon, Edo and Elia, Uri and Peer, Dan},
	month = feb,
	year = {2022},
	pages = {329--336},
}

@article{schwarzer_predictive_2006,
	title = {Predictive simulation of nanoparticle precipitation based on the population balance equation},
	volume = {61},
	issn = {00092509},
	doi = {10.1016/j.ces.2004.11.064},
	language = {en},
	number = {1},
	journal = {Chemical Engineering Science},
	author = {Schwarzer, Hans-Christoph and Schwertfirm, Florian and Manhart, Michael and Schmid, Hans-Joachim and Peukert, Wolfgang},
	month = jan,
	year = {2006},
	pages = {167--181},
}

@article{fee_prediction_2004,
	title = {Prediction of the viscosity radius and the size exclusion chromatography behavior of {PEGylated} proteins},
	volume = {15},
	issn = {1043-1802, 1520-4812},
	doi = {10.1021/bc049843w},
	language = {en},
	number = {6},
	journal = {Bioconjugate Chemistry},
	author = {Fee, Conan J. and Van Alstine, James M.},
	month = nov,
	year = {2004},
	pages = {1304--1313},
}

@article{habicht_predicting_2023,
	title = {Predicting {PC}-{SAFT} pure-component parameters by machine learning using a molecular fingerprint as key input},
	volume = {565},
	issn = {0378-3812},
	doi = {10.1016/j.fluid.2022.113657},
	journal = {Fluid Phase Equilibria},
	author = {Habicht, Jonas and Brandenbusch, Christoph and Sadowski, Gabriele},
	month = feb,
	year = {2023},
	pages = {113657},
}

@article{jeldres_population_2018,
	title = {Population balance modelling to describe the particle aggregation process: {A} review},
	volume = {326},
	issn = {00325910},
	shorttitle = {Population balance modelling to describe the particle aggregation process},
	doi = {10.1016/j.powtec.2017.12.033},
	language = {en},
	journal = {Powder Technology},
	author = {Jeldres, Ricardo I. and Fawell, Phillip D. and Florio, Brendan J.},
	month = feb,
	year = {2018},
	pages = {190--207},
}

@article{ramkrishna_population_2014,
	title = {Population balance modeling: {Current} status and future prospects},
	volume = {5},
	issn = {1947-5438, 1947-5446},
	doi = {10.1146/annurev-chembioeng-060713-040241},
	language = {en},
	number = {1},
	journal = {Annual Review of Chemical and Biomolecular Engineering},
	author = {Ramkrishna, Doraiswami and Singh, Meenesh R.},
	month = jun,
	year = {2014},
	pages = {123--146},
}

@article{sajjadi_population_2009,
	title = {Population balance modeling of particle size distribution in monomer‐starved semibatch emulsion polymerization},
	volume = {55},
	issn = {0001-1541, 1547-5905},
	doi = {10.1002/aic.11917},
	language = {en},
	number = {12},
	journal = {AIChE Journal},
	author = {Sajjadi, Shahriar},
	month = dec,
	year = {2009},
	pages = {3191--3205},
}

@article{karatrantos_polymer_2017,
	title = {Polymer and spherical nanoparticle diffusion in nanocomposites},
	volume = {146},
	issn = {0021-9606, 1089-7690},
	doi = {10.1063/1.4981258},
	language = {en},
	number = {20},
	journal = {The Journal of Chemical Physics},
	author = {Karatrantos, Argyrios and Composto, Russell J. and Winey, Karen I. and Clarke, Nigel},
	month = may,
	year = {2017},
	pages = {203331},
}

@article{bonomi_plumed_2009,
	title = {{PLUMED}: {A} portable plugin for free-energy calculations with molecular dynamics},
	volume = {180},
	issn = {0010-4655},
	shorttitle = {{PLUMED}},
	doi = {10.1016/j.cpc.2009.05.011},
	number = {10},
	journal = {Computer Physics Communications},
	author = {Bonomi, Massimiliano and Branduardi, Davide and Bussi, Giovanni and Camilloni, Carlo and Provasi, Davide and Raiteri, Paolo and Donadio, Davide and Marinelli, Fabrizio and Pietrucci, Fabio and Broglia, Ricardo A. and Parrinello, Michele},
	month = oct,
	year = {2009},
	pages = {1961--1972},
}

@article{chen_phase-field_2002,
	title = {Phase-field models for microstructure evolution},
	volume = {32},
	issn = {1531-7331, 1545-4118},
	doi = {10.1146/annurev.matsci.32.112001.132041},
	language = {en},
	number = {1},
	journal = {Annual Review of Materials Research},
	author = {Chen, Long-Qing},
	month = aug,
	year = {2002},
	pages = {113--140},
}

@article{takaki_phase-field_2014,
	title = {Phase-field modeling and simulations of dendrite growth},
	volume = {54},
	issn = {0915-1559, 1347-5460},
	doi = {10.2355/isijinternational.54.437},
	language = {en},
	number = {2},
	journal = {ISIJ International},
	author = {Takaki, Tomohiro},
	year = {2014},
	pages = {437--444},
}

@article{hayashi_phase_1975,
	title = {Phase transitions of phospholipids in monolayers and surface viscosity},
	volume = {15},
	issn = {00093084},
	doi = {10.1016/0009-3084(75)90043-2},
	language = {en},
	number = {2},
	journal = {Chemistry and Physics of Lipids},
	author = {Hayashi, Makoto and Muramatsu, Toshio and Hara, Ichiro and Seimiya, Tsutomu},
	month = nov,
	year = {1975},
	pages = {209--215},
}

@article{zhou_phase_2006,
	title = {Phase field simulations of early stage structure formation during immersion precipitation of polymeric membranes in {2D} and {3D}},
	volume = {268},
	issn = {03767388},
	doi = {10.1016/j.memsci.2005.05.030},
	language = {en},
	number = {2},
	journal = {Journal of Membrane Science},
	author = {Zhou, Bo and Powell, Adam C.},
	month = jan,
	year = {2006},
	pages = {150--164},
}

@article{guyer_phase_2004,
	title = {Phase field modeling of electrochemistry. {I}. {Equilibrium}},
	volume = {69},
	issn = {1539-3755, 1550-2376},
	doi = {10.1103/PhysRevE.69.021603},
	language = {en},
	number = {2},
	journal = {Physical Review E},
	author = {Guyer, J. E. and Boettinger, W. J. and Warren, J. A. and McFadden, G. B.},
	month = feb,
	year = {2004},
	pages = {021603},
}

@article{wehbe_phase_2022,
	title = {Phase behaviour and {pH}-solubility profile prediction of aqueous buffered solutions of ibuprofen and ketoprofen},
	volume = {560},
	issn = {0378-3812},
	doi = {10.1016/j.fluid.2022.113504},
	journal = {Fluid Phase Equilibria},
	author = {Wehbe, Malak and Haslam, Andrew J. and Jackson, George and Galindo, Amparo},
	month = sep,
	year = {2022},
	pages = {113504},
}

@article{larson_ph-dependent_2022,
	title = {{pH}-dependent phase behavior and stability of cationic lipid–{mRNA} nanoparticles},
	volume = {111},
	issn = {0022-3549},
	doi = {10.1016/j.xphs.2021.11.004},
	number = {3},
	journal = {Journal of Pharmaceutical Sciences},
	author = {Larson, Nicholas R. and Hu, Gang and Wei, Yangjie and Tuesca, Anthony D. and Forrest, M. Laird and Middaugh, C. Russell},
	month = mar,
	year = {2022},
	pages = {690--698},
}

@article{settanni_ph-dependent_2022,
	title = {{pH}-dependent behavior of ionizable cationic lipids in {mRNA}-carrying lipoplexes investigated by molecular dynamics simulations},
	volume = {43},
	issn = {1521-3927},
	doi = {10.1002/marc.202100683},
	language = {en},
	number = {12},
	journal = {Macromolecular Rapid Communications},
	author = {Settanni, Giovanni and Brill, Wolfgang and Haas, Heinrich and Schmid, Friederike},
	year = {2022},
	pages = {2100683},
}

@article{gross_perturbed-chain_2001,
	title = {Perturbed-chain {SAFT}: {An} equation of state based on a perturbation theory for chain molecules},
	volume = {40},
	issn = {0888-5885, 1520-5045},
	shorttitle = {Perturbed-{Chain} {SAFT}},
	doi = {10.1021/ie0003887},
	language = {en},
	number = {4},
	journal = {Ind. Eng. Chem. Res.},
	author = {Gross, Joachim and Sadowski, Gabriele},
	month = feb,
	year = {2001},
	pages = {1244--1260},
}

@article{tkatchenko_performances_2007,
	title = {Performances of {LES} and {RANS} models for simulation of complex flows in a coaxial jet mixer},
	volume = {78},
	issn = {1386-6184, 1573-1987},
	doi = {10.1007/s10494-006-9053-3},
	language = {en},
	number = {2},
	journal = {Flow, Turbulence and Combustion},
	author = {Tkatchenko, Igor and Kornev, Nikolai and Jahnke, Steffen and Steffen, Günter and Hassel, Egon},
	month = feb,
	year = {2007},
	pages = {111--127},
}

@article{bogaerts_parameter_2004,
	title = {Parameter identification for state estimation—application to bioprocess software sensors},
	volume = {59},
	issn = {0009-2509},
	doi = {https://doi.org/10.1016/j.ces.2004.01.066},
	number = {12},
	journal = {Chemical Engineering Science},
	author = {Bogaerts, Ph and Wouwer, A. Vande},
	year = {2004},
	pages = {2465--2476},
}

@article{kutalik_optimal_2004,
	title = {Optimal sampling time selection for parameter estimation in dynamic pathway modeling},
	volume = {75},
	issn = {0303-2647},
	doi = {https://doi.org/10.1016/j.biosystems.2004.03.007},
	number = {1},
	journal = {Biosystems},
	author = {Kutalik, Zoltán and Cho, Kwang-Hyun and Wolkenhauer, Olaf},
	year = {2004},
	pages = {43--55},
}

@article{perala_two-step_2014,
	title = {On the two-step mechanism for synthesis of transition-metal nanoparticles},
	volume = {30},
	issn = {0743-7463, 1520-5827},
	doi = {10.1021/la503199m},
	language = {en},
	number = {42},
	journal = {Langmuir},
	author = {Perala, Siva Rama Krishna and Kumar, Sanjeev},
	month = oct,
	year = {2014},
	pages = {12703--12711},
}

@article{wieland_structural_2021,
	title = {On structural and practical identifiability},
	volume = {25},
	issn = {24523100},
	doi = {10.1016/j.coisb.2021.03.005},
	language = {en},
	journal = {Current Opinion in Systems Biology},
	author = {Wieland, Franz-Georg and Hauber, Adrian L. and Rosenblatt, Marcus and Tönsing, Christian and Timmer, Jens},
	month = mar,
	year = {2021},
	pages = {60--69},
}

@article{grasselli_phase_2023,
	title = {On a phase field model for {RNA}-protein dynamics},
	volume = {55},
	issn = {0036-1410, 1095-7154},
	doi = {10.1137/22M1483086},
	language = {en},
	number = {1},
	journal = {SIAM Journal on Mathematical Analysis},
	author = {Grasselli, Maurizio and Scarpa, Luca and Signori, Andrea},
	month = feb,
	year = {2023},
	pages = {405--457},
}

@article{inguva_numerical_2020,
	title = {Numerical simulation, clustering, and prediction of multicomponent polymer precipitation},
	volume = {1},
	issn = {2632-6736},
	doi = {10.1017/dce.2020.14},
	language = {en},
	journal = {Data-Centric Engineering},
	author = {Inguva, Pavan and Mason, Lachlan R. and Pan, Indranil and Hengardi, Miselle and Matar, Omar K.},
	year = {2020},
	pages = {e13},
}

@article{gupta_nucleic_2021,
	title = {Nucleic acid delivery for therapeutic applications},
	volume = {178},
	issn = {0169409X},
	doi = {10.1016/j.addr.2021.113834},
	language = {en},
	journal = {Advanced Drug Delivery Reviews},
	author = {Gupta, Akash and Andresen, Jason L. and Manan, Rajith S. and Langer, Robert},
	month = nov,
	year = {2021},
	pages = {113834},
}

@article{schall_nucleation_2018,
	title = {Nucleation and growth kinetics for combined cooling and antisolvent crystallization in a mixed-suspension, mixed-product removal system: {Estimating} solvent dependency},
	volume = {18},
	issn = {1528-7483, 1528-7505},
	doi = {10.1021/acs.cgd.7b01528},
	language = {en},
	number = {3},
	journal = {Crystal Growth \& Design},
	author = {Schall, Jennifer M. and Mandur, Jasdeep S. and Braatz, Richard D. and Myerson, Allan S.},
	month = mar,
	year = {2018},
	pages = {1560--1570},
}

@article{yu_nonrandom_2019,
	title = {Nonrandom two-liquid activity coefficient model for aqueous polyelectrolyte solutions},
	volume = {497},
	issn = {0378-3812},
	doi = {10.1016/j.fluid.2019.05.009},
	journal = {Fluid Phase Equilibria},
	author = {Yu, Yue and Li, Yuan and Hossain, Nazir and Chen, Chau-Chyun},
	month = oct,
	year = {2019},
	pages = {1--9},
}

@article{nauman_nonlinear_2001,
	title = {Nonlinear diffusion and phase separation},
	volume = {56},
	issn = {00092509},
	doi = {10.1016/S0009-2509(01)00005-7},
	language = {en},
	number = {6},
	journal = {Chemical Engineering Science},
	author = {Nauman, E.Bruce and He, David Qiwei},
	month = mar,
	year = {2001},
	pages = {1999--2018},
}

@article{yin_non-viral_2014,
	title = {Non-viral vectors for gene-based therapy},
	volume = {15},
	issn = {1471-0056, 1471-0064},
	doi = {10.1038/nrg3763},
	language = {en},
	number = {8},
	journal = {Nature Reviews Genetics},
	author = {Yin, Hao and Kanasty, Rosemary L. and Eltoukhy, Ahmed A. and Vegas, Arturo J. and Dorkin, J. Robert and Anderson, Daniel G.},
	month = aug,
	year = {2014},
	pages = {541--555},
}

@article{besseling_new_2019,
	title = {New unique {PAT} method and instrument for real-time inline size characterization of concentrated, flowing nanosuspensions},
	volume = {133},
	issn = {09280987},
	doi = {10.1016/j.ejps.2019.03.024},
	language = {en},
	journal = {European Journal of Pharmaceutical Sciences},
	author = {Besseling, R. and Damen, M. and Wijgergangs, J. and Hermes, M. and Wynia, G. and Gerich, A.},
	month = may,
	year = {2019},
	pages = {205--213},
}

@article{armstrong_nanosecond_2014,
	title = {Nanosecond lipid dynamics in membranes containing cholesterol},
	volume = {10},
	issn = {1744-683X, 1744-6848},
	doi = {10.1039/c3sm51757h},
	language = {en},
	number = {15},
	journal = {Soft Matter},
	author = {Armstrong, Clare L. and Häußler, Wolfgang and Seydel, Tilo and Katsaras, John and Rheinstädter, Maikel C.},
	year = {2014},
	pages = {2600},
}

@article{kalathi_nanoparticle_2014,
	title = {Nanoparticle diffusion in polymer nanocomposites},
	volume = {112},
	issn = {0031-9007, 1079-7114},
	doi = {10.1103/PhysRevLett.112.108301},
	language = {en},
	number = {10},
	journal = {Physical Review Letters},
	author = {Kalathi, Jagannathan T. and Yamamoto, Umi and Schweizer, Kenneth S. and Grest, Gary S. and Kumar, Sanat K.},
	month = mar,
	year = {2014},
	pages = {108301},
}

@article{roussel_multiscale_2014,
	title = {Multiscale molecular dynamics simulations of sodium dodecyl sulfate micelles: from coarse-grained to all-atom resolution},
	volume = {20},
	issn = {0948-5023},
	doi = {10.1007/s00894-014-2469-0},
	language = {en},
	number = {10},
	journal = {Journal of Molecular Modeling},
	author = {Roussel, Guillaume and Michaux, Catherine and Perpète, Eric A.},
	month = oct,
	year = {2014},
	pages = {2469},
}

@article{da_rosa_multiscale_2018,
	title = {Multiscale modeling and simulation of macromixing, micromixing, and crystal size distribution in radial mixers/crystallizers},
	volume = {57},
	issn = {0888-5885, 1520-5045},
	doi = {10.1021/acs.iecr.8b00359},
	language = {en},
	number = {15},
	journal = {Industrial \& Engineering Chemistry Research},
	author = {da Rosa, Cezar A. and Braatz, Richard D.},
	month = apr,
	year = {2018},
	pages = {5433--5441},
}

@article{schoenmaker_mrna-lipid_2021,
	title = {{mRNA}-lipid nanoparticle {COVID}-19 vaccines: {Structure} and stability},
	volume = {601},
	issn = {03785173},
	doi = {10.1016/j.ijpharm.2021.120586},
	language = {en},
	journal = {International Journal of Pharmaceutics},
	author = {Schoenmaker, Linde and Witzigmann, Dominik and Kulkarni, Jayesh A. and Verbeke, Rein and Kersten, Gideon and Jiskoot, Wim and Crommelin, Daan J.A.},
	month = may,
	year = {2021},
	pages = {120586},
}

@article{trollmann_mrna_2022,
	title = {{mRNA} lipid nanoparticle phase transition},
	volume = {121},
	issn = {0006-3495},
	doi = {10.1016/j.bpj.2022.08.037},
	number = {20},
	journal = {Biophysical Journal},
	author = {Trollmann, Marius F. W. and Böckmann, Rainer A.},
	month = oct,
	year = {2022},
	pages = {3927--3939},
}

@article{nauman_morphology_1994,
	title = {Morphology predictions for ternary polymer blends undergoing spinodal decomposition},
	volume = {35},
	issn = {00323861},
	doi = {10.1016/0032-3861(94)90757-9},
	language = {en},
	number = {11},
	journal = {Polymer},
	author = {Nauman, E.Bruce and He, David Qiwei},
	month = may,
	year = {1994},
	pages = {2243--2255},
}

@incollection{dochain_monitoring_1998,
	address = {Dordrecht},
	title = {Monitoring and {Adaptive} {Control} of {Bioprocesses}},
	isbn = {978-94-015-9111-9},
	booktitle = {Advanced {Instrumentation}, {Data} {Interpretation}, and {Control} of {Biotechnological} {Processes}},
	publisher = {Springer Netherlands},
	author = {Dochain, D. and Perrier, M.},
	editor = {Van Impe, Jan F. M. and Vanrolleghem, Peter A. and Iserentant, Dirk M.},
	year = {1998},
	doi = {10.1007/978-94-015-9111-9_12},
	pages = {347--400},
}

@article{feller_molecular_2000,
	title = {Molecular dynamics simulations of lipid bilayers},
	volume = {5},
	issn = {1359-0294},
	doi = {10.1016/S1359-0294(00)00058-3},
	number = {3},
	journal = {Current Opinion in Colloid \& Interface Science},
	author = {Feller, Scott E.},
	month = jul,
	year = {2000},
	pages = {217--223},
}

@article{rudyak_molecular_2011,
	title = {Molecular dynamics simulation of nanoparticle diffusion in dense fluids},
	volume = {11},
	issn = {1613-4982, 1613-4990},
	doi = {10.1007/s10404-011-0815-4},
	language = {en},
	number = {4},
	journal = {Microfluidics and Nanofluidics},
	author = {Rudyak, Valery Ya. and Krasnolutskii, Sergey L. and Ivanov, Denis A.},
	month = oct,
	year = {2011},
	pages = {501--506},
}

@article{kim_modifications_2022,
	title = {Modifications of {mRNA} vaccine structural elements for improving {mRNA} stability and translation efficiency},
	volume = {18},
	issn = {1738-642X, 2092-8467},
	doi = {10.1007/s13273-021-00171-4},
	language = {en},
	number = {1},
	journal = {Molecular \& Cellular Toxicology},
	author = {Kim, Sun Chang and Sekhon, Simranjeet Singh and Shin, Woo-Ri and Ahn, Gna and Cho, Byung-Kwan and Ahn, Ji-Young and Kim, Yang-Hoon},
	month = jan,
	year = {2022},
	pages = {1--8},
}

@article{valsecchi_modelling_2024,
	title = {Modelling the thermodynamic properties of the mixture of water and polyethylene glycol ({PEG}) with the {SAFT}-γ {Mie} group-contribution approach},
	volume = {577},
	issn = {0378-3812},
	doi = {10.1016/j.fluid.2023.113952},
	journal = {Fluid Phase Equilibria},
	author = {Valsecchi, Michele and Galindo, Amparo and Jackson, George},
	month = feb,
	year = {2024},
	pages = {113952},
}

@article{vonka_modelling_2012,
	title = {Modelling the morphology evolution of polymer materials undergoing phase separation},
	volume = {207-208},
	issn = {13858947},
	doi = {10.1016/j.cej.2012.06.091},
	language = {en},
	journal = {Chemical Engineering Journal},
	author = {Vonka, Michal and Kosek, Juraj},
	month = oct,
	year = {2012},
	pages = {895--905},
}

@article{gasior_modeling_2020,
	title = {Modeling the mechanisms by which coexisting biomolecular {RNA}–protein condensates form},
	volume = {82},
	issn = {0092-8240, 1522-9602},
	doi = {10.1007/s11538-020-00823-x},
	language = {en},
	number = {12},
	journal = {Bulletin of Mathematical Biology},
	author = {Gasior, K. and Forest, M. G. and Gladfelter, A. S. and Newby, J. M.},
	month = dec,
	year = {2020},
	pages = {153},
}

@article{kesler_modeling_2016,
	title = {Modeling size controlled nanoparticle precipitation with the co-solvency method by spinodal decomposition},
	volume = {12},
	issn = {1744-683X, 1744-6848},
	doi = {10.1039/C6SM01198E},
	language = {en},
	number = {34},
	journal = {Soft Matter},
	author = {Keßler, Simon and Schmid, Friederike and Drese, Klaus},
	year = {2016},
	pages = {7231--7240},
}

@article{manzanarez_modeling_2017,
	title = {Modeling phase inversion using {Cahn}-{Hilliard} equations – {Influence} of the mobility on the pattern formation dynamics},
	volume = {173},
	issn = {00092509},
	doi = {10.1016/j.ces.2017.08.009},
	language = {en},
	journal = {Chemical Engineering Science},
	author = {Manzanarez, H. and Mericq, J.P. and Guenoun, P. and Chikina, J. and Bouyer, D.},
	month = dec,
	year = {2017},
	pages = {411--427},
}

@article{liu_modeling_2014,
	title = {Modeling of silver nanoparticle formation in a microreactor: {Reaction} kinetics coupled with population balance model and fluid dynamics},
	volume = {53},
	issn = {0888-5885, 1520-5045},
	doi = {10.1021/ie4031314},
	language = {en},
	number = {11},
	journal = {Industrial \& Engineering Chemistry Research},
	author = {Liu, Hongyu and Li, Jun and Sun, Daohua and Odoom-Wubah, Tareque and Huang, Jiale and Li, Qingbiao},
	month = mar,
	year = {2014},
	pages = {4263--4270},
}

@article{hopp-hirschler_modeling_2018,
	title = {Modeling of pore formation in phase inversion processes: {Model} and numerical results},
	volume = {564},
	issn = {03767388},
	doi = {10.1016/j.memsci.2018.07.085},
	language = {en},
	journal = {Journal of Membrane Science},
	author = {Hopp-Hirschler, Manuel and Nieken, Ulrich},
	month = oct,
	year = {2018},
	pages = {820--831},
}

@article{reschke_modeling_2015,
	title = {Modeling aqueous two-phase systems: {III}. {Polymers} and organic salts as {ATPS} former},
	volume = {387},
	issn = {0378-3812},
	shorttitle = {Modeling aqueous two-phase systems},
	doi = {10.1016/j.fluid.2014.12.011},
	journal = {Fluid Phase Equilibria},
	author = {Reschke, Thomas and Brandenbusch, Christoph and Sadowski, Gabriele},
	month = feb,
	year = {2015},
	pages = {178--189},
}

@article{reschke_modeling_2014,
	title = {Modeling aqueous two-phase systems: {II}. {Inorganic} salts and polyether homo- and copolymers as {ATPS} former},
	volume = {375},
	issn = {0378-3812},
	shorttitle = {Modeling aqueous two-phase systems},
	doi = {10.1016/j.fluid.2014.04.040},
	journal = {Fluid Phase Equilibria},
	author = {Reschke, Thomas and Brandenbusch, Christoph and Sadowski, Gabriele},
	month = aug,
	year = {2014},
	pages = {306--315},
}

@article{li_modeling_2021,
	title = {Modeling aqueous multivalent polyelectrolytes systems with polyelectrolyte {NRTL} model},
	volume = {336},
	issn = {0167-7322},
	doi = {10.1016/j.molliq.2021.116237},
	journal = {Journal of Molecular Liquids},
	author = {Li, Yuan and Yu, Yue and Chen, Chau-Chyun},
	month = aug,
	year = {2021},
	pages = {116237},
}

@article{b_kanwar_modeling_2022,
	title = {Modeling and controller design for enhanced hollow-fiber bioreactor performance},
	volume = {6},
	issn = {2475-1456},
	doi = {10.1109/LCSYS.2021.3050724},
	journal = {IEEE Control Systems Letters},
	author = {{B. Kanwar} and {S. Balakirsky} and {A. Mazumdar}},
	year = {2022},
	pages = {115--120},
}

@article{abt_model-based_2018,
	title = {Model-based tools for optimal experiments in bioprocess engineering},
	volume = {22},
	issn = {22113398},
	doi = {10.1016/j.coche.2018.11.007},
	language = {en},
	journal = {Current Opinion in Chemical Engineering},
	author = {Abt, Vinzenz and Barz, Tilman and Cruz-Bournazou, Mariano Nicolas and Herwig, Christoph and Kroll, Paul and Möller, Johannes and Pörtner, Ralf and Schenkendorf, René},
	month = dec,
	year = {2018},
	pages = {244--252},
}

@article{franceschini_model-based_2008,
	title = {Model-based design of experiments for parameter precision: {State} of the art},
	volume = {63},
	issn = {00092509},
	doi = {10.1016/j.ces.2007.11.034},
	language = {en},
	number = {19},
	journal = {Chemical Engineering Science},
	author = {Franceschini, Gaia and Macchietto, Sandro},
	month = oct,
	year = {2008},
	pages = {4846--4872},
}

@article{duan_model_2020,
	title = {Model reduction of aerobic bioprocess models for efficient simulation},
	volume = {217},
	issn = {0009-2509},
	doi = {10.1016/j.ces.2020.115512},
	journal = {Chemical Engineering Science},
	author = {Duan, Zhaoyang and Wilms, Terrance and Neubauer, Peter and Kravaris, Costas and Cruz Bournazou, Mariano Nicolas},
	month = may,
	year = {2020},
	pages = {115512},
}

@article{meyer_micromixing_2009,
	title = {Micromixing models for turbulent flows},
	volume = {228},
	issn = {00219991},
	doi = {10.1016/j.jcp.2008.10.019},
	language = {en},
	number = {4},
	journal = {Journal of Computational Physics},
	author = {Meyer, Daniel W. and Jenny, Patrick},
	month = mar,
	year = {2009},
	pages = {1275--1293},
}

@article{leung_microfluidic_2015,
	title = {Microfluidic mixing: {A} general method for encapsulating macromolecules in lipid nanoparticle systems},
	volume = {119},
	issn = {1520-6106, 1520-5207},
	doi = {10.1021/acs.jpcb.5b02891},
	language = {en},
	number = {28},
	journal = {The Journal of Physical Chemistry B},
	author = {Leung, Alex K. K. and Tam, Yuen Yi C. and Chen, Sam and Hafez, Ismail M. and Cullis, Pieter R.},
	month = jul,
	year = {2015},
	pages = {8698--8706},
}

@article{canova_mechanistic_2023,
	title = {Mechanistic modeling of viral particle production},
	volume = {120},
	issn = {0006-3592, 1097-0290},
	doi = {10.1002/bit.28296},
	language = {en},
	number = {3},
	journal = {Biotechnology and Bioengineering},
	author = {Canova, Christopher T. and Inguva, Pavan K. and Braatz, Richard D.},
	month = mar,
	year = {2023},
	pages = {629--641},
}

@article{hong_mechanistic_2021,
	title = {Mechanistic modeling and parameter-adaptive nonlinear model predictive control of a microbioreactor},
	volume = {147},
	issn = {0098-1354},
	doi = {10.1016/j.compchemeng.2021.107255},
	journal = {Computers \& Chemical Engineering},
	author = {Hong, Moo Sun and Braatz, Richard D.},
	month = apr,
	year = {2021},
	pages = {107255},
}

@article{thanh_mechanisms_2014,
	title = {Mechanisms of nucleation and growth of nanoparticles in solution},
	volume = {114},
	issn = {0009-2665, 1520-6890},
	doi = {10.1021/cr400544s},
	language = {en},
	number = {15},
	journal = {Chemical Reviews},
	author = {Thanh, Nguyen T. K. and Maclean, N. and Mahiddine, S.},
	month = aug,
	year = {2014},
	pages = {7610--7630},
}

@article{handwerk_mechanism-enabled_2019,
	title = {Mechanism-enabled population balance modeling of particle formation en route to particle average size and size distribution understanding and control},
	volume = {141},
	issn = {0002-7863, 1520-5126},
	doi = {10.1021/jacs.9b06364},
	language = {en},
	number = {40},
	journal = {Journal of the American Chemical Society},
	author = {Handwerk, Derek R. and Shipman, Patrick D. and Whitehead, Christopher B. and Özkar, Saim and Finke, Richard G.},
	month = oct,
	year = {2019},
	pages = {15827--15839},
}

@article{vargas_mechanism_2005,
	title = {Mechanism of {mRNA} transport in the nucleus},
	volume = {102},
	issn = {0027-8424, 1091-6490},
	doi = {10.1073/pnas.0505580102},
	language = {en},
	number = {47},
	journal = {Proceedings of the National Academy of Sciences},
	author = {Vargas, Diana Y. and Raj, Arjun and Marras, Salvatore A. E. and Kramer, Fred Russell and Tyagi, Sanjay},
	month = nov,
	year = {2005},
	pages = {17008--17013},
}

@article{michaud-agrawal_mdanalysis_2011,
	title = {{MDAnalysis}: {A} toolkit for the analysis of molecular dynamics simulations},
	volume = {32},
	issn = {01928651},
	doi = {10.1002/jcc.21787},
	language = {en},
	number = {10},
	journal = {Journal of Computational Chemistry},
	author = {Michaud-Agrawal, Naveen and Denning, Elizabeth J. and Woolf, Thomas B. and Beckstein, Oliver},
	month = jul,
	year = {2011},
	pages = {2319--2327},
}

@article{souza_martini_2021,
	title = {Martini 3: a general purpose force field for coarse-grained molecular dynamics},
	volume = {18},
	issn = {1548-7105},
	shorttitle = {Martini 3},
	doi = {10.1038/s41592-021-01098-3},
	language = {en},
	number = {4},
	journal = {Nature Methods},
	author = {Souza, Paulo C. T. and Alessandri, Riccardo and Barnoud, Jonathan and Thallmair, Sebastian and Faustino, Ignacio and Grünewald, Fabian and Patmanidis, Ilias and Abdizadeh, Haleh and Bruininks, Bart M. H. and Wassenaar, Tsjerk A. and Kroon, Peter C. and Melcr, Josef and Nieto, Vincent and Corradi, Valentina and Khan, Hanif M. and Domański, Jan and Javanainen, Matti and Martinez-Seara, Hector and Reuter, Nathalie and Best, Robert B. and Vattulainen, Ilpo and Monticelli, Luca and Periole, Xavier and Tieleman, D. Peter and de Vries, Alex H. and Marrink, Siewert J.},
	month = apr,
	year = {2021},
	pages = {382--388},
}

@article{kanso_macromolecular_2019,
	title = {Macromolecular architecture and complex viscosity},
	volume = {31},
	issn = {1070-6631, 1089-7666},
	doi = {10.1063/1.5111763},
	language = {en},
	number = {8},
	journal = {Physics of Fluids},
	author = {Kanso, M. A. and Giacomin, A. J. and Saengow, C. and Piette, J. H.},
	month = aug,
	year = {2019},
	pages = {087107},
}

@article{ickenstein_lipid-based_2019,
	title = {Lipid-based nanoparticle formulations for small molecules and {RNA} drugs},
	volume = {16},
	issn = {1742-5247, 1744-7593},
	doi = {10.1080/17425247.2019.1669558},
	number = {11},
	journal = {Expert Opinion on Drug Delivery},
	author = {Ickenstein, Ludger M. and Garidel, Patrick},
	month = nov,
	year = {2019},
	pages = {1205--1226},
}

@article{buck_lipid-based_2019,
	title = {Lipid-based {DNA} therapeutics: {Hallmarks} of non-viral gene delivery},
	volume = {13},
	issn = {1936-0851, 1936-086X},
	doi = {10.1021/acsnano.8b07858},
	language = {en},
	number = {4},
	journal = {ACS Nano},
	author = {Buck, Jonas and Grossen, Philip and Cullis, Pieter R. and Huwyler, Jörg and Witzigmann, Dominik},
	month = apr,
	year = {2019},
	pages = {3754--3782},
}

@article{xu_lipid_2022,
	title = {Lipid nanoparticles for drug delivery},
	volume = {2},
	issn = {2699-9307, 2699-9307},
	doi = {10.1002/anbr.202100109},
	language = {en},
	number = {2},
	journal = {Advanced NanoBiomed Research},
	author = {Xu, Letao and Wang, Xing and Liu, Yun and Yang, Guangze and Falconer, Robert J. and Zhao, Chun-Xia},
	month = feb,
	year = {2022},
	pages = {2100109},
}

@article{kulkarni_lipid_2018,
	title = {Lipid nanoparticles enabling gene therapies: {From} concepts to clinical utility},
	volume = {28},
	issn = {2159-3337, 2159-3345},
	doi = {10.1089/nat.2018.0721},
	language = {en},
	number = {3},
	journal = {Nucleic Acid Therapeutics},
	author = {Kulkarni, Jayesh A. and Cullis, Pieter R. and van der Meel, Roy},
	month = jun,
	year = {2018},
	pages = {146--157},
}

@article{cullis_lipid_2017,
	title = {Lipid nanoparticle systems for enabling gene therapies},
	volume = {25},
	issn = {15250016},
	doi = {10.1016/j.ymthe.2017.03.013},
	language = {en},
	number = {7},
	journal = {Molecular Therapy},
	author = {Cullis, Pieter R. and Hope, Michael J.},
	month = jul,
	year = {2017},
	pages = {1467--1475},
}

@article{soong_lateral_2005,
	title = {Lateral diffusion of {PEG}-lipid in magnetically aligned bicelles measured using stimulated echo pulsed field gradient {1H} {NMR}},
	volume = {88},
	issn = {00063495},
	doi = {10.1529/biophysj.104.043620},
	language = {en},
	number = {1},
	journal = {Biophysical Journal},
	author = {Soong, Ronald and Macdonald, Peter M.},
	month = jan,
	year = {2005},
	pages = {255--268},
}

@article{thompson_lammps_2022,
	title = {{LAMMPS} - a flexible simulation tool for particle-based materials modeling at the atomic, meso, and continuum scales},
	volume = {271},
	issn = {0010-4655},
	doi = {10.1016/j.cpc.2021.108171},
	language = {en},
	journal = {Computer Physics Communications},
	author = {Thompson, Aidan P. and Aktulga, H. Metin and Berger, Richard and Bolintineanu, Dan S. and Brown, W. Michael and Crozier, Paul S. and in 't Veld, Pieter J. and Kohlmeyer, Axel and Moore, Stan G. and Nguyen, Trung Dac and Shan, Ray and Stevens, Mark J. and Tranchida, Julien and Trott, Christian and Plimpton, Steven J.},
	month = feb,
	year = {2022},
	pages = {108171},
}

@article{subramaniam_lagrangianeulerian_2013,
	title = {Lagrangian–{Eulerian} methods for multiphase flows},
	volume = {39},
	issn = {03601285},
	doi = {10.1016/j.pecs.2012.10.003},
	language = {en},
	number = {2-3},
	journal = {Progress in Energy and Combustion Science},
	author = {Subramaniam, Shankar},
	month = apr,
	year = {2013},
	pages = {215--245},
}

@article{hu_kinetic_2019,
	title = {Kinetic control in assembly of plasmid {DNA}/polycation complex nanoparticles},
	volume = {13},
	issn = {1936-0851, 1936-086X},
	doi = {10.1021/acsnano.9b03334},
	language = {en},
	number = {9},
	journal = {ACS Nano},
	author = {Hu, Yizong and He, Zhiyu and Hao, Yue and Gong, Like and Pang, Marion and Howard, Gregory P. and Ahn, Hye-Hyun and Brummet, Mary and Chen, Kuntao and Liu, Heng-wen and Ke, Xiyu and Zhu, Jinchang and Anderson, Caleb F. and Cui, Honggang and Ullman, Christopher G. and Carrington, Christine A. and Pomper, Martin G. and Seo, Jung-Hee and Mittal, Rajat and Minn, Il and Mao, Hai-Quan},
	month = sep,
	year = {2019},
	pages = {10161--10178},
}

@article{ermilova_ionizable_2023,
	title = {Ionizable lipids penetrate phospholipid bilayers with high phase transition temperatures: perspectives from free energy calculations},
	volume = {253},
	issn = {00093084},
	doi = {10.1016/j.chemphyslip.2023.105294},
	language = {en},
	journal = {Chemistry and Physics of Lipids},
	author = {Ermilova, Inna and Swenson, Jan},
	month = jul,
	year = {2023},
	pages = {105294},
}

@article{uebbing_investigation_2020,
	title = {Investigation of {pH}-responsiveness inside lipid nanoparticles for parenteral {mRNA} application using small-angle {X}-ray scattering},
	volume = {36},
	issn = {0743-7463, 1520-5827},
	doi = {10.1021/acs.langmuir.0c02446},
	language = {en},
	number = {44},
	journal = {Langmuir},
	author = {Uebbing, Lukas and Ziller, Antje and Siewert, Christian and Schroer, Martin A. and Blanchet, Clement E. and Svergun, Dmitri I. and Ramishetti, Srinivas and Peer, Dan and Sahin, Ugur and Haas, Heinrich and Langguth, Peter},
	month = nov,
	year = {2020},
	pages = {13331--13341},
}

@article{david_interpretation_1987,
	title = {Interpretation of micromixing effects on fast consecutive-competing reactions in semi-batch stirred tanks by a simple interaction model},
	volume = {54},
	issn = {0098-6445, 1563-5201},
	doi = {10.1080/00986448708911913},
	language = {en},
	number = {1-6},
	journal = {Chemical Engineering Communications},
	author = {David, René and Villermaux, Jacques},
	month = may,
	year = {1987},
	pages = {333--352},
}

@article{zhang_interfacial_2021,
	title = {Interfacial structure and tension of polyelectrolyte complex coacervates},
	volume = {54},
	issn = {0024-9297, 1520-5835},
	doi = {10.1021/acs.macromol.1c01809},
	language = {en},
	number = {23},
	journal = {Macromolecules},
	author = {Zhang, Pengfei and Wang, Zhen-Gang},
	month = dec,
	year = {2021},
	pages = {10994--11007},
}

@article{kessel_interactions_2001,
	title = {Interactions of cholesterol with lipid bilayers: {The} preferred configuration and fluctuations},
	volume = {81},
	issn = {00063495},
	shorttitle = {Interactions of {Cholesterol} with {Lipid} {Bilayers}},
	doi = {10.1016/S0006-3495(01)75729-3},
	language = {en},
	number = {2},
	journal = {Biophysical Journal},
	author = {Kessel, Amit and Ben-Tal, Nir and May, Sylvio},
	month = aug,
	year = {2001},
	pages = {643--658},
}

@article{aliakbarinodehi_interaction_2022,
	title = {Interaction kinetics of individual {mRNA}-containing lipid nanoparticles with an endosomal membrane mimic: {Dependence} on {pH}, protein corona formation, and lipoprotein depletion},
	volume = {16},
	issn = {1936-0851, 1936-086X},
	doi = {10.1021/acsnano.2c04829},
	language = {en},
	number = {12},
	journal = {ACS Nano},
	author = {Aliakbarinodehi, Nima and Gallud, Audrey and Mapar, Mokhtar and Wesén, Emelie and Heydari, Sahar and Jing, Yujia and Emilsson, Gustav and Liu, Kai and Sabirsh, Alan and Zhdanov, Vladimir P. and Lindfors, Lennart and Esbjörner, Elin K. and Höök, Fredrik},
	month = dec,
	year = {2022},
	pages = {20163--20173},
}

@article{hassett_impact_2021,
	title = {Impact of lipid nanoparticle size on {mRNA} vaccine immunogenicity},
	volume = {335},
	issn = {01683659},
	doi = {10.1016/j.jconrel.2021.05.021},
	language = {en},
	journal = {Journal of Controlled Release},
	author = {Hassett, Kimberly J. and Higgins, Jaclyn and Woods, Angela and Levy, Becca and Xia, Yan and Hsiao, Chiaowen Joyce and Acosta, Edward and Almarsson, Örn and Moore, Melissa J. and Brito, Luis A.},
	month = jul,
	year = {2021},
	pages = {237--246},
}

@article{cui_impact_2014,
	title = {Impact of interfacial cholesterol-anchored polyethylene glycol on sterol-rich non-phospholipid liposomes},
	volume = {428},
	issn = {00219797},
	doi = {10.1016/j.jcis.2014.04.031},
	language = {en},
	journal = {Journal of Colloid and Interface Science},
	author = {Cui, Zhong-Kai and Edwards, Katarina and Orellana, Alejandro Nieto and Bastiat, Guillaume and Benoit, Jean-Pierre and Lafleur, Michel},
	month = aug,
	year = {2014},
	pages = {111--120},
}

@article{tadakuma_imaging_2006,
	title = {Imaging of single {mRNA} molecules moving within a living cell nucleus},
	volume = {344},
	issn = {0006291X},
	doi = {10.1016/j.bbrc.2006.03.202},
	language = {en},
	number = {3},
	journal = {Biochemical and Biophysical Research Communications},
	author = {Tadakuma, Hisashi and Ishihama, Yo and Shibuya, Toshiharu and Tani, Tokio and Funatsu, Takashi},
	month = jun,
	year = {2006},
	pages = {772--779},
}

@article{syamlal_hydrodynamics_1985,
	title = {Hydrodynamics of fluidization: {Prediction} of wall to bed heat transfer coefficients},
	volume = {31},
	issn = {0001-1541, 1547-5905},
	shorttitle = {Hydrodynamics of fluidization},
	doi = {10.1002/aic.690310115},
	language = {en},
	number = {1},
	journal = {AIChE Journal},
	author = {Syamlal, M. and Gidaspow, Dimitri},
	month = jan,
	year = {1985},
	pages = {127--135},
}

\end{document}